\documentclass[a4paper,11pt]{article}
\usepackage[top=3cm,bottom=3cm,left=2cm,right=2cm,a4paper]{geometry}
\usepackage{amsfonts,amsmath,amssymb,amsthm,upgreek}
\usepackage{float,fancyhdr,siunitx}
\usepackage[dvipsnames]{xcolor}
\usepackage{setspace}
\usepackage{comment}
\usepackage{lineno}
\usepackage[normalem]{ulem} 
\usepackage[colorlinks]{hyperref}
\hypersetup{
    colorlinks=true, 
    linkcolor={blue!50!black},
    citecolor={green!60!black},
    urlcolor= {red!80!black}
}
\usepackage{enumitem,mdframed}
\usepackage{empheq,longtable}
\usepackage[bottom]{footmisc} 
\usepackage{algorithm2e}
\usepackage{mathbbol}

\usepackage{tikz,pgfplots}
\usepgfplotslibrary{groupplots}
\usetikzlibrary{positioning,arrows,fit,patterns}
\pgfplotsset{compat=newest}
\usepackage{subfig}
\DeclareCaptionLabelFormat{opening}{\textbf{#2)}} 
\makeatletter
\renewcommand*\env@matrix[1][\arraystretch]{%
  \edef\arraystretch{#1}%
  \hskip -\arraycolsep
  \let\@ifnextchar\new@ifnextchar
  \array{*\c@MaxMatrixCols c}}
\makeatother

\usepackage[]{todonotes}
\usepackage{multirow}

\newcommand*\samethanks[1][\value{footnote}]{\footnotemark[#1]}

\newcommand{\ii}          {\mathrm{i}}
\newcommand{\dd}          {\mathrm{d}}
\newcommand{\bx}          {\mathbf{x}}

\newcommand{\bv}          {\boldsymbol{v}}

\newcommand{\bxi}         {\boldsymbol{\xi}}
\newcommand{\bu}          {\boldsymbol{u}}
\newcommand{\bpsi}        {\boldsymbol{\psi}}

\newcommand{\n}           {\boldsymbol{n}} 

\newcommand{\rmax}        {r_{\max}}
\newcommand{\Amat}        {\boldsymbol{A}}
\newcommand{\Bmat}        {\boldsymbol{B}}
\newcommand{\Lliouville}  {\mathcal{L}_{\mathrm{Liouville}}}
\newcommand{\Lliouvillec} {\mathcal{L}_{\mathrm{Liouville-c}}}
\newcommand{\Loriginaldiv}{\mathcal{L}_{\mathrm{odiv}}}

\newcommand{\taua}{\tau_{\mathrm{a}}}
\newcommand{\taub}{\tau_{\mathrm{b}}}
\newcommand{\tauc}{\tau_{\mathrm{c}}}

\newcommand{\mesh}{\mathcal{T}_h}
\newcommand{\ncell}{n_{\mathrm{cell}}}
\newcommand{\nface}{n_{\mathrm{face}}}
\newcommand{\nfaceloc}{n_{\mathrm{face}}^e}

\newcommand{\face}{\mathfrak{f}}
\newcommand{\setface}{\Sigma}
\newcommand{\domain}{D}
\newcommand{\hdgA}{\mathbb{A}}
\newcommand{\hdgC}{\mathbb{C}}
\newcommand{\hdgR}{\mathbb{R}}
\newcommand{\hdgG}{\mathsf{G}}
\newcommand{\hdgB}{\mathbb{B}}
\newcommand{\hdgL}{\mathbb{L}}
\newcommand{\ndof}{n_{\mathrm{dof}}}
\newcommand{\ndofface}{n_{\mathrm{dof}}^\mathfrak{f}}

\newcommand{\rsun}{R_{\odot}}
\newcommand{\robs}{r_{\mathrm{obs}}}
\newcommand{\rsrc}{r_{\mathrm{src}}}

\newcommand{\phifull}{\mathbf{Y}_{\Omega}}
\newcommand{\Rotzero}{\boldsymbol{\Omega}_c}     
\newcommand{\rotzero}{\Omega_c}                  
\newcommand{\rotvec}{\boldsymbol{\Omega}}    
\newcommand{\rot}{\Omega}                    
\newcommand{\pressL}{\delta_p^{\mathrm{L}}}
\newcommand{\pressE}{\delta_p^{\mathrm{E}}}
\newcommand{\Bhyp}{\mathbb{B}_{\mathrm{hyp}}}
\newcommand{\Bell}{\mathbb{B}_{\mathrm{ell}}}

\newcommand{\meshCa}{\texttt{Mesh}$^{001}_{\text{30k}}$}
\newcommand{\meshAa}{\texttt{Mesh}$^{001}_{\text{175k}}$}
\newcommand{\meshAb}{\texttt{Mesh}$^{001}_{\text{600k}}$}
\newcommand{\meshBa}{\texttt{Mesh}$^{01}_{\text{1200k}}$}
\newcommand{\err}{\mathfrak{e}}
\newcommand{\erra}{\err_D}
\newcommand{\errb}{\err_{\mathrm{surf}}}

\DeclareSymbolFont{yhlargesymbols}{OMX}{yhex}{m}{n} 
\DeclareMathAccent{\yhwidehat}{\mathord}{yhlargesymbols}{"62}

\usepackage{stmaryrd}

\newcommand{\modifadd}[1]{#1}

\newtheorem{definition} {Definition}

\newtheorem{remark}     {Remark}

\newtheorem{identity}   {Identity}

\usepackage[capitalize,nameinlink]{cleveref}
\crefname{section}   {Section}   {Sections}
\crefname{subsection}{Subsection}{Subsections}
\Crefname{section}   {Section}   {Sections}
\Crefname{subsection}{Subsection}{Subsections}
\Crefname{figure}    {Figure}    {Figures}

\crefformat{equation}{\textup{#2(#1)#3}}
\crefrangeformat{equation}{\textup{#3(#1)#4--#5(#2)#6}}
\crefmultiformat{equation}{\textup{#2(#1)#3}}{ and \textup{#2(#1)#3}}
{, \textup{#2(#1)#3}}{, and \textup{#2(#1)#3}}
\crefrangemultiformat{equation}{\textup{#3(#1)#4--#5(#2)#6}}%
{ and \textup{#3(#1)#4--#5(#2)#6}}{, \textup{#3(#1)#4--#5(#2)#6}}{, and \textup{#3(#1)#4--#5(#2)#6}}
\Crefformat{equation}{#2Equation~\textup{(#1)}#3}
\Crefrangeformat{equation}{Equations~\textup{#3(#1)#4--#5(#2)#6}}
\Crefmultiformat{equation}{Equations~\textup{#2(#1)#3}}{ and \textup{#2(#1)#3}}
{, \textup{#2(#1)#3}}{, and \textup{#2(#1)#3}}
\Crefrangemultiformat{equation}{Equations~\textup{#3(#1)#4--#5(#2)#6}}%
{ and \textup{#3(#1)#4--#5(#2)#6}}{, \textup{#3(#1)#4--#5(#2)#6}}{, and \textup{#3(#1)#4--#5(#2)#6}}

\crefdefaultlabelformat{#2\textup{#1}#3}

\crefname{proposition}{Proposition}{Propositions}
\Crefname{proposition}{Proposition}{Propositions}
\crefname{definition} {Definition} {Definitions}
\Crefname{definition} {Definition} {Definitions}
\crefname{theorem}    {Theorem}    {Theorems}
\Crefname{theorem}    {Theorem}    {Theorems}
\crefname{remark}     {Remark}     {Remarks}
\Crefname{remark}     {Remark}     {Remarks}
\crefname{assumption} {Assumption} {Assumptions}
\Crefname{assumption} {Assumption} {Assumptions}
\crefname{claim} {Claim} {Claims}
\Crefname{claim} {Claim} {Claims}
\crefname{identity} {Identity} {Identities}
\Crefname{identity} {Identity} {Identities}

\title{
       Numerical simulations of oscillations for axisymmetric 
       solar backgrounds with differential rotation and gravity
      }

\author{
Ha Pham\thanks{Project-Team Makutu, Inria, Universit\'e de Pau et des Pays de l'Adour, 
CNRS UMR 5142, France.}
\and
Florian Faucher\samethanks[1]
\and
Damien Fournier\thanks{Max-Planck-Institut f\"ur 
                       Sonnensystemforschung, Justus-von-Liebig-Weg 3, 
                       37077 G\"ottingen, Germany.}
\and
H\'el\`ene Barucq\samethanks[1]
\and
Laurent Gizon\samethanks[2]
             \thanks{Institut f\"ur Astrophysik und Geophysik, 
                     Georg-August-Universit\"at G\"ottingen, 
                     Friedrich-Hund-Platz 1, 37077 G\"ottingen, Germany.}
\samethanks[1]
}

\numberwithin{equation}{section}

\begin{document}
\maketitle 

\begin{abstract}

Local helioseismology comprises of imaging and inversion techniques
employed to reconstruct the dynamic and interior of the Sun from 
correlations of oscillations observed on the surface, all of which 
require modeling solar oscillations and computing Green's kernels.
In this context, we implement and investigate the robustness of 
the Hybridizable Discontinuous Galerkin (HDG) method  in solving the 
equation modeling stellar oscillations for realistic solar backgrounds 
containing acoustic attenuation, gravity, and differential rotation. 
While a common choice for modeling stellar oscillations is the Galbrun's equation,
our working equations are derived from an equivalent variant, 
involving less regularity in its coefficients,
working with Lagrangian displacement and pressure perturbation 
as unknowns.
Under differential rotation and axisymmetric assumption,
the system is solved in azimuthal decomposition with 
the HDG method. 
Compared to no-gravity approximations,
the mathematical nature of the wave operator is now linked to the profile 
of the solar buoyancy frequency $N$ which encodes gravity, and
leads to distinction into regions of elliptic or hyperbolic behavior 
of the wave operator at zero attenuation.
While small attenuation is systematically included to
guarantee theoretical well-posedness,
the above phenomenon affects the numerical solutions 
in terms of amplitude and oscillation pattern, and 
requires a judicious choice of stabilization.
We investigate the stabilization of the HDG discretization 
scheme, and demonstrate its importance to ensure the accuracy 
of numerical results, which is shown to depend on 
frequencies relative to  $N$, and on the position of 
the Dirac source. 
As validations, the numerical power spectra reproduce accurately 
the observed effects of the solar rotation on acoustic waves.

\end{abstract}
\newpage \tableofcontents 
\section{Introduction}
This work considers the time-harmonic equations modeling 
linear adiabatic stellar oscillations in axisymmetric 
backgrounds containing acoustic attenuation, differential rotation and gravity 
in Cowling's approximation. 
It is carried out in the context of local helioseismology
which aims to reconstruct the interior and dynamics of the Sun 
from observed oscillations in the photosphere, cf. \cite{gizon2010local}. 
The Green's kernels of the associated wave operator play a crucial role
in the forward modeling of helioseismic data and inversion for solar 
parameters, cf. \cite{Birch2004,Boening2017},
and in fact in a
larger family of passive-imaging techniques,
cf. \cite{gouedard2008cross}. 
We compute Green's kernels with realistic solar backgrounds
containing gravity and rotation, providing  validations of the 
numerical results with observables.
Realistic background parameters make appear 
interesting mathematical properties associated with the specificity 
of solar physics, especially due to the profile of its buoyancy 
frequency $N$ which is non-zero when gravity is considered,
and has repercussion on the numerical implementation.

\paragraph{Observing the Sun} The history of helioseismology starts with the now well-known 5-min oscillations of the Sun discovered
by Leighton, Noyes and Simon in 1962 \cite{leighton1962velocity} and observed in the Doppler velocity time series,
see, e.g., the reviews \cite[Figure 2]{gizon2010local}, \cite[Figure 3]{howe2009solar}.
Power spectra of these time series contain rich information of solar oscillation,
cf. \cite{kosovichev2025structure,howe2009solar}:
the power of the spectrum 
is mostly concentrated in the
acoustic\footnote{The dominant restoring force in this frequency range is the 
                  pressure gradient, cf. \cref{hydrodyn::eqn} for the equation 
                  of motion.} 
band (1.5--5.3mHz) along ridges in
the $(\omega,\ell)$-space, cf. \cite[Figure 2,4]{kosovichev2025structure}.
From these ridges which show the standing wave nature of oscillation, 
characteristic frequencies are extracted and
labelled as $\omega_{n\ell m}$ with
radial number $n$, spherical harmonic degree $\ell$, 
and azimuthal order $m$, cf. \cite[Figure 6]{kosovichev2025structure} and \cite[Figure 6]{howe2009solar}.

Comparing to the observation of oscillations, that of solar
rotation has a longer history starting with sun-spot tracking
from which the surface rate of rotation could be estimated;
however it is helioseismology that discovered the structure 
of solar internal rotation \cite{Schou1994}.
The effect of rotation on the acoustic waves 
leads to a dependence on $m$ of the characteristic frequencies, 
called mode $m$ splitting.
This phenomenon can be observed on the $(\omega,m)$-space spectrum 
for a fixed $\ell > 0$, with ridges being curved instead of straight 
(the latter corresponding to no rotation), 
cf. \cite[Figure 7]{howe2009solar} and \cite[Figure 5]{kosovichev2025structure}.
In fact, the curvatures of these ridges characterize quantitatively rotation,
and will be employed to validate our numerical power spectrum with 
the observed one from \cite{Larson2018}.

\paragraph{Modeling oscillations}
Starting with a hydrodynamic description of a stellar body as an 
ideal fluid state, and in adopting a mixed Eulerian-Lagrangian 
perspective\footnote{See \cite{Pham2024stabilization} for explanation 
                     and references which discuss leaving out other effects 
                     such as magnetics, as well as the mixed framework description.}, 
stellar oscillations are encoded in the 3D vector variable $\bxi$ which
 represents small displacements at each Eulerian position, by comparing
the flow in a state of interest considered as perturbed to a (virtual) 
reference state.
The variable $\bxi$ plays an important role in the modeling of solar observables 
including Doppler line-of-sight velocity and intensity \cite{Fournier2025}.
There are several manners to derive the equation of 
motion\footnote{See also \cite{legendre2003rayonnement,rouxelin2021mixed} 
                for a different perspective of derivation.}, 
leading to equivalent variants appearing in literature, e.g., 
\cite{lynden1967stability} and \cite{gough1990effect}. 
The equations considered in \cite{lynden1967stability} are 
also known as Galbrun's equations \cite{Galbrun1931} in 
aeroacoustics, cf. the references in \cite{Pham2024assembling}. 
Although the equations in \cite{lynden1967stability} and \cite{gough1990effect} 
are equivalent to each other for backgrounds in hydrostatic equilibrium, the 
coefficients of the equations in \cite{lynden1967stability}
require more regularity (specifically it uses the Hessian instead of 
gradient of background pressure and gravity potential),
see detailed discussions in \cref{LBOlit::rmk,Goughlit::rmk,Galbrun::rmk}.

In the acoustic frequency range, the contribution of gravity is important 
and needs to be included in the modeling equation to get closer to the observations. It allows to model the surface-gravity modes that are absent in the no-gravity
approximation\footnote{We  recall that gravity is ignored to
                       work with convected Helmholtz-type equation, 
                       see \cite[section 2.1]{Gizon2017},
                       with flow and rotation encoded in the first-order 
                       convection, see also
                       \cite{barucq2023construction,chabassier2016high,Mueller2024} 
                       for numerical discretization, \cite{Pham2020Siam} in radial symmetry, 
                       and \cite{Yang2023,gizon2020meridional} for applications.}
\cite[Figure 10]{Gizon2017},
and to represent more accurately the location of the  peaks in the power spectra \cite[Figures 11 and 12]{Pham2024stabilization}.
The effects of gravity is also shown in this work to give rise to interesting numerical
phenomena\footnote{We refer to \cite{Pham2024stabilization} for further discussion 
                   of effects in terms of absorbing boundary condition.}.
In fact, the inclusion of gravity is standard, 
(see, e.g., the monographs \cite{unno1979nonradial,Gough1993}), 
as included in two reference codes ADIPLS \cite{JCD2008} 
and GYRE \cite{Townsend2013} routinely employed in the stellar 
community. These codes are eigensolvers for radially symmetric background 
and the resulting eigenfunctions can be used to obtain the Green's 
kernels \cite{Birch2004,Boening2017}. This approach is only possible 
when the spectrum is discrete and the background medium is spherically symmetric.

In this work, we consider rotation which results in an axisymmetric background. 
In this case, the problem decouples into 2D equations for each azimuthal order $m$. 
This setup is interesting to study, for example, rapidly rotating stars \cite{Reese2009},
or inertial modes as background rotation needs to be considered in this case \cite{Gizon2021}.
We also mention \cite{chabassier2018solving} that considers the stellar 
oscillation in form of Galbrun's equation, although it employs different 
discretization method and equation, and more importantly does not consider 
realistic solar backgrounds.


\paragraph{Highlights of methodology and results}
Breaking with tradition from our previous works \cite{Pham2024assembling,Pham2021Galbrun}
which employ Galbrun's equations (for radially symmetric background),
our working equation is derived from a formulation closer (and equivalent) 
to that in \cite{gough1990effect}, which has as unknowns the vector 
displacement $\bu$  and the pressure perturbation $w$ and is written as,
see \cref{sect:oscillations},
\begin{subequations}\label{eq:intro}\begin{empheq}[left={\empheqlbrace}]{align}
  \Amat \, \bu \,+\, \boldsymbol{\beta}_{1} \, w \,+\, \nabla w \,=\, \mathbf{g}, \\
  \nabla\cdot \, \bu \,+\, \boldsymbol{\beta}_{2} \,\cdot\,\bu \,+\,\varrho \, w \,=\, h.
\end{empheq}\end{subequations}
Here, the operator $\Amat$ contains two orders of material derivative 
$(\mathbf{v}_0\cdot \nabla)$ with $\mathbf{v}_0$ representing the 
reference background flow velocity. The vector-valued functions 
$\boldsymbol{\beta}_{1,2}$ and scalar function $\varrho$ are 
multiplication operators.
Their value and property are determined by a choice of background model. 
The final working equation is obtained after a Liouville change of variable 
to handle solar stratification near the surface, cf. \cref{Liouville::subsec}.
Under axisymmetric assumptions, the problem is solved in azimuthal 
decomposition (called 2.5D problem) with the Hybridizable Discontinuous 
Galerkin (HDG) method. Our choice of discretization method follows the 
HDG framework of \cite{Faucher2020adjoint,Pham2024stabilization} in the 
perspective of Cockburn et al. \cite{Cockburn2009}. 
We refer to \cite{Pham2024stabilization} and the references listed therein 
for discussion of the advantages of the method for large-scale problems.

On each azimuthal mode $m$, the differential operator $\Amat$ 
acts simply as a multiplication operator and is denoted by $\Amat_m$.
With $\nabla_m\cdot$ and $\nabla_m$ representing the action of the 
divergence and the gradient in this decomposition, there is a connection 
with a second-order scalar equation whose highest order terms 
is  $\nabla_m\cdot \Amat_m^{-1}\nabla_m$, 
cf. \cref{subsection:scalar-pde,cylinexpansion_GK::subsec}.
This crucial fact facilitates carrying over the HDG discretization 
framework of \cite{Faucher2020adjoint,Pham2024stabilization,nguyen2009implicit}
to the current equation, specifically in formulating the local 
boundary value problems and their boundary data, both of which 
constitute key steps in formulating the HDG problem, 
cf. \cref{section:axi-symmetric-hdg}.
While the generic form of the working system 
of equations resembles that of the scalar convection--diffusion 
equation (e.g., \cite{nguyen2009implicit}),
its nature is different when the coefficients follow realistic 
solar models, specifically due to gravity which is represented 
in the non-zero buoyancy frequency, $N$, \cite{christensen2014lecture}.

This frequency plays a role in determining the invertibility 
of $\Amat$ and $\Amat_m$ (or ellipticity in view of the scalar PDE) 
at zero attenuation,  e.g., ellipticity is lost when $N^2 > 0$ and 
for frequencies above $N$, and is restored at non-zero attenuation.
While the numerical resolutions are  always carried out with small attenuation 
assumption, the solutions in these region behave differently, featuring zigzag 
patterns associated with propagation of singularities, cf. discussion in \cref{sect:numerical}.
The work will also show that, in all cases,  
the solutions in zero-gravity approximation are very different,
both in magnitude and oscillation patterns, and do not have this 
relation with $N$ (as $N=0$ without gravity).
We also note that the well-posedness of the equation of motion in $\bxi$ 
is proved under the assumption of attenuation in \cite{halla2021well} 
in form of Galbrun's equation and for generic settings, in particular,
without indicating the link between the frequency $\omega$ and $N$.

\paragraph{Organization}
The discussion of the investigation is organized as follows.
The oscillation equations are derived from first principles in \cref{sect:oscillations} and the working formulation is given in \cref{workingbackgrounds::subsec}. 
The axisymmetric equations are obtained in \cref{section:axi-symmetric}, 
and the discretization using the HDG method is carried out in \cref{section:axi-symmetric-hdg}. We solve numerically for the solar backgrounds in \cref{sect:numerical} and provide 
a qualitative analysis on the choice of the stabilization. 
Finally, the numerical solutions are converted onto helioseismic observables 
and validated in this context in \cref{section:numerical-helio}.
The experiments are carried out with software \texttt{Hawen}~\cite{Hawen2021}. 
For reproducibility, the setups for the numerical experiments, including the 
solar background models employed as well as the parameter files and scripts, 
are available in the online Zenodo repository, \url{https://doi.org/10.5281/zenodo.21297984}.


\section{Time-harmonic oscillation equations} 
\label{sect:oscillations}
Linear solar oscillations are modeled as small-amplitude oscillatory motions
in a moving fluid which are caused by perturbations in the state of the fluid
compared to a reference configuration. 
The derivation of the oscillation equation starts with 
the hydrodynamic description of the Sun as an 
adiabatic fluid in self-gravitation and 
rotation around axis $\mathbf{e}_z$, listed in \cref{refstate::subsec}.
Perturbations are then introduced in a mixed Eulerian-Lagrangian 
perspective, cf. \cref{perturb::subsec}.
Under appropriate assumptions, linearization is performed
to obtain linear equations describing small perturbations, cf. \cref{perturb::subsec}.
From these linear equations, the oscillation equations are derived in  \cref{eqnmot::subsec}, 
in terms of the displacement 
$\boldsymbol{\upxi}$ defined at a fixed position (i.e., Eulerian) which
 represents the small difference in position of a particle 
carried by the perturbed flow versus the reference one.
After introducing the working equations of oscillations, 
we will also introduce the scalar Green's kernel $G_p$
in \cref{scalarppGKgen}, which forms a key ingredient in modeling observables.

There are several approaches to arrive at the final oscillation equations,
for a review we refer to \cite[Section 1.7]{legendre2003rayonnement}.
We note two approaches commonly found in literature:
The first one is to introduce Eulerian perturbations in the
conservation laws written in Eulerian form and
then linearize, as was considered in \cite{godin1997reciprocity}, see also \cite{hagg2021well}
for a more mathematical discussion.
In the second approach, one works with Lagrangian perturbations of the conservation 
laws on which form linearization is performed; after this, to obtain the equations 
satisfied by the Eulerian perturbations and displacement, the 
relation between Lagrangian and Eulerian perturbations are employed, cf. \cite{brazier2001eulerian,legendre2003rayonnement}.
Another approach which works with Lagrangian perturbations is given 
in \cite{lynden1967stability}, and adopted 
in \cite[Section 2.2]{halla2021well} to provide a derivation of the oscillation equations.
In our work, we adopt a hybrid approach: we follow closely the rigorous derivation given in \cite{legendre2003rayonnement}
for the perturbation of the conservation of mass, energy and equation of state, while for the conservation 
of momentum, we adopt the first approach, i.e. working with Eulerian perturbations and linearization, for its simplicity.

Although the first approach is much simpler and provide readily
linear equations via Euler perturbation analysis of the conservation of mass, momentum and energy, 
it does require assumption in integrating the time material derivative to arrive at two crucial ingredients 
needed for the equation of oscillations: the relation between the perturbations 
of the pressure and density in terms of the displacement, 
listed below in \cref{relxideltaE_rhop}. This is discussed in \cite{hagg2021well} regarding the conservation of mass and \cite{godin1997reciprocity} concerning conservation of energy, see also
discussion in \cite{rouxelin2021mixed}. We discuss this briefly in \cref{Nores::rmk}.
We also note that the derivation of the 
oscillation equation are mostly obtained in physics and astrophysics literature for non-moving background (i.e with
$\mathbf{v}_0 = 0$ see notation in \cref{backgroundpar}), cf. \cite{unno1979nonradial, priest2014magnetohydrodynamics,pringle2007astrophysical,christensen2014lecture}. 
In this case, the material time derivative is simply the time derivative which 
allows for normal integration and avoid the problem listed in \cref{Nores::rmk}, 
known as the no-resonance assumption.
A second problem with adopting a pure Eulerian
approach arises in the perturbation analysis of the equation of state which provides relations between the thermodynamics quantities (density, pressure, entropy, temperature etc), cf. \cref{eqn_state_Eul} and \cref{eqn_state_Lag}.
These are Lagrangian in nature, which 
 means that the perturbations of the equation of state are first obtained in terms of Lagrangian 
perturbations, cf. \cite[Section 13.4 p.98]{unno1979nonradial}.

The notation and symbols used 
throughout the document are summarized in \cref{table:notation}.

\begin{table}[ht!] 
\begin{center} 
\caption{List of the main symbols and notations used in this work.}
\vspace*{-0.50em} 
\label{table:notation} 

\renewcommand{\arraystretch}{1.20} 

\begin{minipage}[t]{17em}
\vspace*{0pt}

\begin{tabular}{|>{\centering\arraybackslash}p{1.2em}|
                 >{\arraybackslash}p{15em}|
                 }
\hline
$\bxi$     &  Lagrangian displacement\\
$\pressL$  &  Lagrangian pressure perturbation\\
$\pressE$  &  Eulerian   pressure perturbation\\
$\rho_0$   & Density\\
$\phi_0$   & Gravitational potential\\
$c_0$      & Adiabatic sound speed\\
$p_0$      & Fluid pressure\\
$\Gamma_1$ & Adiabatic index\\
$\omega$   & Angular frequency\\
$\gamma_{\mathrm{att}}$ & Attenuation coefficient\\
$\sigma$   &  Complex frequency, such that \newline
              $\sigma^2(\bx)=\omega^2 +2\ii\omega\gamma_{\mathrm{att}}(\bx)$\\
\hline
\end{tabular}
\end{minipage}\hspace*{2em}\begin{minipage}[t]{22em}
\vspace*{0pt}

\begin{tabular}{|>{\centering\arraybackslash}p{1.2em}|
                 >{\arraybackslash}p{20em}|
                 }
\hline
$\bxi_L$ & Liouville unknown $\bxi_L = \sqrt{\rho_0} \, \bxi$\\
$w_L$    & Liouville unknown $w_L  = \pressE/\rho_0$\\
$\bxi_c$ & Liouville-c unknown $\bxi_c = c_0 \sqrt{\rho_0}\,  \bxi$\\
$w_c$    & Liouville-c unknown $w_c  = \pressE/(c_0 \rho_0)$\\
$\Rotzero$ &  Angular velocity vector of the rotating reference frame, 
              $\Rotzero = \rotzero \, \mathbf{e}_z$\\ 
$\rotvec$  &  Internal solar rotation in an inertial frame, $\rotvec= \rot(r,\theta)\, \mathbf{e}_z$\\
$\mathbf{v}_0$ & Background flow velocity, 
                $\mathbf{v}_0 = (\rotvec - \Rotzero) \times \mathbf{x}$\\
$N$        & Buoyancy frequency/Brunt-V\"ais\"al\"a frequency\\
$\phifull$ & Auxiliary quantity defined in \cref{phifull::def_pre} and \cref{phifull::deffullrot}\\
$\Bmat_0^{\Omega}$ & Auxiliary quantity defined in \cref{Bmat::def} \\
$\boldsymbol{\alpha}_\bullet$ & inverse scale-height vector: for a scalar function $g$, 
                                $\boldsymbol{\alpha}_g=-\nabla g / g$ \\
\hline
\end{tabular}
\end{minipage}
\end{center}
\end{table}

\medskip

\paragraph{Notations} 
We introduce the notations that appear frequently in subsequent discussion.
\begin{itemize}[leftmargin = *]
\item All of the equations in this work are considered on a bounded 
domain denoted by $\mathbb{B}_{\odot}$ which represents 
the Sun up to a certain height in the photosphere.
\item 
The material derivative operator acting on tensors and in the direction of vector $\mathbf{w}$ is denoted by
$(\mathbf{w}\cdot \nabla)$ (or simply written
without the parenthesis). We list its action on scalar functions and vectors: for a  scalar function $h$, this is the directional derivative 
along $\mathbf{w}$, and is equivalently defined in terms of the gradient operator $\nabla$,
\begin{equation}
(\mathbf{w}\cdot \nabla) h = \mathbf{w}\cdot (\nabla h).
\end{equation} 
For a vector $\tilde{\mathbf{w}}$, its definition in Cartesian basis is given by, 
\begin{equation}\label{matdervec::def} [ (\mathbf{w}\cdot  \nabla) \tilde{\mathbf{w}}]_{\text{Cartesian}}
 = ( \mathbf{w}\cdot \nabla \tilde{w}_i )_{i=\text{`x',`y',`z'}}, \quad \text{where} \hspace*{0.3cm} [\tilde{\mathbf{w}}]_{\text{Cartesian}}=(\tilde{w}_i)_{i=\text{`x',`y',`z'}}.
\end{equation}
Another equivalent definition of \cref{matdervec::def} is in terms of a product between a matrix 
$\nabla \tilde{\mathbf{w}}$ and a vector $\mathbf{v}$:
\begin{equation}
(\mathbf{w}\cdot\nabla) \tilde{\mathbf{w}} =
(\nabla \tilde{\mathbf{w}}) \cdot \mathbf{w}, \quad \text{ where} \hspace*{0.3cm} [\nabla \tilde{\mathbf{w}} ]_{\text{Cartesian}}
 =  \left(\partial_j\tilde{w}_{i}\right)_{i,j=\text{`x',`y',`z'}}.
\end{equation}

\item We introduce the \emph{inverse scale height} operator, 
defined for a scalar quantity $g\neq 0$, as:
\begin{equation}\label{scaleheight::def}
  \boldsymbol{\alpha}_g(\mathbf{x}) :=  - \dfrac{ \nabla g(\mathbf{x})}{g(\mathbf{x})} \,, \qquad \qquad \text{inverse scale height vector}.
\end{equation}
For functions only depending on the radial variable $r$, 
we work with the scalar inverse scale height $\alpha$, with $\mathbf{e}_r$ the radial vector,
\begin{equation}\label{scaleheightradial}\boldsymbol{\alpha}_g =  \alpha_g \,\mathbf{e}_r, \quad \text{with } 
\alpha_g(r):=-\dfrac{\partial_r g(r)}{g(r)}.
\end{equation}


\end{itemize}
\subsection{Hydrodynamics description: perturbed vs reference states}\label{refstate::subsec} 

\paragraph{Perturbed state}
The mathematical description of a perturbed fluid state is
characterized by a set of (state) parameters:
\begin{equation}\label{idealstate} \begin{aligned}
& \makebox[13em][l]{$- \,\, \text{the fluid density }  \rho$,}
\makebox[13em][l]{$- \,\, \text{the flow velocity } \mathbf{v}$,} 
\makebox[13em][l]{$- \,\, \text{the fluid pressure } p$,} \\
& \makebox[13em][l]{$- \,\, \text{the specific entropy } \frak{s}$,} 
\makebox[13em][l]{$- \,\, \text{the gravitational potential } \Phi$.}
\end{aligned} \end{equation}
In a rigidly co-rotating observation frame rotating
at rate $\rotzero$, these parameters \cref{idealstate} satisfy the conservation 
of mass, momentum and energy equations, cf. \cite[Eqs (13.1--13.3)]{unno1979nonradial}
\begin{subequations}\label{hydrodyn::eqn}
\begin{empheq}[left={\empheqlbrace\,}]{align}
 & \partial_t \rho + \nabla\cdot (\rho\,  \mathbf{v}) \, = \,  0  \,, \label{hydrodyn_mass::eqn}\\[0.5em]
 & \rho \left( \partial_t \mathbf{v}\, + \, (\mathbf{v} \cdot \nabla)  \mathbf{v} \right) \, = \, -\nabla p \,- \, \rho \nabla \Phi \, - \, 2 \, \rho \, \Rotzero \times \mathbf{v}
 \, - \, \rho \, \Rotzero \times \Rotzero \times \mathbf{x}
 \, + \, \mathfrak{F} \,,\label{cons_momentum::eqn}\\[0.5em]
&   \partial_t  \frak{s} + (\mathbf{v} \cdot \nabla)   \frak{s}  \, =\, 0 . \label{cons_energy_pert}
\end{empheq}
\end{subequations}
The potential $\Phi$ is related to $\rho$
via $\Delta \Phi = 4\pi G \rho$ where $G$ is the gravitational constant.
The thermodynamic quantities $\rho$, $p$, $\frak{s}$ are related by the equation of state
which is of Lagrangian in nature and is listed below in \cref{eqn_state_Lag}.
In \cref{hydrodyn::eqn}, $\Rotzero = \rotzero \, \mathbf{e}_z$.
The term $ \Rotzero \times \Rotzero \times \mathbf{x}$ is 
called the centrifugal force, while 
$\Rotzero \times \mathbf{v}$ is the Coriolis force.
The external source term is represented by $\mathfrak{F}$.
The equation \cref{cons_energy_pert} defines an adiabatic flow, cf. \cite[Eq. (2.5)]{landau1987fluid}, \cite[Eq. (13.3)]{unno1979nonradial}.

\begin{remark}[Centrifugal potential]\label{Centripot::rmk}
  The centrifugal term $ \Rotzero \times \Rotzero \times \mathbf{x}$
  can be written 
  as\footnote{This identity also holds for non-constant rotation.  
  For a general differential rotation 
  rate $\rotvec=\rotvec(r,\theta) = \rot(r,\theta) \mathbf{e}_z$ 
  around axis $\mathbf{e}_z$, we have that
  $\rotvec \times\rotvec  \times \mathbf{x} = \rotvec  (\rotvec \cdot \mathbf{x})
 \,- \,\mathbf{x} \,\lvert\rotvec \rvert^2\, =\,  (\rot z)\,\rotvec\,- \,\rot^2 \,\mathbf{x}$. 
This follows from the identity: for vector $\mathbf{a}$, $\mathbf{b}$, and $\mathbf{c}$,
we have
$\mathbf{a}\times (\mathbf{b}\times \mathbf{c})  = \mathbf{b} (\mathbf{a}\cdot \mathbf{c}) - \mathbf{c} ( \mathbf{a}\cdot \mathbf{b})$. 
}

\begin{equation}
\Rotzero \times \Rotzero \times \mathbf{x} = \rotzero (\rotzero\cdot \mathbf{x}) -  \rotzero^2\mathbf{x}.
\end{equation}
As $\rotzero$ is constant, the centrifugal force $ \Rotzero \times \Rotzero \times \mathbf{x}$
can be written as a gradient of a scalar potential: 
\begin{equation}\label{centripot::def}
\Rotzero \times \Rotzero \times \mathbf{x}\,\, =\,\,\nabla_{\mathbf{x}} \psi^S, \hspace*{1cm} \text{with } \quad \psi^S := - \dfrac{1}{2} \left( \rotzero^2 \lvert \mathbf{x}\rvert^2  - (\Rotzero\cdot \mathbf{x})^2 \right). 
\end{equation}
\end{remark}

\begin{remark}[Effective potential]\label{effpot::rmk}
With centrifugal potential $\psi^S$ introduced in \cref{Centripot::rmk}, 
one refers to $(\phi_0 + \psi^S)$ as the effective potential, and we can 
write $\nabla\phi_0 \,+\,\Rotzero\times\Rotzero\times\bx
                    = \nabla(\phi_0 + \psi^S)$ .
\end{remark}

\paragraph{Reference state}
The parameters characterizing an adiabatic reference state are denoted 
with a subscript $0$:
\begin{equation}\label{backgroundpar}
\begin{aligned}
& \makebox[13em][l]{$- \,\, \text{the density } \rho_0$,}
\makebox[13em][l]{$- \,\, \text{the flow velocity } \mathbf{v}_0$,} 
\makebox[13em][l]{$- \,\, \text{the pressure } p_0$,} \\
& \makebox[13em][l]{$- \,\, \text{the specific entropy }\frak{s}_0$.}
\makebox[13em][l]{$- \,\, \text{the gravitational potential }\phi_0$.}
\end{aligned} 
\end{equation}
As in \cref{idealstate},  $(\rho_0, \mathbf{v}_0,\frak{s}_0)$ satisfy 
the conservation of mass, momentum and energy,
\begin{subequations}\label{hydrodyn::eqn_background}
\begin{empheq}[left={\empheqlbrace\,}]{align}
 & \partial_t \rho_0 + \nabla\cdot (\rho_0\,  \mathbf{v}_0) \, = \,  0  \,,\label{hydrodyn_mass0::eqn} \\[0.5em]
 & \rho_0 \left( \partial_t \mathbf{v}_0\, + \, (\mathbf{v}_0 \cdot \nabla)  \mathbf{v}_0 \right) \, = \, -\nabla p_0 \,- \, \rho \nabla \Phi_0 \, - \, 2 \, \rho_0 \, \Rotzero \times \mathbf{v}_0
 \, - \, \rho_0 \, \Rotzero \times \Rotzero \times \mathbf{x}
 \, + \, \mathbf{f}_\mathrm{ghost} \,,\\[0.5em]
&   \partial_t  \frak{s}_0 + (\mathbf{v}_0 \cdot \nabla)   \frak{s}_0  \, =\, 0 . \label{cons_energy_pert0}
\end{empheq}
\end{subequations}
This is coupled with the equation of state, cf. \cref{eqn_state_Lag}. The potential $\phi_0$ satisfies the Poisson equation with source $\rho_0$, listed below in \cref{poisson_eqn_phi}. 
Here, we allow for the contribution of an external 
source $\mathbf{f}_\mathrm{ghost}$, cf. also 
\cref{fghost,rolefghost::rmk} for further discussion.
The coefficients of the oscillation equations also employ adiabiatic soundspeed $c_0$ and adiabatic index $\Gamma_1$ emerging from the thermodynamic relations
and perturbation of the equation of states which are introduced below in \cref{adiaba}. 

\subsection{Small perturbation analysis}
\label{perturb::subsec}

Perturbations can be formulated from an Eulerian or a Lagrangian point of view.
Eulerian approach measures changes between the two states at each fixed coordinate, 
while the Lagrangian approach describes changes associated
with a fixed collection of virtual particles in the course of their movement by the flow.
In the following discussion, we follow the notation and formalism in \cite[Chapter 1]{legendre2003rayonnement};
 the same idea could be found in \cite[Section 4.3]{pringle2007astrophysical} or \cite{brazier2001eulerian}.

\begin{figure}[ht!]
\begin{center}
\includegraphics[scale=0.85]{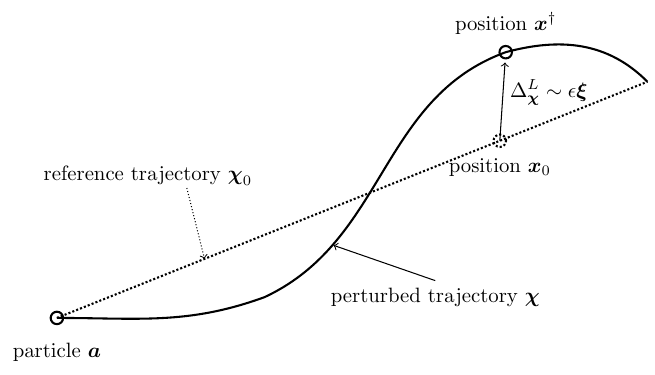}
\caption{Representation of the reference flow $\boldsymbol{\chi}_0$ and perturbed flow $\boldsymbol{\chi}$ and displacement $\Delta^L_{\boldsymbol{\chi}}$ defined in \cref{Ldisp::def}.}\label{EL::fig}
\end{center}
\end{figure}

\noindent $\bullet$ 
Denote the trajectory of a virtual particle $\mathbf{a}$ in the reference flow by $\boldsymbol{\chi}_0$, 
and the perturbed one by $\boldsymbol{\chi}$, represented in \cref{EL::fig}. 
The position of the particle $\mathbf{a}$ at time $t$
on the reference flow $\boldsymbol{\chi}_0$ and 
on the perturbed flow $\boldsymbol{\chi}$ are denoted respectively by, 
\begin{equation}\label{ref_pert_pos}
\mathbf{x}_0=\boldsymbol{\chi}_0( \mathbf{a},t) , \quad \mathbf{x}^{\dagger}=\boldsymbol{\chi}( \mathbf{a},t).
\end{equation} 
The velocities of the two trajectories are respectively denoted by,
\begin{equation}\label{ref_pert_vel}
\mathbf{v}_0(\mathbf{a},t) := \partial_t \boldsymbol{\chi}_0( \mathbf{a},t), \quad
\mathbf{v}(\mathbf{a},t) := \partial_t \boldsymbol{\chi}( \mathbf{a},t).
\end{equation}

\noindent $\bullet$  Following \cite{legendre2003rayonnement}, the notation $\tilde{\cdot}$ 
is employed to rewrite a quantity 
in Lagrangian coordinates $(\mathbf{a},t)$ to Eulerian coordinates $(\mathbf{x},t)$ 
which are related by corresponding flows. For example, for a quantity $f$ and 
in association with the reference flow $\boldsymbol{\chi}_0$, we write
\begin{equation}
f(\mathbf{x},t) = \tilde{f}(\mathbf{a},t), \quad \text{ where } \mathbf{x} = \boldsymbol{\chi}_0(\mathbf{a},t).
\end{equation}
The equation of state relating pressure, density and entropy of the perturbed states \cref{idealstate} and reference states \cref{backgroundpar} are given in Lagrangian coordinates, i.e. cf. \cite[Eq. (1.12), (1.17)]{legendre2003rayonnement}, 
\begin{equation}\label{eqn_state_Eul}
p = \mathcal{P}(\rho , \frak{s};\mathbf{a}) , \hspace*{0.3cm} p_0 = \mathcal{P}(\rho_0 , \frak{s}_0;\mathbf{a}), 
\end{equation}
or equivalently, cf. \cite[Eq. (1.28), (1.32)]{legendre2003rayonnement}, 
\begin{equation}\label{eqn_state_Lag}
\tilde{p} = \mathcal{P}(\tilde{\rho}, \tilde{\frak{s}};\mathbf{a}), \hspace*{0.3cm} \tilde{p}_0 = \mathcal{P}(\tilde{\rho}_0, \tilde{\frak{s}}_0;\mathbf{a}).
\end{equation}

\noindent $\bullet$ Displacement vector:  Associated with a virtual particle $\mathbf{a}$, 
the Lagrangian displacement $\tilde{\Delta}^L_{\boldsymbol{\chi}}$ measures the difference between the two positions in \cref{ref_pert_pos}, while the 
difference in the particle velocities in \cref{ref_pert_vel} are denoted by $\tilde{\Delta}^L_{\mathbf{v}}$, see also \cref{EL::fig}, 
\begin{equation}\label{Ldisp::def}
\tilde{\Delta}^L_{\boldsymbol{\chi}}(\mathbf{a},t) := \boldsymbol{\chi}( \mathbf{a},t) - \boldsymbol{\chi}_0( \mathbf{a};t)  , 
\hspace*{0.5cm} \tilde{\Delta}_{\mathbf{v}}^L(\mathbf{a},t) := \mathbf{v}(\mathbf{a},t)  - \mathbf{v}_0(\mathbf{a},t).
\end{equation}
We introduce the quantity $\boldsymbol{\upxi}$ defined at a fixed position (i.e., Eulerian) in the 
observation frame to represent small amplitude $\tilde{\Delta}^L_{\boldsymbol{\chi}}$:
\begin{equation}\label{dispxi::def}
\begin{gathered}
\tilde{\Delta}^L_{\boldsymbol{\chi}}(\mathbf{a},t) \, =\, \Delta^L_{\boldsymbol{\chi}} (\mathbf{x},t) \, =\,  \mathbf{x}^{\dagger}- \mathbf{x}
\, =\,  \epsilon\,  \boldsymbol{\upxi} (\mathbf{x}_0,t) \, + \, \mathsf{O}(\epsilon^2) ,\\ 
\hspace*{1cm} \text{ where } \quad
\mathbf{a} = \boldsymbol{\chi}_0^{-1}(\mathbf{x},t), \quad \mathbf{x}^{\dagger} = \boldsymbol{\chi}(\mathbf{a},t).
\end{gathered}
\end{equation}
In the same manner, we introduce the small Lagrangian perturbation vector denoted by $\updelta^L_{\mathbf{v}}$, 
\begin{equation}
\tilde{\Delta}_{\mathbf{v}}^L(\mathbf{a},t) = \Delta_{\mathbf{v}}^L(\mathbf{x},t) 
 = \epsilon \updelta^L_\mathbf{v}(\mathbf{x},t) + \mathsf{O}(\epsilon^2). 
\end{equation}
We have  the relation between $\updelta^L_{\mathbf{v}}$ and $\boldsymbol{\upxi}$ as, cf. \cite[Eq. (1.60)]{legendre2003rayonnement}, 
\begin{equation}\label{Lvxi}
 \updelta^L_\mathbf{v}(\mathbf{x},t)  = ( \partial_t \, + \, (\mathbf{v}_0\cdot \nabla)) \boldsymbol{\upxi}  + \mathsf{O}(\epsilon). 
\end{equation}

\noindent $\bullet$ {Lagrangian perturbations}: 
The Lagrangian perturbation associated with a parameter $f$ describes the change of $f$ compared to a reference $f_0$ measured on the corresponding flows, 
\begin{equation}
\widetilde{\Delta^L_f}(\mathbf{a},t) := f(\boldsymbol{\chi}(\mathbf{a},t) - f_0 (\boldsymbol{\chi}_0(\mathbf{a},t)). 
\end{equation}
The small Lagrangian perturbation associated with $f$ is denoted by $\updelta^L_f$, cf. \cite[Eq. (13.20)]{unno1979nonradial},
\begin{equation}
\begin{gathered}
 \widetilde{\Delta^L_f}(\mathbf{a},t) =\Delta^L_f(\mathbf{x},t) =
f(\mathbf{x}, t)  - f_0 (\mathbf{x}_0, t) \, = \, \epsilon \, \delta^L_f( \mathbf{x}_0, t) \, +\,  \mathsf{O}(\epsilon^2),\\
\text{ where} \hspace*{0.3cm}   \mathbf{x}_0 = \boldsymbol{\chi}_0(\mathbf{a},t), \quad
\mathbf{a} = \boldsymbol{\chi}_0^{-1}(\mathbf{x}_0,t), \quad \mathbf{x} = \boldsymbol{\chi}(\mathbf{a},t) = \mathbf{x}_0 + \epsilon \, \bxi + \mathsf{O}(\epsilon^2).
\end{gathered}
\end{equation}

\noindent $\bullet$ {Eulerian perturbations}: 
For a quantity $f$, the Eulerian perturbation at position $(\mathbf{x}_0,t)$
is denoted $\updelta^E_f$, cf. \cite[Eq. (13.19)]{unno1979nonradial},
\begin{equation}
    f(\mathbf{x}_0, t)  = f_0 (\mathbf{x}_0, t) \, + \, \epsilon \, \updelta^E_f( \mathbf{x}_0, t) \, +\,  \mathsf{O}(\epsilon^2).
\end{equation}
The relation between the small Eulerian and Lagrangian perturbations are given by, 
cf. \cite[Eq. (1.63)]{legendre2003rayonnement}, or \cite[Eq. (13.21)]{unno1979nonradial}, 
\begin{equation}\label{LE_rel}
\updelta^L_f( \mathbf{x}_0, t) = \updelta^E_f( \mathbf{x}_0, t)\,  +\, \boldsymbol{\upxi}(\mathbf{x}_0,t) \cdot \nabla f_0 (\mathbf{x}_0,t) + 
\mathsf{O}(\epsilon).
\end{equation}
Use this relation \cref{LE_rel} and \cref{Lvxi} which gives 
Lagrangian perturbation $\updelta^L_\mathbf{v}$, we obtain an
expression of the Eulerian vector perturbation $\updelta^E_\mathbf{v}$ in terms of
$\boldsymbol{\upxi}$, 
\begin{equation}
\updelta^E_\mathbf{v}   = \left(\partial_t \, + \, (\mathbf{v}_0\cdot \nabla)\right) \boldsymbol{\upxi}   \, - \, (\boldsymbol{\xi} \cdot \nabla )\mathbf{v}_0.
\end{equation}


\paragraph{Perturbations of conservation laws}
We carry out Eulerian perturbations of the momentum equations and then linearize.
For the remaining conservation laws, we follow the approach detailed in \cite{legendre2003rayonnement}.
Therein, the conservation laws are rewritten in Lagrangian form and/or coordinates, the Lagrangian perturbations are introduced and linearization is performed on the resulting equations.
For the energy equation, the formalism of \cite{legendre2003rayonnement} is the same as \cite{unno1979nonradial}, from where we cite the relation among the rates of change of the thermodynamical quantities $\rho$, $p$ and $\frak{s}$. 

\medskip

\noindent $\bullet$ {Eulerian perturbation of the momentum equation}: We replace
$\mathbf{v}$, $\rho$, $p$ and source $\frak{F}$ in expansion of small Eulerian perturbations,  in
the conservation of momentum \cref{cons_momentum::eqn},
\begin{gather}
\mathbf{v} = \mathbf{v}_0 + \epsilon\, \updelta^E_{\mathbf{v}} + \mathsf{O}(\epsilon^2),
\hspace*{0.5cm} \rho = \rho_0 + \epsilon\, \updelta^E_{\rho} + \mathsf{O}(\epsilon^2),
\hspace*{0.5cm} p = p_0 + \epsilon\, \updelta^E_{p} + \mathsf{O}(\epsilon^2),\\
\mathfrak{F}(\mathbf{x},t) 
\,=\,  \textcolor{black}{\mathbf{f}_\mathrm{ghost}}(\mathbf{x},t) 
\, + \, \epsilon\, \mathsf{F}(\mathbf{x},t) + \mathsf{O}(\epsilon^2) .\label{fghost}
\end{gather}
By keeping only terms of order $\epsilon$,  we obtain the following linear equation, 
\begin{equation}\label{lin_pert_Emomentum}
   \rho_0   \left(    \partial_t   \updelta^E_\mathbf{v}  \,+\, ( \mathbf{v}_0 \cdot \nabla) \updelta^E_\mathbf{v}
  + (\updelta^E_\mathbf{v} \cdot \nabla) \mathbf{v}_0
 \, + 2\,  \Rotzero \times \updelta^E_\mathbf{v}  
  \right) \, = \, -\nabla  \updelta^E_p
    \, \textcolor{black}{-} \,\updelta^E_\rho\, \phifull  \, + \, \mathsf{F} ,
\end{equation}
where $\phifull$ denotes
\begin{equation}\label{phifull::def_pre}
\phifull \,:=\, \nabla\phi_0 \, +\,  \Rotzero\times\Rotzero\times\bx
                   \,+\,  (\mathbf{v}_0\cdot \nabla) \mathbf{v}_0
                   \,+\, 2\, \Rotzero \times \mathbf{v}_0\,.
 \end{equation}

\noindent $\bullet$ {Perturbations of the conservation of mass}: 
Following \cite{legendre2003rayonnement}, we first rewrite in Lagrangian form
the conservation of mass, currently listed in Eulerian form in \cref{hydrodyn_mass::eqn} and \cref{hydrodyn_mass0::eqn}.
Denote by $J$ and $J_0$ the Jacobien of the diffeomorphisms $\boldsymbol{\chi}$ and $\boldsymbol{\chi}_0$ , cf. \cite[Eq. (1.18)]{legendre2003rayonnement}, we have
\begin{subequations}
\begin{align}
\mathbf{x} = \boldsymbol{\chi}(\mathbf{a},t), \quad  F = \left(\dfrac{ \partial x_i }{\partial a_{\alpha}} \right), \,\, \alpha,i = 1,2,3, \quad
J = \mathrm{det }\, F , \\
\mathbf{x}_0 = \boldsymbol{\chi}_0(\mathbf{a},t), \quad  F_0 = \left(\dfrac{ \partial x_i }{\partial a_{\alpha}} \right), \,\, \alpha,i = 1,2,3, \quad
J_0 = \mathrm{det } \,F_0 .
\end{align}
\end{subequations}
Then, the conservation of mass can be written as in Lagrangian form, 
cf. \cite[Eq. (1.23), (1.33)]{legendre2003rayonnement} or \cite[Section 1.1 p.11]{chorin1990mathematical}, 
\begin{equation}
J\, \tilde{\rho} \,= \,\tilde{\rho_{\mathbf{a}}} \,=\, J_0\, \tilde{\rho}_0 ,
\end{equation}
where $\tilde{\rho}$ is the mass of the collection of particles $\mathbf{a}$ at time $t$, and $\tilde{\rho_{\mathbf{a}}}$ at an initial time. We introduce the Lagrangian perturbation associated with $\rho$ and the Jacobian $J$, to obtain,
\begin{equation}\label{Lrho_xi}
\begin{aligned}
& \widetilde{\updelta^L_{\rho} } = - \widetilde{\uprho_0} \dfrac{ \updelta_J^L}{J_0} + \mathsf{O}(\epsilon) , \hspace*{0.3cm} 
\text{cf. \cite[Eq. (1.45)]{legendre2003rayonnement}}\\
\Rightarrow  \hspace*{0.5cm}& \updelta_{\rho}^L = - \rho_0 \nabla \cdot \boldsymbol{\upxi} + \mathsf{O}(\epsilon) , \hspace*{0.3cm}
\text{cf. \cite[Eq. (1.57)]{legendre2003rayonnement}}.
\end{aligned}
\end{equation}
Employing \cref{LE_rel}, we obtain, \cite[Eq. (1.66)]{legendre2003rayonnement},
\begin{equation}\label{Erho_xi}
\updelta_{\rho}^E = - \rho_0 \nabla \cdot \boldsymbol{\upxi} + \boldsymbol{\upxi}\cdot \nabla \rho_0  .
\end{equation}

\noindent $\bullet$ {Perturbations of the conservation of energy}: 
The conservation of energy is written in Lagrangian coordinates $(\mathbf{a},t)$ using $\tilde{\cdot}$ notation with $\tilde{\frak{s}}(\mathbf{a},t)$, $\tilde{\frak{s}_0}(\mathbf{a},t)$, 
\begin{equation}\label{pert_Lenergy}
\begin{aligned}
\partial_t \,\tilde{\frak{s}}(\mathbf{a},t) = 0 , \hspace*{0.5cm} & 
\partial_t \,\tilde{\frak{s}}_0 (\mathbf{a},t) = 0 ,  \hspace*{0.5cm}   \text{cf. \cite[Eqs (1.27, 1.31)]{legendre2003rayonnement} }\\[0.3em]
 \Rightarrow \hspace*{0.2cm} \partial_t\, \widetilde{\delta^L_{\frak{s}}} (\mathbf{a},t) = \mathsf{O}(\epsilon) \hspace*{0.5cm} &  \Rightarrow \hspace*{0.3cm} \delta^L_{\frak{s}}=\mathsf{O}(\epsilon), 
\hspace*{0.5cm}  \text{cf. \cite[Eqs (1.47, 1.48)]{legendre2003rayonnement} }.
\end{aligned}
\end{equation}

\noindent $\bullet$ {Perturbations of the equation of state}: This computation is important in providing
the relation between the perturbations of pressure and density. Recall from \cref{eqn_state_Lag} that
\begin{equation*}
\tilde{p} = \mathcal{P}(\tilde{\rho}, \tilde{\frak{s}};\mathbf{a}), \hspace*{0.5cm} \tilde{p}_0 = \mathcal{P}(\tilde{\rho}_0, \tilde{\frak{s}}_0;\mathbf{a}).
\end{equation*}
The derivation in \cite{unno1979nonradial} and \cite[Eqs (1.48, 1.49)]{legendre2003rayonnement} lead to the same result.
The perturbations of the equation of state are carried out in terms of Lagrangian perturbation, 
cf. \cite[Section 13.4 p.98]{unno1979nonradial}. In particular, we have,  cf. \cite[Eq. (13.79)]{unno1979nonradial}, modulo $\mathsf{O}(\epsilon)$, 
\begin{equation}
\dfrac{\updelta^L_{\rho}}{\rho_0} = \dfrac{1}{\Gamma_1} \dfrac{\updelta^L_p}{p_0} + \nabla_{\mathrm{ad}}  \dfrac{\rho_0 T_0}{p_0} \updelta_{\frak{s}}^L, 
\end{equation}
where the adiabatic gradient $\nabla_{\mathrm{ad}}$, the adiabatic index $\Gamma_1$ and sound speed $c_0^2$ are defined as, cf \cite[Eq (13.44)]{unno1979nonradial},
\begin{equation}\label{adiaba}
\nabla_{\mathrm{ad}} = \left(\dfrac{\partial \ln T_0}{\partial \ln p_0} \right)_{\frak{s}},\hspace*{2em} \Gamma_1 = \left(\dfrac{\partial \ln p_0}{\partial \ln \rho_0} \right)_{\frak{s}},\hspace*{2em} c_0^2: = \dfrac{\Gamma_1 p_0}{\rho_0} = \partial_{\rho_0} \mathcal{P}.
\end{equation}
The second expression of $c_0^2$ is given by \cite[Eq. (149)]{legendre2003rayonnement} and agrees with the definitions in \cite{unno1979nonradial}, 
\begin{equation*}
c_0^2 := \partial_{\rho_0} \mathcal{P} = \left( \dfrac{ \partial p_0}{ \partial \rho_0 }\right)_{\frak{s}}
 = \left( \dfrac{ \partial \ln p_0}{  \partial \ln \rho_0 }\right)_{\frak{s}} \dfrac{ p_0}{\rho_0} = \Gamma_1 \dfrac{p_0}{\rho_0}, 
 \quad \text{ with } \Gamma_1:= \left( \dfrac{ \partial \ln p_0}{  \partial \ln \rho_0 }\right)_{\frak{s}}.
\end{equation*}

 If we assume that the perturbation does not introduce change in entropy for 
each virtual particle, i.e. $\updelta^L_{\frak{s}}=0$, 
which is given by \cref{pert_Lenergy}, then we obtain a direct relation between $\delta^L_p$ and $\delta^E_p$, 
\begin{equation}\label{pert_entro}
\updelta_{\frak{s}}^L=0\hspace*{0.4cm} \Rightarrow \hspace*{0.6cm}
\updelta^L_{\rho} \, = \,  \frac{\updelta^L_p}{c_0^2} .
\end{equation}
In employing \cref{Lrho_xi} and \cref{LE_rel}, we obtain, cf. \cite[Eq. (1.68)]{legendre2003rayonnement},
\begin{equation}
\updelta_p^E  =  \updelta_p^L  - (\boldsymbol{\upxi} \cdot \nabla) p_0 = - \rho_0 c_0^2 \nabla \cdot \boldsymbol{\upxi} - (\boldsymbol{\upxi} \cdot \nabla) p_0.
\end{equation}
From the same identities, we can obtain the relation
\begin{equation}
 \updelta^E_{p} = c_0^2 \updelta^E_{\rho} + \boldsymbol{\upxi} \cdot  \left(    - \nabla p_0 + c_0^2 \nabla \rho_0 \right).
\label{relEp_xi_entro}
\end{equation}
It can also be found in \cite[Eq. (14.1)]{unno1979nonradial} 
and \cite[Eq. (4.32)-(4.34)]{pringle2007astrophysical}
for the radial symmetric background.

\begin{remark}[No-resonance assumption]\label{Nores::rmk} As mentioned above, another approach
to arrive at the oscillation equations is to introduce the Eulerian perturbations
to the conservation laws, and then linearize, as was done in \cite{godin1997reciprocity}. 
The small Eulerian perturbations
$\updelta^E_{\rho}$, $\updelta^E_p$, and $\updelta_\mathbf{v}^E$, satisfy the system of linear equations,
\begin{equation}\label{LEE_full}
\begin{dcases}
\partial_t   \updelta_{\rho}^E + \nabla\cdot ( \updelta_\rho^E  \mathbf{v}_0 +  \updelta_\mathbf{v}^E \rho_0 )  \, = \,  0,\\[0.3em]
    \rho_0   \left(    \partial_t   \updelta_\mathbf{v}^E  \,+\, ( \mathbf{v}_0 \cdot \nabla) \updelta_\mathbf{v}^E
  + (\updelta_\mathbf{v}^E \cdot \nabla) \mathbf{v}_0
 \, + 2\, \Rotzero \times \updelta_\mathbf{v}^E
  \right) \, = \, -\nabla  \updelta_p^E
    \, \textcolor{black}{-} \,\updelta_\rho^E\, \phifull  \, + \, F .
\end{dcases}
\end{equation}
In taking $D_t = \partial_t + (\mathbf{v}_0\cdot \nabla)$, 
the first equation can be written as
\begin{equation}\label{alter_lin_EP_mass}
\rho_0 D_t \left(  \dfrac{  \updelta_\rho^E + \nabla \cdot ( \rho_0 \cdot \bxi)}{\rho_0}  \right) =0 \,.
\end{equation}
Employing the identity which gives the commutator 
between the material derivative $\bxi\cdot \nabla$ and $D_t$, 
cf. \cite[Eq. (15)]{godin1997reciprocity}, we get, 
\begin{equation}
\mathbf{w}\cdot \nabla Q = D_t (  \bxi\cdot \nabla  Q  ) - ( \bxi \cdot \nabla) D_t Q.
\end{equation}
From \cref{alter_lin_EP_mass}, the relation \cref{Erho_xi} is imposed as an assumption, 
which is referred to as the no-resonance assumption in \cite{hagg2021well}, cf. Assumption~27 therein.

Regarding the conservation of energy, 
Eulerian perturbation was employed in \cite{godin1997reciprocity},
to obtain $D_t \delta^E_\frak{s} + \delta_\mathbf{v}^E\cdot \nabla_{\frak{s}_0} = 0$, 
which was then rewritten as, cf. \cite[Eq. (22)]{godin1997reciprocity},
\begin{equation*}
D_t \left( \delta^E_{\frak{s}} + \boldsymbol{\upxi}\cdot \frak{s}_0  \right) = 0 .
\end{equation*}
From the above equation to the identity \cref{pert_Lenergy}, $\delta^E_{\frak{s}} + \boldsymbol{\upxi}\cdot \frak{s}_0=0$ was then taken as an assumption in \cite[Eq. (22)]{godin1997reciprocity}. It is here that the name `no resonance' assumption was introduced, 
as noted in \cite{hagg2021well}. 
\end{remark}


\subsection{Time-harmonic equations of oscillations in Cowling's approximation}\label{eqnmot::subsec} 
We now employ equation \cref{lin_pert_Emomentum} and the relation between 
the displacement and Eulerian perturbations to derive the oscillation equations.

\paragraph{Adiabatic time-invariant background}
We assume that the background is time-invariant, which includes the source:
\begin{equation}
\begin{gathered}
\rho_0(\mathbf{x},t) = \rho_0(\mathbf{x}), \quad p_0(\mathbf{x},t) = p_0(\mathbf{x}), \quad \mathbf{v}_0(\mathbf{x},t) = \mathbf{v}_0(\mathbf{x}),\\
\phi_0(\mathbf{x},t) = \phi_0(\mathbf{x}),  \quad  \textcolor{black}{\mathbf{f}_\mathrm{ghost}}(\mathbf{x},t) = \textcolor{black}{\mathbf{f}_\mathrm{ghost}}(\mathbf{x}).
\end{gathered} \end{equation}
The conservation of mass and momentum in \cref{hydrodyn::eqn_background} simplify to, 
\begin{subequations}\label{genbackground}
\begin{empheq}[left={\empheqlbrace\,}]{align}
& \hspace*{1cm} \nabla \cdot (\rho_0 \mathbf{v}_0)  = 0 \, ,\\[0.5em]
 & \rho_0\,   (\mathbf{v}_0 \cdot \nabla )\mathbf{v}_0 +\rho_0  \nabla \phi_0 +2 \rho_0 \,  \Rotzero \times \mathbf{v}_0 +\rho_0 \Rotzero \times \Rotzero \times \mathbf{x} \,=\, -\nabla p_0+ \textcolor{black}{\mathbf{f}_\mathrm{ghost}} .
  \label{l0_eqn2}
\end{empheq}
\end{subequations}
The gravitational potential $\phi_0$ satisfies, 
\begin{equation}\label{poisson_eqn_phi}
\Delta \phi_0  = 4\pi G \rho_0, \quad \text{with } \quad \phi_0 \rightarrow 0 \text{ as } \lvert \mathbf{x}\rvert \rightarrow \infty.
\end{equation}
We also recall the adiabatic sound speed $c_0^2(\mathbf{x})$ and adiabatic index $\Gamma_1(\mathbf{x})$, from \cref{adiaba}, 
\begin{equation}
c_0^2 \, \rho_0 \, = \, \Gamma_1\, p_0 .
\end{equation}
We have introduced $\phifull$ in \cref{phifull::def_pre}, which now denotes the left-hand side of \cref{l0_eqn2} divided by $\rho_0$,
\begin{equation}\label{phifull::def}
\phifull \,:=\, \nabla\phi_0 \, +\,  \Rotzero\times\Rotzero\times\bx
                   \,+\,  (\mathbf{v}_0\cdot \nabla) \mathbf{v}_0
                   \,+\, 2\, \Rotzero \times \mathbf{v}_0\,.
 \end{equation}
The background constraint \cref{l0_eqn2} can thus be written as, 
 \begin{equation}\label{backgroundcon_Y0}
   \phifull\,= \,- \dfrac{\nabla p_0}{\rho_0}   \, + \, \dfrac{\textcolor{black}{\mathbf{f}_\mathrm{ghost}}}{\rho_0} \,.
 \end{equation}

\begin{remark}
The value of $\phifull$ and thus of both sides of background constraint \cref{backgroundcon_Y0} 
are independent of the observation frame, specifically of $\rotzero$. 
We show this fact for backgrounds with only differential rotation 
in \cref{rotback::subsubsec}, \cref{inderotrate::identity}. 
\end{remark} 

We assume that the forces and solutions are time-harmonic, 
following convention, 
\begin{equation}\label{FourierCon}
\begin{gathered}\mathsf{F}(\mathbf{x},t)= F(\mathbf{x}) \,e^{-\ii \omega t}, \quad
\boldsymbol{\upxi}(\mathbf{x},t) = \bxi(\mathbf{x})\, e^{-\ii \omega t} , \quad 
\updelta^E_{\mathbf{v}}(\mathbf{x},t) =\updelta^E_{\mathbf{v}}(\mathbf{x})\, e^{-\ii \omega t}, \\[0.3em]
\updelta^E_{\rho}(\mathbf{x},t) = \delta^E_{\rho}(\mathbf{x}) \, e^{-\ii \omega t} , \quad 
\updelta^E_{p}(\mathbf{x},t) = \delta^E_{p}(\mathbf{x}) \, e^{-\ii \omega t} .
\end{gathered}\end{equation}
Recall from \cref{lin_pert_Emomentum,relEp_xi_entro,Erho_xi}, one has obtained 
the following relation between $\bxi$ and the Eulerian perturbations 
(listed in the frequency domain), cf. \cite[p.30]{legendre2003rayonnement}, 
\cite{christensen2014lecture},
\begin{subequations}\label{relxideltaE}
\begin{empheq}[left={\empheqlbrace\,}]{align}
&\hspace*{1cm} \delta^E_\mathbf{v}   = (-\ii \omega ) \boldsymbol{\upxi} + (\mathbf{v}_0 \cdot \nabla )\boldsymbol{\xi}   - (\boldsymbol{\xi} \cdot \nabla )\mathbf{v}_0,   \label{relxideltaE_v}\\[0.3em]
& \delta^E_\rho  =  - \nabla\cdot (\rho_0 \boldsymbol{\xi}) , \hspace*{1cm}
\delta_p^E  =   - \rho_0 c_0^2 \nabla \cdot \boldsymbol{\xi} - (\boldsymbol{\xi} \cdot \nabla) p_0 .  \label{relxideltaE_rhop}
 \end{empheq}  \end{subequations}
We will employ these in the linearized Eulerian perturbation of momentum \cref{lin_pert_Emomentum} relisted below,
\begin{equation}\label{lin_pert_Emomentum_rep}
   \rho_0   \left(   -\ii \omega \,  \delta^E_\mathbf{v}  \,+\, ( \mathbf{v}_0 \cdot \nabla) \delta^E_\mathbf{v}
  + (\delta^E_\mathbf{v} \cdot \nabla) \mathbf{v}_0
 \, + 2\,  \boldsymbol{\Upomega}_0 \times \delta^E_\mathbf{v}  
  \right) \, = \, -\nabla  \delta^E_p
    \, \textcolor{black}{-} \,\delta^E_\rho\, \phifull  \, + \, F .
\end{equation}
To obtain the oscillation equations, we 
need \cref{keyid_derosceqn::id} to rewrite 
the left-hand sied of \cref{lin_pert_Emomentum_rep} in terms of $\bxi$, which 
allows to rewrite \cref{lin_pert_Emomentum_rep} as
\begin{equation}\label{lin_pert_Emomentum_wxi}
\begin{gathered}
 \rho_0 \bigg[ 
         \big(-\ii \omega  + \mathbf{v}_0\cdot\nabla\big)^2
      +2\, \Rotzero \times 
              (-\ii \omega + \mathbf{v}_0 \cdot \nabla)
      -2 \,\Rotzero \times \nabla \mathbf{v}_0  -\nabla  (\mathbf{v}_0\cdot \nabla \mathbf{v}_0) 
      \bigg] \bxi \\
       \, = \, -\nabla  \delta^E_p
    \, \textcolor{black}{-} \,\delta^E_\rho\, \phifull  \, + \, F .
\end{gathered}
\end{equation}

\begin{identity}\label{keyid_derosceqn::id}
We have the following equality,
\begin{equation}\label{keyid_derosceqn}
\begin{aligned}
&\text{Left-hand side of \cref{lin_pert_Emomentum_rep}} \\
 & \hspace*{0.5cm} =  \rho_0 \bigg[ 
         \big(-\ii \omega  + \mathbf{v}_0\cdot\nabla\big)^2
      +2\, \Rotzero \times 
              (-\ii \omega + \mathbf{v}_0 \cdot \nabla)
      -2 \,\Rotzero \times \nabla \mathbf{v}_0  -\nabla  (\mathbf{v}_0\cdot \nabla \mathbf{v}_0) 
      \bigg] \bxi .
\end{aligned}\end{equation}
Here, the last term is written as a matrix-vector product between the 
matrix $\nabla  (\mathbf{v}_0\cdot \nabla \mathbf{v}_0)$ and the 
vector $\bxi$; this is equivalently written as two material derivatives, 
\begin{equation}\label{xigradvgradv}
\big[\nabla  (\mathbf{v}_0\cdot \nabla \mathbf{v}_0) 
      \big] \bxi   \, = \, (\bxi \cdot \nabla) (\mathbf{v}_0 \cdot \nabla) \mathbf{v}_0.
\end{equation}
\end{identity}
\begin{proof}
We need to compute the commutator of two material derivatives. 
To simplify the notation in the derivations, the parenthesis in the notation of material derivatives is dropped where there is no ambiguity, 
    \begin{equation}
(\mathbf{v}\cdot \nabla \mathbf{w}) = (\mathbf{v}\cdot \nabla) \mathbf{w}.
    \end{equation}
    In Cartesian basis, we recall the $(i,j)$-component of $\nabla V$ for vector $V =(v_i)$ , 
      \begin{equation} (\nabla V)_{ij} = \partial_j v_i , \quad \text{ with $i$ the row index and $j$ the column index}.
      \end{equation}
      
\noindent $\bullet$      Consider the vectors $\mathbf{v}$, $\mathbf{w}$ and $G$ with their Cartesian components
$\mathbf{v} = (v_i)$, $\mathbf{w}=(w_i)$, and $G=(g_i)$. 
Two material derivatives acting on the vector $G$, first in the direction of $\mathbf{w}$ 
then along $\mathbf{v}$, give
\begin{equation}\label{dertwomatder}
    \begin{aligned}
   \big( \,(\mathbf{v}\cdot \nabla ) (\mathbf{w}\cdot \nabla) G\,\big)_i
&    \,\, =\,\,  \sum_k v_k\partial_k  \Big(\sum_j  w_j \partial_j \,g_i \Big)
    \,\, =\,\, \sum_{k,j} v_k (\partial_k w_j)\, \partial_j\,g_i
      \,+\, \sum_{k,j}  w_j \,v_k\, \partial_j\,\partial_{k} g_i\\
     &= \sum_{k,j} v_k (\partial_k w_j) \,\partial_jg_i \, + \, \sum_{k,j}  w_j \partial_j \,(v_k \partial_{k} \,g_i)
   \,- \,\sum_{k,j} (\partial_k g_i) (\partial_j v_k) w_j .
  \end{aligned}
      \end{equation}
    From the above expressions, we obtain
      \begin{equation}\label{commutator2matder}
    (\mathbf{v}\cdot \nabla ) (\mathbf{w}\cdot \nabla) G   = 
    \underbrace{\big[(\mathbf{v}\cdot \nabla \mathbf{w})\, \cdot\, \nabla\big] G }_{ 
    (\nabla G)\, \cdot\, \left(\mathbf{v}\cdot \nabla \mathbf{w}   \right) }
   \,\,    +\,\, (\mathbf{w}\cdot \nabla)  (\mathbf{v}\cdot \nabla) G    -   
       \underbrace{ \big[(\mathbf{w}\cdot\nabla \mathbf{v}) \, \cdot\,  \nabla \big]G }_{(\nabla G) \cdot \big(\mathbf{w}\cdot\nabla \mathbf{v} \big)}.
    \end{equation}
In the expressions under the under-braces, the first and third term are equivalently rewritten as matrix-vector products with matrix $\nabla G$.

\medskip     
     
  \noindent $\bullet$  Secondly, from the first equality in \cref{dertwomatder}, we can also rewrite two material derivatives as a matrix-vector product, 
    \begin{equation}\label{2matder_matvec}
      (\mathbf{v}\cdot \nabla ) (\mathbf{w}\cdot \nabla) F  \,\, = \,\,
      \underbrace{ \left[  \nabla \left(       \mathbf{w}\cdot\nabla \,  F \right) \right]}_{
      \begin{array}{c}\text{ matrix given by gradient}\\
      \text{ of vector $ (\mathbf{w}\cdot \nabla) F$} \end{array}} \mathbf{v}  .
    \end{equation}

 \noindent $\bullet$   We now return to left-hand side 
           of \cref{lin_pert_Emomentum_rep}. 
           Replacing $\delta_\mathbf{v}^E$ by \cref{relxideltaE_v},
   after some rearrangement, we obtain 
    \begin{equation}
     \begin{aligned} \text{Lhs of \cref{lin_pert_Emomentum_rep}} \, &= \, (-\ii \omega)  \left[(-\ii \omega) \bxi + \mathbf{v}_0 \cdot \nabla \bxi  - \bxi\cdot \nabla \mathbf{v}_0\right]  \,+\,  \mathbf{v}_0 \cdot \nabla \left[(-\ii \omega) \bxi + \mathbf{v}_0 \cdot \nabla \bxi  - \bxi \cdot \nabla \mathbf{v}_0\right]\\
 & \hspace*{1cm} + \left[(-\ii \omega) \bxi+ \mathbf{v}_0 \cdot \nabla \bxi   - \bxi\cdot \nabla \mathbf{v}_0\right] \cdot \nabla \mathbf{v}_0
 \, + 2\,  \boldsymbol{\Upomega}_0 \times  \left[(-\ii \omega) \bxi \,\,+\,\, \mathbf{v}_0 \cdot \nabla \bxi \,\,  -\,\, \bxi \cdot \nabla \mathbf{v}_0\right] \\
 &    = \left[(-\ii \omega) + \mathbf{v}_0 \cdot \nabla\right]^2 \bxi
     \,\, + \, \,  2\,  \boldsymbol{\Upomega}_0 \times \left[ (-\ii \omega) \bxi
     \,\,+\, \,  \mathbf{v}_0 \cdot \nabla \bxi   \,\,-\,\, \bxi \cdot \nabla \mathbf{v}_0\right] \\
 &  \hspace*{1cm}   \underbrace{-\,   (\mathbf{v}_0 \cdot \nabla) \left(   \bxi \cdot \nabla \mathbf{v}_0\right) 
\, +\, \left[ \left( \mathbf{v}_0\cdot \nabla  \bxi \right)\,\cdot\, \nabla\right]\mathbf{v}_0
\, - \,\left[ \left( \bxi \cdot \nabla\mathbf{v}_0\right) \cdot \nabla \right]\mathbf{v}_0}_{ -(\bxi\cdot \nabla)  (\mathbf{v}_0\cdot \nabla) \mathbf{v}_0}. 
 \end{aligned}
          \end{equation}
           The simplification of the last three terms is obtained by employing the identity \cref{commutator2matder}, 
     \begin{equation}
-   (\mathbf{v}_0\cdot \nabla ) (\bxi\cdot \nabla) \mathbf{v}_0
 = -[(\mathbf{v}_0\cdot \nabla) \bxi\, \cdot\, \nabla ]\mathbf{v}_0
 \, -\, (\bxi\cdot \nabla)  (\mathbf{v}_0\cdot \nabla) \mathbf{v}_0    \, +  \,
 [ (\bxi\cdot \nabla) \mathbf{v}_0 \,\cdot\, \nabla ] \mathbf{v}_0  .
     \end{equation}  
     To arrive at the final expression \cref{keyid_derosceqn}, we use identity \cref{2matder_matvec} which leads to \cref{xigradvgradv}.
  \end{proof}
   

\paragraph{Oscillation equations}
The system of equations of motion are given by \cref{lin_pert_Emomentum_wxi}
together with the relations  \cref{relxideltaE_rhop} between $\bxi$ and 
the Eulerian perturbations of pressure and density. We refer to this as the mixed formulation.
 We also introduce an acoustic attenuation 
term $\gamma_{\mathrm{att}}$, as was done in \cite{Pham2021Galbrun}.
The working system  in terms of unknown $(\bxi,\pressE)$ is given
on $\mathbb{B}_{\odot}$ as, 
\begin{subequations} \label{eq:main:init}\begin{empheq}[left={\empheqlbrace}]{align}
   &    \rho_0 \bigg[ 
         \big(-\ii \omega  + \mathbf{v}_0\cdot\nabla\big)^2
      +2\, \Rotzero \times 
              (-\ii \omega + \mathbf{v}_0 \cdot \nabla)
      -2 \,\Rotzero \times \nabla \mathbf{v}_0 
      -\nabla  (\mathbf{v}_0\cdot \nabla \mathbf{v}_0) \,-\,2\ii\omega\gamma_{\mathrm{att}}
      \bigg] \bxi \label{eq:main:init:a} \\
  & \hspace*{10em} 
    \,-\, \phifull \, \nabla\cdot(\rho_0 \bxi)    \,+\, \nabla \pressE 
  \,=\, \mathbf{g}\,, \, \nonumber \\
  & \pressE
    \,+\, \bxi \cdot\nabla p_0
\,+\, \rho_0 c_0^2\nabla\cdot\bxi \,=\, 0 \,. \label{eq:main_rel2:init}
\end{empheq} \end{subequations}
The system \cref{eq:main:init} is closed by imposing on $\partial \mathbb{B}_{\odot}$ a vanishing 
Lagrangian perturbation in pressure\footnote{The original form of the condition is, $\pressL \,=\, 0$. The working form \cref{vLpBC_gen} is obtained by employing the relation between Lagrangian and Eulerian perturbation, $\pressL = -\rho c_0^2 \nabla \cdot \bxi =
                \pressE + \bxi \cdot \nabla p_0$.}, \cite{christensen2014lecture},
\begin{equation}\label{vLpBC_gen}
  \pressE + \bxi \cdot \nabla p_0 \,=\, 0\,,\quad      \text{on }  \partial  \mathbb{B}_{\odot}\,,  
  \hspace*{1em} \text{Vacuum boundary condition.}
\end{equation}
Recall the term $\mathbf{v}_0\cdot \nabla$ in \cref{eq:main:init:a} is a material derivative, cf. \cref{matdervec::def}, and $[\nabla (\mathbf{v}_0\cdot \nabla \mathbf{v}_0) ]\bxi $ has a second interpretation in \cref{xigradvgradv}.

\paragraph{Original-div formulation $\Loriginaldiv$}
By using equation \cref{eq:main_rel2:init}, 
$\nabla\cdot\bxi$ is expressed as a linear 
combination of $\bxi$ and $\pressE$:
\begin{equation}
\nabla\cdot( \rho_0 \bxi) = \bxi \cdot \left( \nabla\rho_0 -  \dfrac{\nabla p_0}{c_0^2}\right)
          - \dfrac{\pressE}{c_0^2}\,.
          \end{equation}
By employing this identity to replace\footnote{This is seen as follows: 
from \cref{eq:main_rel2:init}, we have 
\begin{equation*}
\begin{aligned}
\nabla\cdot( \rho_0 \bxi) 
  & = \bxi \cdot  \nabla\rho_0 + \rho_0 \nabla\cdot\bxi 
    = \bxi \cdot \nabla\rho_0 + \rho_0 
    \Big( - \dfrac{\pressE}{\rho_0 c_0^2} 
                             - \bxi \cdot \dfrac{\nabla p_0}{\rho_0 c_0^2} 
                       \Big) 
  = \bxi \cdot \nabla\rho_0 - \dfrac{\pressE}{c_0^2} 
          - \bxi \cdot \dfrac{\nabla p_0}{c_0^2}\\
 \Rightarrow \,-\, \phifull \,\nabla\cdot(\rho_0 \bxi) 
 & = 
      \dfrac{\phifull}{c_0^2}   \pressE
   \,+\, \rho_0 \Big[ \phifull \otimes  \Big(\boldsymbol{\alpha}_\rho + \dfrac{\nabla p_0}{\rho_0 c_0^2} \Big)
         \Big] \bxi \,.
          \end{aligned}\end{equation*}} 
the divergence of $\bxi$ in \cref{eq:main:init:a}, we arrive at an equivalent 
system: 
\begin{subequations} \label{eq:main:original-div}
\begin{empheq}[  
left={\empheqlbrace}]{align}
   &    \rho_0  \Amat_0^{\Omega}\,\bxi
  \,\,+\,\, \dfrac{\phifull}{c_0^2}  \, \pressE  
  \,\,+\,\, \nabla \pressE 
  \,=\, \mathbf{g}\,, \, \nonumber \\
  & \dfrac{\pressE}{\rho_0 c_0^2} 
    \,+\, \bxi \cdot {\color{black}{\dfrac{\nabla p_0}{\rho_0 c_0^2}}}
\,+\, \nabla\cdot\bxi \,=\, 0 \,,
\end{empheq} \end{subequations}
where $\phifull$ is defined in \cref{phifull::def} and operator $\Amat_0^{\Omega}$ is defined as 
\begin{gather}
\Amat_0^{\Omega} : = \Bmat_0^{\Omega} \,-\,2\ii\omega\gamma_{\mathrm{att}}\,+\, \phifull \otimes
      \Big(\boldsymbol{\alpha}_\rho \,+\, \dfrac{\nabla p_0}{\rho_0 c_0^2}\Big); \label{Amat::def}\\
\text{with} \hspace*{0.5cm}      \Bmat_0^{\Omega} :=    \big(-\ii \omega  + \mathbf{v}_0\cdot\nabla\big)^2
      +2 \Rotzero \times 
              (-\ii \omega + \mathbf{v}_0 \cdot \nabla)
      -2 \Rotzero \times \nabla \mathbf{v}_0 
      -\nabla  (\mathbf{v}_0\cdot \nabla \mathbf{v}_0).  \label{Bmat::def}
\end{gather}
We refer to this system \cref{eq:main:original-div} with $\Loriginaldiv$,
where we consider a more general form containing a non-zero scalar source $h$:
\begin{equation}\label{origdiv_form}
\Loriginaldiv \begin{pmatrix}\bxi \\[0.2em]\pressE \end{pmatrix} 
      \,=\, \begin{pmatrix}\mathbf{g}\\[0.2em] h \end{pmatrix}, 
      \qquad \text{ with } \qquad 
      \Loriginaldiv =\begin{pmatrix} \rho_0 \,\Amat_0^{\Omega} \quad & \quad \boldsymbol{\beta}_{a 0}^{\Omega} \,+\, \nabla  \\[0.4em] \nabla\cdot  \,+ \,\boldsymbol{\beta}_{b 0}^{\Omega} \cdot\quad & \quad   \varrho_0^{\Omega}
\end{pmatrix}
\end{equation}
\begin{equation}
\hspace*{3cm} \text{ where} \hspace*{0.5cm}\boldsymbol{\beta}_{a 0}^{\Omega} = \dfrac{\phifull}{c_0^2}, \hspace*{0.5cm}  \boldsymbol{\beta}_{b 0}^{\Omega} = 
 \dfrac{\nabla p_0}{\rho_0 c_0^2}   \overset{\cref{adiaba}}{=} - \dfrac{\boldsymbol{\alpha}_p}{\Gamma_1} , \hspace*{0.5cm}  \varrho_0^{\Omega} = \dfrac{1}{\rho_0 c_0^2}\,.
\end{equation}

\medskip
\paragraph{Scalar Green's kernel to scalar source} 
We are interested in the scalar response 
given by $\delta^E_p$ to model helioseismic observables, 
cf. \cref{syndata::subsec}.  We thus introduce the following Green's kernel.
\begin{definition}\label{scalarppGKgen}
The scalar pressure perturbation Green's kernel,  $G_p(\mathbf{x},\mathbf{s})$
is defined as follows,
\begin{equation}\label{scalarppGKgen::def}
w(\cdot, \mathbf{s}) := G_p(\mathbf{x},\mathbf{s}) \,\, \text{ is a distributional solution to } \,\,\Loriginaldiv \begin{pmatrix} \bu \\ w \end{pmatrix} = \begin{pmatrix} 0 \\ \delta(\mathbf{x}-\mathbf{s})\end{pmatrix} .
\end{equation}
\end{definition}
The Green's kernel provides the solution 
to $\Loriginaldiv $ with only a scalar source 
$h$, i.e., 
\begin{equation}\label{scalarppGKgen_sol}
w = \int_{\mathbb{B}_{\odot}} G_p(\mathbf{x},\mathbf{s}) \, h(\mathbf{s})\, d\mathbf{s}   \quad \text{ solves } \quad 
\Loriginaldiv \begin{pmatrix} \bu \\ w \end{pmatrix} = \begin{pmatrix} 0 \\ h\end{pmatrix}  \,. 
\end{equation}

In the following remarks, we relate Equation \cref{eq:main:init} to existing forms in literature.
\begin{remark}\label{Goughlit::rmk}
  Equation \cref{eq:main:init} is the same as \cite[(2.1)--(2.4)]{gough1990effect} by Gough \& Thompson for differential rotation flow $\mathbf{v}_0$, without magnetics' effects and in Cowling's approximation.  Here the constraint \cref{backgroundcon_Y0} with $\mathbf{f}_{\mathrm{ghost}} = 0$
  is employed to yield $\phifull =  -\dfrac{\nabla p_0 }{\rho_0}$, and the equation of motion \cref{eq:main:init:a} is equivalently written as, 
  \begin{subequations}
  \begin{align}
 \cref{eq:main:init:a} \hspace*{0.3cm}& \Longleftrightarrow \hspace*{0.3cm} \rho_0  \Bmat_0^{\Omega}  \bxi = - \nabla \pressE  + \delta^E_\rho\dfrac{\nabla p_0}{\rho_0} + \mathbf{g} \\
 \hspace*{0.3cm}& \Longleftrightarrow \hspace*{0.3cm} 
  \rho_0  \Bmat_0^{\Omega}  \bxi = - \nabla \pressE - (\nabla \cdot \bxi) \nabla p_0 - (\bxi \cdot \nabla \rho_0) \dfrac{\nabla p_0}{\rho_0} + \mathbf{g}.\label{Gough}
 \end{align}
  \end{subequations}
 
\end{remark}

\begin{remark}\label{LBOlit::rmk}
Here we relate equation \cref{eq:main:init} with the one derived by Lynden-Bell \& Ostriker \cite[Equation 30]{lynden1967stability}
under the same assumption as in \cref{Goughlit::rmk}. 
We first need the following identities with material derivative cf. \cref{matdervec::def}: following from product rule, 
\begin{equation}\label{LBOid1}
(\bxi\cdot \nabla)\left( \dfrac{1}{\rho_0} \nabla p_0\right) = \dfrac{1}{\rho_0} (\bxi\cdot  \nabla)\nabla p_0 - 
 \dfrac{(\bxi\cdot \nabla \rho_0 )}{\rho_0^2 } \nabla p_0,
\end{equation}
and from the fact that $\rotzero $ is constant,
\begin{equation}\label{LBOid2}
\begin{aligned}
(\bxi\cdot \nabla)\left( \rotzero \times \mathbf{v}_0 \right) &= \rotzero \times (\bxi\cdot\nabla)\mathbf{v}_0 ,\\
(\bxi\cdot \nabla)\left( \rotzero \times \rotzero \times\mathbf{x} \right)&= 
\rotzero \times \rotzero \times  (\bxi\cdot \nabla)\mathbf{x} = \rotzero \times \rotzero \times \bxi.
 \end{aligned}
\end{equation}

Consider background constraint \cref{l0_eqn2} with $\mathbf{f}_{\rm ghost}=0$, to both sides of 
this equality, divide by $\rho_0$ and apply material derivative $\bxi\cdot \nabla$ to both sides.
Employ identity \cref{LBOid1} and \cref{LBOid2},  and after some rearrangement, we obtain the following form of background constraint, 
\begin{equation}\label{LBidentity}
\begin{gathered}
- \dfrac{ (\bxi \cdot \nabla) \rho_0 }{ \rho_0^2 } \nabla p_0 + 2 \rotzero \times  (\bxi \cdot \nabla) \mathbf{v}_0 
+ (\bxi\cdot  \nabla)  ( \mathbf{v}_0 \cdot \nabla) \mathbf{v}_0 \\
=   -\bxi \cdot \mathrm{H}_{\phi_0} -  \bxi \cdot \dfrac{\mathrm{H}_{p_0}}{\rho_0} - \rotzero \times \rotzero \times \bxi.
\end{gathered}
\end{equation}
Here $\mathrm{H}_g$ denotes Hessian applied to scalar function $g$. 

Define operator $\mathcal{P}$ as
\begin{equation}
\mathcal{P}\bxi :=\underbrace{-\nabla ( \Gamma_1 p_0 \nabla \cdot \bxi ) - \nabla( \bxi \cdot \nabla p_0 )}_{\nabla \pressE}\,  +\, (\nabla p_0) ( \nabla\cdot \bxi)\, +\, \bxi \cdot \mathrm{H}_{p_0} \label{LBpop}.
\end{equation}
Employ the above identity \cref{LBidentity} in the equation \cref{Gough}, then we have the equivalence,
\begin{equation}\label{LBO_Gough_equiv}
\begin{aligned}
&(\text{Current work} ) : \hspace*{0.2cm} \cref{eq:main:init:a}   \text{ with } \mathbf{f}_{\rm ghost}=0, \,\,\phifull =  -\tfrac{\nabla p_0 }{\rho_0}\\[0.5em]
 & \Longleftrightarrow \hspace*{0.4cm}
\underset{\substack{\\[0.2em]\text{GT-variant}}}{\cref{Gough}}\hspace*{0.4cm} \Longleftrightarrow \hspace*{0.4cm} 
(\substack{\text{LBO}\\\text{variant}}): \hspace*{0.cm} \left\{
\begin{array}{l}
\rho_0 (- \ii \omega + \mathbf{v}_0\cdot\nabla)^2 \bxi + 2\rho_0 \rotzero \times (-\ii \omega  + 
\mathbf{v}_0\cdot\nabla) \bxi \\[0.4em]
 \hspace*{1cm} =-  \rho_0 \rotzero\times \rotzero \times \bxi
   - \mathcal{P}\bxi  - \rho_0 \bxi \cdot \mathrm{H}_{\phi_0} + \mathbf{g} \end{array} \right. , 
\end{aligned}
\end{equation}
Next employ identity,    $
p_0 \nabla \nabla\cdot \bxi = \nabla ( p_0 \nabla\cdot \bxi) - (\nabla p_0) (\nabla \cdot \bxi)$, to rewrite the third term on the rhs of \cref{LBpop}, and we arrive at form of $\mathcal{P}$ as appeared in \cite[Equation 25]{lynden1967stability},
\begin{equation}
\mathcal{P} \bxi =  \nabla\left[ (1 - \Gamma_1) p_0 \nabla \cdot \bxi \right] - p_0 \nabla ( \nabla\cdot \bxi) 
- \nabla ( \bxi\cdot \nabla p_0 )  + \bxi \cdot\mathrm{H}_{p_0}.
\end{equation}

In previous work \cite{Pham2021Galbrun,Pham2024assembling}, we employed 
the LBO variant \cref{LBO_Gough_equiv}
without flow and rotation, cf. \cref{main_eqn_Galbrun_unscaled} in \cref{Galbrun::rmk}.
In the latter remark, we show that the mixed-formulation associated with this form
coincides with the mixed formulation \cref{eq:main:original-div} (associated with \cref{eq:main:init:a}).
  
\end{remark}

\section{Equations of oscillations with differential rotation}
\label{workingbackgrounds::subsec}

For the remaining of the investigation, we consider \cref{eq:main:init} 
for backgrounds where the only nonzero contribution to the flow $\mathbf{v}_0$
is due to rotation. More specifically, we consider two types of backgrounds: 
non-rotating backgrounds ($\rotzero=\rot=0$, i.e., $\mathbf{v}_0=0$), 
and differentially rotating ones defined in 
\cref{difflowbackground::def}, 
which yields axisymmetric systems. 
As another property, for non-rotating background, 
$\Amat_0^{\Omega=0}$ in \cref{origdiv_form} is a multiplicative operator;
and it can be decomposed to such for differentially rotating backgrounds.
This allows to introduce the Liouville change of variable, which
yields the variants in \cref{Liouville::subsec} employed for the numerical computations.
These types of backgrounds also allow to relate the system of 
equations with a second order scalar equation, which helps shed 
light on the properties of solutions, cf. \cref{subsection:scalar-pde}.

\subsection{Differential rotation background}\label{rotback::subsubsec}

The cylindrical coordinate system used, $(\eta,\phi,z)$,
is related to the Cartesian coordinates $\mathbf{x}=(x,y,z)$, 
and radial distance $r$, as
\begin{equation}\label{cylcoord}
\eta^2 = x^2 + y^2 \,, \hspace*{2em} 
x\,=\,\eta \cos\phi \,,\hspace*{2em} 
y\,=\,\eta \sin\phi\, , \quad r^2 = \eta^2 + z^2.
\end{equation}
The corresponding orthonormal basis is 
$(\mathbf{e}_\eta,\mathbf{e}_\phi, \mathbf{e}_z)$, 
and we denote by $\mathbf{e}_r$ the radial vector.

\begin{mdframed}
\begin{definition}[Differential flow backgrounds]\label{difflowbackground::def}
 By a differential flow background, 
 we refer to background parameters 
 satisfying the following properties:
\begin{itemize}[leftmargin = *]
\item In addition to \cref{genbackground}, the background parameters, 
      $\rho_0$, $p_0$, $\Gamma_1$ and $\Phi_0$ are independent 
      of $\phi$. 
\item The body is in rotation around axis $\mathbf{e}_z$ 
      with angular rate $\rot(r,\theta)$. 
\item The background flow $\mathbf{v}_0$ in \cref{genbackground} 
      is only induced by rotation, which means in the observation 
      frame rotating rigidly at constant rate $\rotzero$, $\mathbf{v}_0$ is 
      the difference between the full rotation and the frame rotation:
\begin{equation}\label{purediffrot}
\mathbf{v}_0 = (\rotvec - \Rotzero) \times \mathbf{x} ,  \hspace*{2em} 
\text{where} \hspace*{3em} 
\rotvec= \rot(r,\theta)\, \mathbf{e}_z \,, 
\quad \text{and} \quad \Rotzero = \rotzero \mathbf{e}_z \,.
\end{equation}
\end{itemize} 

\end{definition}
\end{mdframed}

In our investigation, we further distinguish cases depending on the constancy of $\rot$:
\begin{itemize} 
\item \emph{Differential rotation}: When $\rot$ is non-constant, 
       we refer to this case as the differential rotation.
\item \emph{Uniform rotation}: When $\rot$ is constant and nonzero,
      $\rot=\rot_u$, with $\rot_u\neq 0$, the fluid body is in uniform rotation. 
      In this case, the observation frame is usually chosen 
      to be co-rotating at the same rate, i.e. $\rotzero:=\rot_u$, 
      in which choice, there is no background flow: 
\begin{equation}\label{unifrot_cotrot::def}
\rot =\rot_u \text{ constant } \hspace*{0.2cm}\text{and} \hspace*{0.2cm}  \rotzero := \rot_u \hspace*{0.3cm}  \Rightarrow  \hspace*{0.3cm}\mathbf{v}_0 = 0  \,.
\end{equation}
\item \emph{Non-rotating Sun in a rigidly co-rotating frame}: In this case, $\rot=0$ but 
    $\rotzero$ is not. It gives $\mathbf{v}_0= -\Rotzero\times\bx$. This is used 
    in the experiments to validate the correct behaviour of our solver in \cref{heliorot::subsec}  .  
    
\end{itemize}
We next note the independence of the background constraints \cref{l0_eqn2} 
on the choice of the observation frame.

\begin{identity}\label{inderotrate::identity}
For the background flow $\mathbf{v}_0$ given in \cref{purediffrot}, we have the identity
\begin{gather}
\mathbf{v}_0  \,=\, \tilde{\rot} \, \mathbf{e}_z \times \mathbf{x}\,\, \overset{\cref{v0phi0_app}}{=}\,\, \eta \,\tilde{\rot}\, \mathbf{e}_\phi, \qquad \text{where }\quad \tilde{\rot} := \rot - \rotzero;\\
\text{and} \hspace*{0.3cm} \mathbf{v}_0 \cdot \nabla \mathbf{v}_0  
  \,+\, 2 \Rotzero \times \mathbf{v}_0 \,+\, 
  \, \Rotzero \times \Rotzero \times \mathbf{x} 
  \, = \, \rotvec \times \rotvec \times \mathbf{x}
  \, = - \rot^2 \, \eta \, \mathbf{e}_\eta  \,.\label{phifullexp}
\end{gather}
\end{identity}
\begin{proof}
With $\tilde{\Omega} := \rot - \rotzero$, we have
\begin{equation}
\mathbf{v}_0 \cdot \nabla \mathbf{v}_0 \overset{\cref{matder_cyl}}{=} -\eta \tilde{\rot}^2 \mathbf{e}_\eta; 
\qquad   2 \Rotzero \times \mathbf{v}_0 \overset{\cref{rotmatrix}}{=} -2 \,\rotzero\,  \tilde{\rot} \,\eta\,
\mathbf{e}_\eta; 
\qquad \Rotzero \times \Rotzero \times \mathbf{x} \overset{\cref{phifullcyl}}{=}  - \eta \rotzero^2 \mathbf{e}_\eta.
\end{equation}
Summing up these three equalities, we obtain,
\begin{equation}
\mathbf{v}_0 \cdot \nabla \mathbf{v}_0  
  \,+\, 2 \Rotzero \times \mathbf{v}_0 \,+\,
  \, \Rotzero \times \Rotzero \times \mathbf{x}  \,\,= \,\,- (\tilde{\rot} + \rotzero )^2 \eta \mathbf{e}_\eta = - \rot^2 \,\eta\, \mathbf{e}_\eta.
\end{equation}
\end{proof}


\paragraph{Background constraints for differential rotation}
 
As a result of \cref{inderotrate::identity},
for a differential flow background given \cref{difflowbackground::def}, 
we have another equivalent definition of $ \phifull$, 
explicitly independent of $\rotzero$:
\begin{equation}\label{phifull::deffullrot}
\phifull :=    \nabla \phi_0 \, + \, \rotvec\times \rotvec\times \mathbf{x}  \, \overset{\cref{phifullexp}}{=} \,
\nabla \phi_0 \,  -\, \rot^2\, \eta\, \mathbf{e}_\eta    \,. 
\end{equation}
The background constraint \cref{backgroundcon_Y0} (or \cref{l0_eqn2}) 
can be written independently of $\rotzero$:
\begin{equation}\label{backgroundcon_Y0difrot}
\left(\phifull \hspace*{0.1cm}:= \right) \hspace*{0.3cm}   \nabla \phi_0 \, + \, \rotvec\times \rotvec\times \mathbf{x} \,= \, 
-\dfrac{\nabla p_0}{\rho_0}   \, + \, \dfrac{\textcolor{black}{\mathbf{f}_\mathrm{ghost}}}{\rho_0}   .
\end{equation}
The quantity $\mathbf{f}_\mathrm{ghost}$ measures 
the deviation from standard background states 
which have $\mathbf{f}_\mathrm{ghost} =0$, with the 
two following situations for the cases with and without 
rotation.

\begin{enumerate}
  \item Rotational hydrostatics: For uniform rotation 
        in co-rotating frame, the constraint \cref{backgroundcon_Y0difrot}
        with $\mathbf{f}_\mathrm{ghost}=0$ is referred to
        as \emph{rotational hydrostatics}: 
        \begin{equation}\label{backgroundcon_Y0difrot_stand}
        \left(\phifull \hspace*{0.1cm}:= \right) 
        \nabla \phi_0 \, + \, \rotvec\times \rotvec\times \mathbf{x} \,= \, -\dfrac{\nabla p_0}{\rho_0}\,,
        \qquad \text{Rotational hydrostatics.}
        \end{equation}        
        It can also be written in terms of the centrifugal 
        potential $\psi^s$ defined in \cref{centripot::def},
        replacing the left-hand side by $\nabla (\phi_0 + \psi^S)$.
 
  \item Hydrostatic equilibrium: 
        For non-rotating backgrounds with $\mathbf{v}_0=0$, $\rot=0$ and $\Rotzero=0$,
        the relation \cref{backgroundcon_Y0difrot_stand} with $\mathbf{f}_\mathrm{ghost}=0$ 
        is referred to as \emph{hydrostatic equilibrium}. It writes as,
        \begin{equation}\label{steq}
        \nabla p_0 \,=\, -\rho_0 \nabla \phi_0   \hspace*{0.25cm} \Leftrightarrow \hspace*{0.25cm} 
                           \dfrac{\boldsymbol{\alpha}_{p_0}}{\Gamma_1} 
                           \,=\,\dfrac{\nabla \phi_0}{c_0^2}  \,,
        \hspace*{1cm} \text{Hydrostatic equilibrium}\,. 
        \end{equation}
        The definition of $\phifull$ \cref{phifull::def} reduces to,
        \begin{equation}
          \mathbf{Y}_0 := \mathbf{Y}_{\rotvec=0} \,=\, \nabla \phi_0 \,.
        \end{equation}
\end{enumerate}

\begin{remark}\label{rolefghost::rmk}

For numerical resolution of the equation of motions \cref{eq:main:original-div}, 
one needs the values of the parameters
\begin{equation}\label{partocompute}
\begin{array}{c}
\rotvec, \, \Rotzero, \, \rho_0 ,\,\phi_0 ,\\ 
 c_0, \,   p_0, \, \Gamma_1, \text{  and } \phifull\end{array}
 \qquad \text{ satisfying } \qquad  \begin{dcases}
 \quad \text{ adiabaticity constraint \cref{adiaba}}, \\
 \quad \text{ background constraint } \cref{backgroundcon_Y0difrot} ,\\
 \quad \, \phi_0 \text{ determined by } \rho_0 \text{ via } \cref{poisson_eqn_phi},\\
 \quad \, \phifull \text{ defined by } \phi_0 \text{ and }\rotvec \text{ by } \cref{phifull::deffullrot}.
 \end{dcases}
\end{equation}

The constraint \cref{backgroundcon_Y0difrot} is employed
either to obtain the value of $\nabla p_0$, or to 
compensate for the deviation from using a standard 
background \cref{backgroundcon_Y0difrot_stand}.
Specifically, from inputs $\rotvec$, $\Rotzero$, $\rho_0$ (which determines $\phi_0$), 
taking $\mathbf{f}_\mathrm{ghost} = 0$ means that the parameters must be constructed 
to satisfy the rotational hydrostatics \cref{backgroundcon_Y0difrot_stand}. 
This approach is taken, e.g., in \cite[Equations (1--3)]{reese2006acoustic} with a 
polytropic relation for $\rho_0$ and $p_0$ in order to construct an axially 
symmetrical background with rotation.

One way to obtain a complete set of parameters in \cref{partocompute},  
for purpose of numerical resolution, is to employ
an existing background model for $p_0$ and $c_0$. In this approach, 
including rotation can result in a deviation from rotational hydrostatics given 
by $\mathbf{f}_\mathrm{ghost}$.
For instance, in our experiments, we employ $p_0$ and $c_0$ 
from  the standard solar 
Model-S, \cite{christensen1996current} that gives quantities depending only on $r$
and
verifying the hystrostatic equilibrium \cref{steq} and adiabaticity \cref{adiaba}. 
Incorporating rotation results in 
$\mathbf{f}_\mathrm{ghost} = \rho_0 \, \rotvec \times \rotvec \times \bx \neq 0$,
although this quantity is not directly implemented in \cref{eq:main:original-div}. 
In the solar case, the deviation is the greatest near the center and is at most 3\%.

As an example, for the case of 
uniform rotation in co-rotating frame \cref{unifrot_cotrot::def}, $\rot=\rot_u=\rot_c =$ constant and $\mathbf{v}_0=0$, 
the important terms in \cref{eq:main:original-div} take the following values,
\begin{equation}\label{modS_unirot}
  \text{Under \cref{unifrot_cotrot::def}:} 
\begin{dcases}
&\dfrac{\phi_0'(r)}{c_0^2(r)} = \dfrac{\alpha_p(r)}{\Gamma_1(r)}, \hspace*{0.7cm} \phifull: = c_0^2(r) \left(- \rot_u^2\dfrac{ \eta}{c_0^2(r) }  \mathbf{e}_\eta \, + \, \dfrac{\phi_0'(r)}{c_0^2(r)} \mathbf{e}_r\right),\\
&\phifull\otimes (\boldsymbol{\alpha}_\rho - \dfrac{\boldsymbol{\alpha}_p}{\Gamma_1} ) = N^2(r) \mathbf{e}_r\otimes \mathbf{e}_r 
 - \rot_u^2 \, \eta \left(  \alpha_\rho(r) - \dfrac{\phi_0'(r)}{c_0^2(r)} \right) \mathbf{e}_\eta \otimes \mathbf{e}_r.
  \end{dcases}
\end{equation}
Here the buoyancy frequency $N$ is defined in \cref{Nfreq::def}, 
with $c_0$, $\rho_0$, $\phi_0$ given by e.g. Model-S.

\end{remark}

\subsection{Liouville variants of equations of motion}
\label{Liouville::subsec}
We next consider a change of variable defined by a 
factor $\mathfrak{f}\neq 0$, that commutes with 
operator $\Amat$ of \cref{origdiv_form}. 
This property is satisfied by 
the solar backgrounds described in \cref{workingbackgrounds::subsec} 
and for factor $\mathfrak{f}$ independent of $\phi$.
We will need the following properties satisfied by the inverse scale height $\boldsymbol{\alpha}_f$ in \cref{scaleheight::def}: for constant $\gamma\in \mathbb{R}$, and $f$, $g\neq 0$: 
\begin{equation}\label{scaleheightid}
\boldsymbol{\alpha}_{g^{\gamma}}= \gamma \boldsymbol{\alpha}_g , \quad \boldsymbol{\alpha}_{ fg} = \boldsymbol{\alpha}_f + \boldsymbol{\alpha}_g.
\end{equation}
In particular, we have
 \begin{equation}\label{alpha_identity}
  \boldsymbol{\alpha}_{\sqrt{\rho}} =\boldsymbol{\alpha}_{\rho}/2,  \quad   \boldsymbol{\alpha}_{\sqrt{\rho} c} = \boldsymbol{\alpha}_{\rho}/2 + \boldsymbol{\alpha}_{c}.
  \end{equation}
   
\begin{identity}\label{gencov::identity} For $\mathfrak{f}\neq 0$, 
that commutes with operator $\Amat$, i.e., 
$\Amat\, \mathfrak{f} \,= \, \mathfrak{f} \Amat$. 
For $( \bu, w )$, and $( \bu_\mathfrak{f}, w_\mathfrak{f})$ related by
\begin{equation}\label{gencov}
  \mathbf{u}_\mathfrak{f} := \mathfrak{f} \,\mathbf{u} , \hspace*{4em}
    w_\mathfrak{f} := \mathfrak{f}^{-1} \, w \,,
\end{equation}
we have the equivalence, 
\begin{equation}\label{gencov_1stsystem}
\left\lbrace \begin{aligned}
  \Amat  \mathbf{u}
  +  \boldsymbol{\beta}_1\,w
  + \nabla w = \mathbf{g}, \\[.2em]
  \nabla\cdot   \mathbf{u} + \boldsymbol{\beta}_2 \cdot \mathbf{u}
  +\varrho \,w = h,
\end{aligned} \right. \quad \overset{\cref{gencov}}{\Longleftrightarrow} \quad 
\left\lbrace \begin{aligned}
  \dfrac{\Amat}{\mathfrak{f}^2} \,\mathbf{u}_\mathfrak{f}
  \,+\, \left(\boldsymbol{\beta}_1 - \boldsymbol{\alpha}_{\mathfrak{f}} \right) \, w_\mathfrak{f}
  \,+\, \nabla w_\mathfrak{f}\,=\, \dfrac{\mathbf{g}}{\mathfrak{f}}, \\
  \nabla\cdot   \mathbf{u}_\mathfrak{f} \,+\, \left( \boldsymbol{\beta}_2 
  \,+\, \boldsymbol{\alpha}_\mathfrak{f} \right) \cdot \mathbf{u}_\mathfrak{f}
  \,+\, \varrho\,\mathfrak{f}^2\, w_\mathfrak{f} \,=\, \mathfrak{f} \,h.
\end{aligned} \right.
\end{equation}
\end{identity}
\begin{proof}
Employing 
\begin{align*}
\nabla w &= \nabla ( \mathfrak{f} w_\mathfrak{f} ) = \mathfrak{f} \left(\nabla w_\mathfrak{f} -  w_\mathfrak{f}\boldsymbol{\alpha}_\mathfrak{f} \right),\\
\nabla\cdot \mathbf{u} &= \nabla \cdot ( \mathfrak{f}^{-1} \mathbf{u} _\mathfrak{f}) =\mathfrak{f}^{-1} \left( \nabla \cdot  \mathbf{u} _\mathfrak{f}   - \mathbf{u} _\mathfrak{f}\cdot \boldsymbol{\alpha}_{1/\mathfrak{f}} \right)
 = \mathfrak{f}^{-1} \left( \nabla \cdot  \mathbf{u} _\mathfrak{f}   + \mathbf{u} _\mathfrak{f}\cdot \boldsymbol{\alpha}_{\mathfrak{f}} \right)\,,
\end{align*}
we can rewrite the left-hand sides of the system in variable $(w,\mathbf{u} )$ as
\begin{equation}
\begin{aligned}
\Amat\, \mathbf{u} 
  + \boldsymbol{\beta}_1 \,w
  + \nabla w &= \Amat(\mathfrak{f}^{-1} \mathbf{u}_\mathfrak{f} )  + \boldsymbol{\beta}_1 \, ( \mathfrak{f} w_\mathfrak{f} ) 
  + \nabla (\mathfrak{f} w_\mathfrak{f} )   =  \mathfrak{f} \left( \dfrac{\Amat}{\mathfrak{f}^2} \, \bxi_\mathfrak{f}
  + \left(\boldsymbol{\beta}_1 - \boldsymbol{\alpha}_{\mathfrak{f}} \right) \, w_\mathfrak{f}  + \nabla w_\mathfrak{f} \right),\\
   \nabla\cdot  \mathbf{u}  + \boldsymbol{\beta}_2 \cdot\mathbf{u} 
  +\varrho w &=  \nabla\cdot  (\mathfrak{f}^{-1} \mathbf{u} _\mathfrak{f} ) + \boldsymbol{\beta}_2 \cdot(\mathfrak{f}^{-1} \mathbf{u}_\mathfrak{f} ) 
  +\varrho \mathfrak{f} w_\mathfrak{f} = \mathfrak{f}^{-1} \left(  \nabla\cdot \mathbf{u}_\mathfrak{f} + \left( \boldsymbol{\beta}_2 
  + \boldsymbol{\alpha}_\mathfrak{f} \right) \cdot\mathbf{u}_\mathfrak{f}  + \varrho\mathfrak{f}^2\, w_\mathfrak{f}   \right) .
\end{aligned}
\end{equation}
The equivalent system in variable $(\mathbf{u}_\mathfrak{f}, w_{\mathfrak{f}} )$ is obtained by multiplying the first equation by $\mathfrak{f}^{-1}$ and second by $\mathfrak{f}$. 
\end{proof}

\paragraph{Liouville and Liouville-c change of variables} 
We employ \cref{gencov::identity} with factors $\mathfrak{f}_{\mathrm{L}} = \sqrt{\rho_0}$ and
$\mathfrak{f}_{\mathrm{Lc}} = c_0 \sqrt{\rho_0}$ in \cref{gencov},  to obtain variants,
respectively called the Liouville and Liouville-c variants, 
equivalent to the original equation \cref{origdiv_form} .
Specifically, we introduce the Liouville variables $(\bxi_L,w_L)$, and the 
Liouville-c variables $(\bxi_c,w_c)$, 
\begin{subequations} \label{eq:liouville-unknowns}
\begin{align}
\text{Liouville unknowns}: \qquad &  \bxi_L \,=\, \sqrt{\rho_0} \bxi \,, 
  \qquad 
  w_L \,=\, \dfrac{1}{\sqrt{\rho_0}} \pressE ;   \label{L_cov}\\
\text{Liouville-c unknowns}: \qquad &  \bxi_c \,=\, c_0 \sqrt{\rho_0} \bxi \,, 
  \qquad 
  w_c \,=\, \dfrac{1}{c_0 \sqrt{\rho_0}} \pressE . \label{Lc_cov}
\end{align} \end{subequations}
The operators of Liouville $\Lliouville$ and Liouville-c variants $\Lliouvillec$, are defined respectively as 
\begin{equation}
   \Lliouville := \begin{pmatrix} 
   \mathbf{A}_{\mathrm{L}}^{\Omega} \quad & \quad \boldsymbol{\beta}_{a \mathrm{L}}^{\Omega} \,+\, \nabla  \\[0.2em]  
                         \nabla\cdot  + \boldsymbol{\beta}_{b \mathrm{L}}^{\Omega} \cdot\quad & \quad  
                         \varrho_\mathrm{L}^{\Omega}
   \end{pmatrix} \hspace*{1.5em}\text{and}\hspace*{1.5em}
   \Lliouvillec \,:=\,\begin{pmatrix} 
   \mathbf{A}_{\mathrm{Lc}}^{\Omega} \quad & \quad \boldsymbol{\beta}_{a \mathrm{Lc}}^{\Omega} \,+\, \nabla  \\[0.2em]  
                         \nabla\cdot  + \boldsymbol{\beta}_{b \mathrm{Lc}}^{\Omega} \cdot\quad & \quad  
                         \varrho_\mathrm{Lc}^{\Omega}
   \end{pmatrix} \,,
\end{equation}
with their coefficients related to those of $\Loriginaldiv$ \cref{origdiv_form} as follows,
\begin{equation}\label{eq:formulation-coeffs}
\begin{array}{l l l l }
\Amat_{\mathrm{L}}^{\Omega} =   \Amat_{0}^{\Omega}, \hspace*{0.5cm}& \boldsymbol{\beta}_{a \mathrm{L}}^{\Omega}  = \boldsymbol{\beta}_{a 0}^{\Omega} - \dfrac{\boldsymbol{\alpha}_{\rho_0}}{2}  ,\hspace*{0.5cm} & \boldsymbol{\beta}_{b \mathrm{L}}^{\Omega}  =  \boldsymbol{\beta}_{b 0}^{\Omega} 
  + \dfrac{\boldsymbol{\alpha}_{\rho_0}}{2}, \quad &   \varrho_\mathrm{L}^{\Omega}= \varrho^{\Omega}_0\rho_0=\dfrac{1}{c_0^2},\\
  \Amat_{\mathrm{Lc}}^{\Omega} = \dfrac{\Amat_0^{\Omega}}{c_0^2}, \quad &\boldsymbol{\beta}_{a \mathrm{Lc}}^{\Omega} 
=\boldsymbol{\beta}_{a 0}^{\Omega} -\boldsymbol{\alpha}_{c_0} - \dfrac{\boldsymbol{\alpha}_{\rho_0}}{2}  , \quad 
&\boldsymbol{\beta}_{b \mathrm{Lc}}^{\Omega}    = \boldsymbol{\beta}_{b 0}^{\Omega} + \boldsymbol{\alpha}_{c_0} +\dfrac{\boldsymbol{\alpha}_{\rho_0}}{2} ,
&\varrho_\mathrm{Lc}^{\Omega}=\varrho_0^{\Omega} c_0^2 \rho_0=1.
\end{array}
\end{equation}
Applying \cref{gencov::identity,alpha_identity} gives the equivalence among the 
three variants:
\begin{equation}
\begin{aligned}
&\Loriginaldiv \begin{pmatrix}\bxi \\[0.2em]\pressE \end{pmatrix} 
= \begin{pmatrix}\mathbf{g}\\[0.2em] h \end{pmatrix}
\hspace*{0.5cm}\Longleftrightarrow \hspace*{0.5cm}   \Lliouville\begin{pmatrix} \bxi_L\\  w_L\end{pmatrix}=\dfrac{1}{\sqrt{\rho_0} } \begin{pmatrix}\mathbf{g}\\[0.2em] \rho_0 \,h \end{pmatrix} \\
&\hspace*{0.5cm}\Longleftrightarrow  \hspace*{0.5cm}  \Lliouvillec\begin{pmatrix} \bxi_c\\  w_c\end{pmatrix}=\dfrac{1}{c_0\sqrt{\rho_0} } \begin{pmatrix}\mathbf{g}\\[0.2em] c_0^2 \, \rho_0 \,h \end{pmatrix}.
\end{aligned}\end{equation}

\begin{remark}[Background models used depending on the formulation]
  \label{remark:motivation-liouville}
  The motivation of the Liouville variants $\Lliouville$ and $\Lliouvillec$
  is that the background model density $\rho_0$ does not appear explicitly
  in the system of equations anymore, hence avoiding to numerically handle 
  the exponential decay in the atmosphere, cf., \cite{Pham2020Siam,Pham2024assembling,Pham2021Galbrun}.
  Instead, the Liouville variants rely on the inverse density scale height 
  $\alpha_{\rho}$  which is a constant when $\rho_0$ is exponentially decreasing.
  Comparing the Liouville variants, we note that $\Lliouvillec$ needs the inverse 
  velocity scale height $\alpha_{c}$, hence the derivatives of $c_0$, which might 
  complicate the consideration of non-radial velocity perturbations.
  See further~\cref{subsection:numerical-comparison-formulation} for
  numerical comparisons.
\end{remark}

%
%
%

%

\paragraph{Unification of variants} The three systems introduced above, $\Loriginaldiv$, $\Lliouville$ and $\Lliouvillec$, 
coupled with vanishing $\pressL$ boundary condition \cref{vLpBC_gen} are unified 
in the following notation with unknowns $(\bu,w)$, 
\begin{subequations}\label{uniformulation}\begin{empheq}[left={\empheqlbrace}]{align}
& \Amat \, \bu \,+\, \boldsymbol{\beta}_{1} \, w \,+\, \nabla w \,=\, \mathbf{g},  
  \hspace*{3.75em} \text{ in } \mathbb{B}_{\odot}, \label{uniformulation_eq1}\\
& \nabla\cdot    \, \bu \,+\, \boldsymbol{\beta}_{2} \,\cdot\,\bu \,+\,\varrho\, w \,=\, h, 
  \hspace*{2.5em} \text{ in } \mathbb{B}_{\odot},\label{uniformulation_eq2}\\
& w + \boldsymbol{\alpha}_{\mathrm{bc}} \cdot \bu   = 0, \hspace*{7.75em} \text{ on } \partial \mathbb{B}_{\odot} . \label{vLpBC_genvar}
\end{empheq}\end{subequations}
We denote the operator in system \cref{uniformulation} as
\begin{equation}\label{calLbullet::def}
  \mathcal{L}_\bullet := \begin{pmatrix} \Amat \quad & \quad \boldsymbol{\beta}_1 + \nabla  \\ \nabla\cdot +\boldsymbol{\beta}_2 \cdot \quad &  \varrho\end{pmatrix} \,. 
\end{equation}
 
The explicit definitions of the coefficient in \cref{uniformulation} 
vary depending on the variants, \cref{eq:formulation-coeffs}.
The coefficient $\boldsymbol{\alpha}_{\mathrm{bc}}$ of the 
boundary condition \cref{vLpBC_genvar} takes the following 
values,
\begin{equation}
\begin{array}{c|c|c} 
    \Loriginaldiv   &  \Lliouville  &  \Lliouvillec \\[.20em] \hline  & & \\[-0.750em]
  \makebox[07em][c]{$\boldsymbol{\alpha}_{\mathrm{bc}}^{0} \,=\, \nabla p_0$}
& \makebox[09em][c]{$\boldsymbol{\alpha}_{\mathrm{bc}}^{\mathrm{L}} \,=\, \nabla p_0 / \rho_0$}
& \makebox[10em][c]{$\boldsymbol{\alpha}_{\mathrm{bc}}^{\mathrm{Lc}} \,=\,\nabla p_0 / (c_0^2\,\rho_0)$}
\end{array}
\end{equation}


\subsection{Explicit expressions for non-rotating backgrounds in inertial frames}
%


A model commonly employed for non-rotating backgrounds
in the standard solar 
Model-S, \cite{christensen1996current} which is radially symmetric 
and satisfies the 
hydrostatic equilibrium \cref{steq}.
When the background parameters only depend on the radial distance $r$,
it is useful to work with the buoyancy frequency $N(r)$,
with its squared defined as, \cite{christensen2014lecture,Pham2024assembling,Pham2021Galbrun},
\begin{equation}\label{Nfreq::def}
N^2 : = \phi_0' \left( \alpha_{\rho_0} - \dfrac{\alpha_{p_0}}{\Gamma_1} \right) \, \qquad
\text{Buoyancy frequency squared}.
\end{equation}
Here $\alpha$ is the radial inverse scale height \cref{scaleheightradial}. 

In non-rotating background in inertial frame,
$\rot=\rotzero=0$, $\mathbf{v}_0=0$, and we have,
\begin{equation} 
\mathbf{Y}_{\Omega = 0} := \nabla\phi_0, \hspace*{0.5cm} \Amat_{0}^{\Omega=0} =
\rho_0 \bigg[ -\omega^2 \,-\,2\ii\omega\gamma_{\mathrm{att}}
             \,+\, \nabla\phi_0 \otimes
      \Big(\boldsymbol{\alpha}_\rho \,-\, \dfrac{\boldsymbol{\alpha}_p}{\Gamma_1} 
      \Big)       \bigg] .
\end{equation}
If we assume radial symmetry, it gives:
\begin{equation}\label{radsym::background}
    \text{Radial symmetry}: \quad  \left\{\begin{array}{c}  \Amat_0^{\Omega=0} =\Amat_0^0 =
\rho_0  \left(-\omega^2 -2\ii\omega\gamma_{\mathrm{att}}   + N^2 \,\mathbf{e}_r \otimes \mathbf{e}_r\right),\\[0.3em]
\text{For the three variants}: \quad \Amat_{\bullet}^0 = (\Amat_{\bullet}^0)^t, \quad \boldsymbol{\beta}^0_{1\bullet} \parallel \boldsymbol{\beta}^0_{2\bullet} \parallel \mathbf{e}_r \end{array} \right.\,. 
\end{equation}

Under hydrostatic equilibrium \cref{steq}, and sufficient
regularity\footnote{The form of $\Amat_0^0$ in \cref{symcoeff_norot} employs the gradient version of \cref{steq}. 
Taking the gradient on both sides of \cref{steq}, with $H$ denoting the Hessian operator, we obtain, 
\begin{equation}\label{gradHE0}H_{p_0} = -\nabla (\rho_0 \nabla\phi_0).\end{equation}
We next rewrite the right-hand side, 
\begin{equation}
\nabla (\rho_0 \nabla\phi_0) = \rho_0 H_{\phi_0} + \nabla\phi_0 \otimes \nabla \rho_0 \overset{\cref{steq}}{=}  \rho_0 H_{\phi_0} - \dfrac{\nabla p_0}{\rho_0} \otimes \nabla \rho_0 
 = \rho_0 \left( H_{\phi_0} - c_0^2 \dfrac{\boldsymbol{\alpha}_{p_0} }{\Gamma_1} \otimes \boldsymbol{\alpha}_{\rho_0} \right). \end{equation}
 In the last expression, the adiabatic relation \cref{adiaba}, $c_0^2 \rho_0 = p_0 \Gamma_1$, is employed, and the definition of $\boldsymbol{\alpha}$ \cref{scaleheight::def}. After some rearrangement, we thus obtain,
 \begin{equation} \cref{steq}+
 \text{ regularity in } p_0,\phi_0  \hspace*{0.5cm} \Rightarrow \hspace*{0.5cm} 
H_{\phi_0}
 + \dfrac{H_{p_0}}{\uprho_0} =   c_0^2\, \dfrac{\boldsymbol{\alpha}_{p_0} }{\Gamma_1}  \otimes \boldsymbol{\alpha}_{\rho_0}
  = c_0^2\,\boldsymbol{\alpha}_{\rho_0}\odot \dfrac{\boldsymbol{\alpha}_{p_0} }{\Gamma_1} .\label{steq:grad}
 \end{equation}
Here $\odot$ denotes the symmetric tensor product $\mathbf{u}\odot\mathbf{v} =  (\mathbf{u}\otimes \mathbf{v}+\mathbf{v}\otimes \mathbf{u})/2$. The third expression is due to the symmetry of the Hessians. This also means that, under hydrostatic equilibrium \cref{steq},
 \begin{equation} \nabla \rho_0 \parallel \nabla p_0  \parallel \nabla \phi_0 \hspace*{0.5cm} \text{ or equivalently } \hspace*{0.5cm} \boldsymbol{\alpha}_{\rho_0} \parallel \boldsymbol{\alpha}_{p_0} \parallel \nabla \phi_0.
 \end{equation}
} for $p_0$ and $\phi_0$, we have the following structure of the coefficients in $\mathcal{L}_\bullet$ \cref{calLbullet::def}, 
\begin{equation}\label{symcoeff_norot}
\text{\cref{steq}}:  \quad\left\{ \begin{array}{c}\Amat_0^0 =
\rho_0 \left(-\omega^2 -2\ii\omega\gamma_{\mathrm{att}}
             + \nabla\phi_0 \odot
      \left(\boldsymbol{\alpha}_\rho \,-\, \dfrac{\nabla \phi_0}{c_0^2} 
      \right)    \right),\\[0.5em]
\text{For 3 variants} : \quad\Amat_{\bullet}^0 = (\Amat_{\bullet}^0)^t , \hspace*{0.7cm}  \boldsymbol{\beta}^0_{1\bullet}= - \boldsymbol{\beta}^0_{2\bullet}   . \end{array}\right.
\end{equation}
We note that the symmetry properties in \cref{symcoeff_norot} no longer hold when rotation is included.

\begin{remark}[Mixed formulation obtained from Galbrun's equation]\label{Galbrun::rmk}
In \cref{LBOlit::rmk}, we have shown the relation between our working equation 
and the equation by Lynden-Bell \& Ostriker \cite{lynden1967stability}, referred 
to as the LBO variant, cf. \cref{LBO_Gough_equiv}.
Here we show that our mixed system \cref{eq:main:original-div} associated with \cref{eq:main:init:a}, 
coincides with the mixed system directly derived from LBO variant \cref{LBO_Gough_equiv}.
Without rotation and flow, the LBO variant \cref{LBO_Gough_equiv} simplifies to,
\begin{equation}\label{main_eqn_Galbrun_unscaled}
\begin{aligned}
 \boldsymbol{\mathfrak{L}} \, \bxi   =   \mathbf{g} \, \quad \text{ in } \mathbb{R}^3\,\,,\text{with} \quad   \boldsymbol{\mathfrak{L}} \,  \bxi  \,=\,  - \rho_0  \big( \omega^2 + 2\,\mathrm{i} \omega\, \gamma_{\mathrm{att}} \big)\,\bxi
  \,- \,  \nabla \big[ \Gamma_1 \, p_0 \nabla\cdot  \bxi  \big] 
   + (\nabla\, p_0) (\nabla \cdot \bxi )\\[0.3em]
   \hspace*{5cm}- \nabla \,(  \bxi \cdot \nabla \,p_0 ) \,+\, ( \bxi \cdot \nabla)\nabla\, p_0
  \, + \, \rho_0 \, ( \bxi  \cdot \nabla) \nabla\, \phi_0\,.
 \end{aligned}
\end{equation}
This equation was employed in our previous works \cite{Pham2021Galbrun,Pham2024assembling} and is also called Galbrun's equation in aero-acoustics.
We next rewrite this as a mixed system,
\begin{gather}
\cref{main_eqn_Galbrun_unscaled} \hspace*{0.5cm}
 \Leftrightarrow \hspace*{0.5cm} \begin{dcases}  \rho_0 \left( \mathsf{B}  - \sigma^2     \right) \bxi + \dfrac{\boldsymbol{\alpha}_p}{\Gamma_1}  \,\pressE
         + \nabla \pressE =  \mathbf{g}, \\
         \dfrac{\pressE}{\rho_0 c_0^2} 
    \,+\, \bxi \cdot {\color{black}{\dfrac{\nabla p_0}{\rho_0 c_0^2}}}
\,+\, \nabla\cdot\bxi \,=\, 0,
         \end{dcases}\label{Galbrun_equiv}\\
  \text{where} \hspace*{0.5cm}\mathsf{B}     \, := \, \dfrac{\mathrm{H}_{\mathsf{p}_0}}{\rho_0} +  \mathrm{H}_{\phi_0}\, -\,  \dfrac{ \nabla p_0 \otimes  \nabla p_0  }{\Gamma_1 p_0 \rho_0 } 
   =  \dfrac{\mathrm{H}_{p_0}}{\rho_0} +  \mathrm{H}_{\phi_0}\, -\, c_0^2\,\dfrac{\boldsymbol{\alpha}_{p} }{\Gamma_1}\odot \dfrac{\boldsymbol{\alpha}_{p} }{\Gamma_1}\,. \label{matB::def}
  \end{gather}
Employing identity \cref{steq:grad}, which replaces
$H_{\phi_0}
 + \tfrac{H_{p_0}}{\rho_0} $ by 
  $c_0^2\,\boldsymbol{\alpha}_{\rho}\odot \tfrac{\boldsymbol{\alpha}_{p} }{\Gamma_1} $, we can rewrite 
$\mathsf{B}$ as,
 \begin{equation}
\mathsf{B} \overset{ \cref{steq:grad} }{=} c_0^2\,\boldsymbol{\alpha}_{\rho}\odot \dfrac{\boldsymbol{\alpha}_{p} }{\Gamma_1}
 \, -\, c_0^2\,\dfrac{\boldsymbol{\alpha}_{p} }{\Gamma_1}\odot \dfrac{\boldsymbol{\alpha}_{p} }{\Gamma_1}
  = c_0^2   \dfrac{\boldsymbol{\alpha}_{p} }{\Gamma_1}\odot \left( \boldsymbol{\alpha}_{\rho} - \dfrac{\boldsymbol{\alpha}_{p} }{\Gamma_1}\right) \overset{\cref{steq}}{=} \nabla \phi_0 \odot \left( \boldsymbol{\alpha}_{\rho} - \dfrac{\boldsymbol{\alpha}_{p} }{\Gamma_1}\right).
 \end{equation}
This leads to the coefficients in this work, cf. \cref{symcoeff_norot}.

\end{remark}

\subsection{Relation to scalar PDEs}
\label{subsection:scalar-pde}

The relation of the mixed system \cref{uniformulation} 
with a scalar PDE is established below.
We focus on the backgrounds described in \cref{workingbackgrounds::subsec}, 
in which operator $\Amat$ is either a multiplication 
operator without rotation, or can be reduced 
to one with differential rotation. 
The details for the latter case are 
further developed in \cref{cylinexpansion_GK::subsec}.
The equivalence with a scalar PDE is obtained readily 
when there is only a scalar source, cf. \cref{eqscalar_novsrc::identity}.
Here we limit our discussion to a one-way relation for vector source.

\medskip

\paragraph{With scalar source} 
We consider  \cref{uniformulation} when 
there is only a scalar source input, i.e. $\mathbf{g}=0$. 
We introduce a scalar differential operator of anisotropic 
convected Helmholtz type,
\begin{equation}\label{calLomega_op::def}
L^{\Omega} w:= 
-\nabla \cdot \Amat^{-1}  \nabla w - (\Amat^{-t}\boldsymbol{\beta}_2 +  \Amat^{-1}\boldsymbol{\beta}_1 ) \cdot \nabla w +w \left[\varrho - \boldsymbol{\beta}_2 \cdot \Amat^{-1}\cdot  \boldsymbol{\beta}_1 -\nabla \cdot (  \Amat^{-1}\boldsymbol{\beta}_1)\right]  \,. 
\end{equation}
In the case $\rot=\rotzero=0$, the above operator simplifies to,
\begin{equation}\label{Lscalar_op::def}L w:= 
-\nabla \cdot \Amat^{-1}  \nabla w  + \left(\varrho + \boldsymbol{\gamma}\cdot \Amat^{-1} \cdot\boldsymbol{\gamma}  +  \nabla \cdot (\Amat^{-1}  \boldsymbol{\gamma})\right)w; \quad \boldsymbol{\gamma} := \boldsymbol{\beta}_2^0 = -\boldsymbol{\beta}_1^0.
\end{equation}
\begin{identity}\label{equiv2nd1st3d}
For generic invertible matrix-valued $\Amat$, vector-valued 
$\boldsymbol{\beta}$ and scalar-valued $\varrho$ functions, we have the following equivalence,
 
 \begin{minipage}{0.48\linewidth}
\begin{subequations} \label{1stsystem}
\begin{empheq}[left={ \empheqlbrace\,}]{align}
\Amat \mathbf{u} + \boldsymbol{\beta}_1 w + \nabla w & = 0 \,, \label{1stsystem1} \\
\nabla\cdot \mathbf{u} + \boldsymbol{\beta}_2 \cdot\mathbf{u} + \varrho w &= h \,.\label{1stsystem2}
\end{empheq}
\end{subequations}\end{minipage}\hspace*{0.2cm}
\begin{minipage}{0.45\linewidth}
\begin{equation}\label{2ndscalarpde} 
\Longleftrightarrow \hspace*{0.6cm}L^{\Omega} w\, = \,h.\end{equation}
 \end{minipage}

\end{identity}

\begin{proof}
We employ \cref{1stsystem1} as the defining expression for 
$\mathbf{u} =- A^{-1}(\boldsymbol{\beta}_1 w +  \nabla w)$. 
Substitute this in the second equation \cref{1stsystem2}, employing $\nabla \cdot (A^{-1}\boldsymbol{\beta}_1 w) = w \nabla \cdot (  A^{-1}\boldsymbol{\beta}_1) + (A^{-1}\boldsymbol{\beta}_1) \cdot \nabla w$, we get
\begin{equation}
\begin{aligned}
\nabla\cdot \mathbf{u} & + \boldsymbol{\beta}_2 \cdot\mathbf{u} + \varrho\, w
  = -\nabla \cdot [ \Amat^{-1}(\boldsymbol{\beta}_1 w +  \nabla w )]- \boldsymbol{\beta}_2 \cdot \Amat^{-1}(\boldsymbol{\beta}_1 w +  \nabla w )+ \varrho \,w\\
&=    -\nabla \cdot \Amat^{-1}  \nabla w +w \left[\varrho - \boldsymbol{\beta}_2 \cdot \Amat^{-1}\cdot  \boldsymbol{\beta}_1 -\nabla \cdot (\Amat^{-1}\boldsymbol{\beta}_1)\right] - (\Amat^{-t}\boldsymbol{\beta}_2 +  \Amat^{-1}\boldsymbol{\beta}_1 ) \cdot \nabla w \,.
\end{aligned}
\end{equation}
\end{proof}

As a corollary, we obtain the following identity for the case without rotation.
 \begin{identity}\label{eqscalar_novsrc::identity}
Without rotation and flow, i.e., $\rot=\rotzero=0$, we have for all three variants, 
 \begin{equation}
  \Amat^0 = (\Amat^0)^t, 
  \quad \boldsymbol{\beta}_2^0 = -\boldsymbol{\beta}_1^0 =: \boldsymbol{\gamma} \,.
\end{equation}
The system \cref{uniformulation} with $\mathbf{g}= 0$ is equivalent to an anisotropic Helmholtz equation,
\begin{equation}\label{eqscalar_notrot}
\begin{aligned}  \begin{dcases}
\Amat^0 \mathbf{u} -\boldsymbol{\gamma} w + \nabla w  = 0 \\
\nabla\cdot\mathbf{u} + \boldsymbol{\gamma} \cdot \mathbf{u} + \varrho \,w = h \\
\end{dcases} \hspace*{0.5cm} \Longleftrightarrow \hspace*{0.5cm}L w = h, \hspace*{0.2cm} \text{ with } L \text{ defined in }\cref{Lscalar_op::def}.
 \end{aligned}\end{equation}
\end{identity}


\paragraph{With vector source}
Employing essentially the same proof, with $L^{\Omega}$ defined in \cref{calLomega_op::def}, for a source $\mathbf{g}$ with enough regularity, we can still arrive at a scalar equation (here without declaring an equivalence),
\begin{equation}
\mathcal{L}\begin{pmatrix}  \bu \\ w\end{pmatrix} = \begin{pmatrix}  \mathbf{g}\\ h\end{pmatrix}
\quad \Longrightarrow \quad \mathcal{L}^{\Omega}  w = h - g_{\mathrm{scalar}} , \quad \text{where }\hspace*{0.2cm} g_{\mathrm{scalar}}:=( \nabla \cdot + \boldsymbol{\beta}_2\cdot )\mathbf{A}^{-1} \mathbf{g}.
\end{equation}

\subsection{Elliptic and hyperbolic regions for radial backgrounds at zero attenuation}
The equivalence in \cref{eqscalar_notrot} helps give insight in the nature of the vector operator. 
Since the nature of the PDE is the same for all three variants, for convenience, we consider the Liouville variant $\Lliouville$. 
For a radially symmetric background, from \cref{radsym::background}, we have 
\begin{gather}
 \Amat_L^0 =-\sigma^2\mathbb{Id} + N^2 \,\mathbf{e}_r \otimes \mathbf{e}_r, \quad \sigma^2:= \omega^2 +2\ii\omega\gamma_{\mathrm{att}} \nonumber \\
  \Rightarrow \quad (\Amat_L^0)^{-1} = \dfrac{1}{N^2-\sigma^2} \mathbf{e}_r\otimes \mathbf{e}_r
 -\dfrac{1}{\sigma^2} \left( \mathbf{e}_{\theta}\otimes \mathbf{e}_{\theta} + \mathbf{e}_{\phi}\otimes \mathbf{e}_{\phi}  \right) \,.
\end{gather}
The highest-order term in $\cref{eqscalar_notrot} $ takes the following form
without attenuation ($\gamma_{\mathrm{att}}= 0$):
\begin{equation}\nabla\cdot (\Amat_L^0)^{-1} \nabla 
\,\, = \,\, \dfrac{1}{r^2} \partial_r\dfrac{r^2}{N^2-\omega^2} \partial_r   - \dfrac{1}{\omega^2 r^2 } \Delta_{\mathbb{S}^2} \,.
\end{equation}
Then, the operator $L$ \cref{Lscalar_op::def}
is elliptic when the coefficient of $\partial_r^2$ and $\Delta_{\mathbb{S}^2}$ have the same sign, and is of hyperbolic type when 
they are of opposite signs. 
This observation leads to the following definitions,
\begin{equation}\label{ellhyp::def}
\begin{aligned}
& \Bell:=\{ \omega^2 > N^2(r)\}  \hspace*{1.15cm} \text{(Elliptic region)}, \\
& \Bhyp:=\{\omega^2 < N^2(r)\}   \hspace*{1.00cm} \text{(Hyperbolic region)}.
\end{aligned}
\end{equation}

\begin{remark}
 Throughout the work, we work with nonzero-attenuation, which guarantees 
 the ellipticity of the problem. 
 However, at small levels of attenuation, at frequency $\omega$ and position 
 of Dirac source $\mathbf{s}$  such that $\mathbf{s} \in \Bhyp$, we will see 
 that the solutions manifest behaviors reminiscent of those of wave operators, 
 in particular propagation of singularities along line-cone structures, 
 cf. \cref{numerical-hdg:qualitative}.
 These notions \cref{ellhyp::def} will also be employed to provide 
 interpretations for the choice of the HDG stabilization 
 in \cref{subsection:hdg-stabilization}. 
\end{remark}

\section{Cylindrical decomposition for axially symmetric backgrounds} 
\label{section:axi-symmetric}

In this section, we exploit the axial symmetry
of the background parameters to derive the system in so-called 2.5D.
We recall the generic system of equations \cref{uniformulation} 
with unknowns $(\bu,w)$ and backgrounds that verify \cref{difflowbackground::def},
\begin{subequations}\label{eq:main-section-axisym}
\begin{empheq}[left={\empheqlbrace}]{align}
  \Amat \, \bu \,+\, \boldsymbol{\beta}_1 \, w \,+\, \nabla w \,=\, \mathbf{g}, \\
  \nabla\cdot \, \bu \,+\, \boldsymbol{\beta}_2 \,\cdot\,\bu \,+\,\varrho\, w \,=\, h,
\end{empheq}\end{subequations}
with $\mathbf{g}$ and $h$ supported away from the rotation axis $\mathbf{e}_z$. 
For simplification in terms of boundary condition \cref{vLpBC_gen}, we assume that,
$\nabla p_0$ is parallel to the (outward-pointing) normal vector of $\mathbb{B}_{\odot}$ denoted
by $\mathbf{n}_{\mathbb{B}_{\odot}}$, i.e.
\begin{equation}\label{assump:simpLgrad}
\nabla p_0  \parallel  \mathbf{n}_{\mathbb{B}_{\odot}}  , \quad  \text{on }  \partial \mathbb{B}_{\odot} .
\end{equation} 

We will decouple the above system  which is axially symmetric 
with respect to the z-axis $\mathbf{e}_z$ into modal problems 
defined for each cylindrical/azimuthal mode
$m$ (or equivalently Fourier series in $\phi$) 
in \cref{modalprob::subsec}. 
In \cref{cylinexpansion_GK::subsec}, the cylindrical expansion of the 
scalar Green's kernel $G_p$ is introduced.

\subsection{Azimuthal modal problem}\label{modalprob::subsec}

\begin{figure}[ht!] \centering
  \includegraphics[scale=0.65]{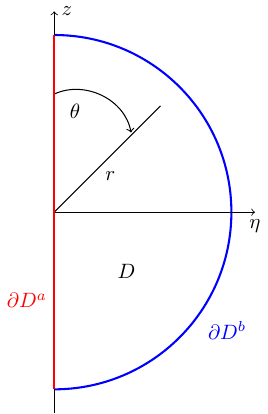}
  \caption{Illustration of the numerical domain for axisymmetry 
           which corresponds to the meridional half-disk. 
           The boundary is separated between $\partial D^a$ on
           axis $\mathbf{e}_z$, and $\partial D^b$ for the exterior.}
  \label{figure:sketch-domain-2.5d}
\end{figure}

Due to the symmetry, we work in cylindrical coordinate system $(\eta,\phi,z)$, in 
which the coefficients of operator $\Amat$,
$\boldsymbol{\beta}_{1}$, $\boldsymbol{\beta}_2$ and $\varrho$ are independent of $\phi$.
We list some general notations required to state the problem:
\begin{itemize}[leftmargin = *]
\item 
We denote by the square bracket $[\cdot]$ the components of a tensor in the basis $(\mathbf{e}_\eta,\mathbf{e}_\phi, \mathbf{e}_z)$, specifically for a vector $\bv$, 
\begin{equation}
    [\bv] =  ( v_\eta , v_\phi, v_z)^t , \quad \text{ for } \bv = v_\eta \mathbf{e}_\eta + v_{\phi} \mathbf{e}_\phi + v_z \mathbf{e}_z.
\end{equation}

\item The modal problem is defined on the restriction of $\mathbb{B}_{\odot}$ 
      to the meridional half-plane. For simplicity we called this geometry 
      the meridional half-disk $D$, illustrated in \cref{figure:sketch-domain-2.5d}.

\item We denote the boundary of the meridional half-disk $D$ 
by $\partial D$, and partition this boundary into the 
exterior one, denoted by $\partial^b D$, and the portion 
lying in the rotation axis $\mathbf{e}_z$, 
denoted by $\partial^a D$, see \cref{figure:sketch-domain-2.5d}. 
In terms of $(\eta,\phi,z)$, $D$ and its boundary 
are parametrized by $(\eta,z) \in \mathbb{R}^+ \times \mathbb{R}$,
\begin{equation}
\begin{aligned}
&   D = \{ (\eta, z) \in \mathbb{R}^+ \times \mathbb{R}\, \big| \,(\eta, z, \phi) \in \mathbb{B}_{\odot}\} \hspace*{1cm}  \partial D = \partial D^a \cup \partial D^b,\\\
 & \hspace*{1cm} \text{where} \,\, \partial D^a =\{( 0, z)  \in D \}, \quad \partial D^b = \partial D \setminus \partial D^a\,.
  \end{aligned}
\end{equation}

\item 
We denote the cylindrical components of the outward pointing normal vector of $D$ in the meridional 
plane by $\tilde{\n}$, in relation to the normal vector 
$\mathbf{n}_{\mathbb{B}_{\odot}}$ of $\mathbb{B}_{\odot}$ 
in 3D, we have
\begin{equation}\label{eq:normals-axisym}
  \tilde{\n} \,=\, (n_\eta,n_z)^t, \quad  \n:=[\n_{\mathbb{B}_{\odot}}] = (n_\eta,0,n_z)^t \,.
\end{equation}
\end{itemize}

We note that the discussion in this section works for 
a general geometry of $\mathbb{B}_{\odot}$ that is 
rotationally invariant, such as an oblate spheroid.

\begin{remark} When $\mathbb{B}_{\odot}$ is a sphere, 
$\mathbb{B}_{\odot} = \{ \mathbf{x} \in \mathbb{R}^3 \, | \, \lvert \mathbf{x}\rvert \leq  \rmax\} $, the domain $D$ and its boundary can be more explicitly described as, 
\begin{equation}
\begin{aligned}
&   D = \{ (\eta, z) \in \mathbb{R}^+ \times \mathbb{R}\, \big| \, \eta^2 + z^2  \leq  \rmax^2 \}, \hspace*{1cm}  \partial D = \partial D^a \cup \partial D^b,\\\
 & \hspace*{1cm} \text{where} \,\, \partial D^a =\{( 0, z) \, |\, \lvert z\rvert \leq \rmax\}, \quad \partial D^b = \{  (\eta, z) \in \mathbb{R}^+ \times \mathbb{R} \, \big|\, \eta^2 + z^2  = \rmax^2\}\,.
  \end{aligned}
\end{equation}
\end{remark}

\paragraph{Cylindrical expansion of the unknowns and sources}
For $\mathbf{x}$ away from the rotation axis $\mathbf{e}_z$, 
we write the scalar unknown $w$ and scalar source $h$ in cylindrical expansion, which is the Fourier series expansion in $\phi $:
\begin{equation}
w (\eta,\phi,z)=  \sum_{m\in \mathbb{Z}}  w_m(\eta,z) \, e^{\mathrm{i}m\phi} , \hspace*{2em}
h (\eta,\phi,z)=  \sum_{m\in \mathbb{Z}}  h_m(\eta,z) \, e^{\mathrm{i}m\phi} \,.
\end{equation}
For the vector quantities, i.e., the source $\mathbf{g}$ and unknown $\bu$, 
the cylindrical expansion is applied to their components in $(\mathbf{e}_\eta,\mathbf{e}_\phi, \mathbf{e}_z)$ in cylindrical expansion, such that
\begin{equation}
   \begin{aligned}
   [\bu](\eta,\phi,z)\, &=\, \sum_{m\in \mathbb{Z}} \bu_m(\eta,z)\,   e^{\mathrm{i}m \phi}
\,, \quad \text{where} \,\, 
 \bu_m(\eta,z) = (u_{\eta m}, u_{\phi m}, u_{z m} )^t, \\
[\mathbf{g}](\eta,\phi,z)\, &= \,  
\sum_{m\in \mathbb{Z}} \mathbf{g}_m(\eta,z)\,  e^{\mathrm{i}m\phi} 
 \,, \quad \text{where} \,\,  \mathbf{g}_m(\eta,z) = (g_{\eta m}, g_{\phi m}, g_{z m} )^t\,.
  \end{aligned} 
 \end{equation}
We also denote the restriction to the $(\eta,z)$ component by
\begin{equation} 
  \widetilde{\bu_m} \,=\, (u_{m\eta}, u_{mz})^\mathrm{t}.
\end{equation}

\paragraph{Useful identities}
Below are some identities needed to derive the modal problem.
\begin{enumerate}[leftmargin = *]
\item
Introducing the operator $ [\nabla]_m$ acting on a scalar function $w$, 
and $[\nabla\cdot]_m$ acting on a vector $\bv = (v_1,v_2,v_3)^t$, we have,
\begin{equation}\label{modalgraddiv}
 [\nabla]_m w := ( \partial_\eta w, \tfrac{\ii\,m}{\eta} w,\partial_z w)^t\,, 
 \qquad 
 [\nabla\cdot]_m \bv
  =  ( \partial_\eta + \tfrac{1}{\eta} ) v_1
  \,+\,   \tfrac{\ii\,m}{\eta} v_2 \,+\, \partial_z v_3  \,.
\end{equation}
They describe the action of the gradient 
and divergence operators on each azimuthal 
mode $m$:
\begin{equation}\label{eq:axisym-id-grad-div}
\nabla w (\eta,\phi,z)= \sum_{m\in \mathbb{Z}}[\nabla]_m w_m(\eta,z) \,e^{\ii\, m \, \phi}, \quad \nabla \cdot \bv (\eta,\phi,z)= \sum_{m\in \mathbb{Z}}
[\nabla\cdot]_m  \bv_m  \,e^{\ii\, m \, \phi} \,.
\end{equation}

\item
Due to the assumption on the rotation,
the differential operator $\Amat$ is reduced  on each mode $m$ to a 
multiplication operator with $3\times 3$ matrix-valued functions 
denoted by $\Amat_m(\eta,z)$,
\begin{equation}\label{actionAmat::def}
[\Amat \bu](\eta,\phi,z)  = \sum_{m \in \mathbb{Z}}
\Amat_m(\eta,z)  \bu_m(\eta,z)  \,e^{\ii\, m \, \phi}\,.
\end{equation}
The explicit expression of $\Amat_m(\eta,z)$ 
with differential rotation is further given in \cref{appendix:rotation-axisym}. 

\item
Under assumption \cref{assump:simpLgrad}, 
the boundary condition becomes:
\begin{equation}\label{bcLgradial} w + \alpha_{\mathrm{bc}} \bu \cdot \n_{\mathbb{B}_{\odot}} = 0 , \quad \text{ where } \quad \alpha_{\mathrm{bc}}: =\boldsymbol{\alpha}_{\mathrm{bc}}  \cdot \n_{\mathbb{B}_{\odot}} \,.
\end{equation}
We also note that $\n_{\mathbb{B}_{\odot}} = \mathbf{e}_r$, 
 \begin{equation}
\bu \cdot \n_{\mathbb{B}_{\odot}} = [\bu] \cdot \n = \sum_{m\in \mathbb{Z}} \widetilde{\bu_m} \cdot \tilde{\n}
 \, e^{\ii m \phi}= \sum_{m\in \mathbb{Z}}  \bu_m \cdot \n\, e^{\ii m \phi}.
\end{equation}
 The cylindrical decomposition of \cref{bcLgradial} readily gives, for each mode $m$, 
\begin{equation}
w_m +  \alpha_{\mathrm{bc}} \,\bu_m \cdot \n = 0 , \quad \text{ on } \partial D^b .
\end{equation}
\end{enumerate}

\paragraph{Boundary value problems on azimuthal mode $m$} 
By employing the above identities and upon equating coefficients 
of $e^{\ii m \phi}$ on both sides of \cref{eq:main-section-axisym}, 
we obtain an equation for each mode $m$. 
With assumption \cref{assump:simpLgrad}, in terms of 
unknowns $(\bu_m(\eta,z) , w_m(\eta,z))$ defined on $D$, we have,
\begin{subequations}\label{eq:main-section-axisym:mode_pre}
\begin{empheq}[left={\empheqlbrace}]{align}
  \Amat_m \, \bu_m 
  \,+\, [\boldsymbol{\beta}_1] \, w_m 
  \,+\, [\nabla]_m w_m \,=\, \mathbf{g}_m, \\
          [\nabla\cdot]_m \bu_m
    \,+\, [\boldsymbol{\beta}_2] \cdot\bu_m \,+\,\varrho\, w_m \,=\, h_m.
\end{empheq}\end{subequations}
The system is coupled with vanishing Lagrangian pressure 
on the exterior boundary $\partial D^b$ and symmetry 
condition on the rotation axis $\partial D^a$:
\begin{equation}\label{bc:main-section-axisym:mode_pre}
  \bu_m \cdot \n \,=\, 0\,,\quad \text{on } \partial D^a \,; \hspace*{3em}
 w_m +  \alpha_{\mathrm{bc}} \,\bu_m \cdot \n  \,=\, 0\,,\quad      \text{on } \partial D^b \,.
\end{equation}

We denote the operator in the modal problem \cref{eq:main-section-axisym:mode_pre} as $\mathcal{L}_m$:
\begin{equation}\label{calLM::def}\mathcal{L}_m  \begin{pmatrix} \bu_m \\ w_m \end{pmatrix} = \begin{pmatrix} 
 \Amat_m  \,\,&  \,\,
 [\boldsymbol{\beta}_1]  
  + [\nabla]_m   \\ 
          [\nabla\cdot]_m  \, + \,[\boldsymbol{\beta}_2] \cdot \,\, & \,\, 
 +\varrho \end{pmatrix}  \begin{pmatrix} \bu_m \\ w_m \end{pmatrix} \,.
\end{equation}

\paragraph{Weak solution} 
Denote the cylindrical components $(\eta,z)$ by $\tilde{\mathbf{x}}$, 
\begin{equation}
\tilde{\mathbf{x}} := (\eta,z) , \quad d\tilde{\mathbf{x}} = d\eta dz \,.
\end{equation} 
The weak solution to \cref{eq:main-section-axisym:mode_pre} is defined with respect to weighted $L^2(D,\eta d\tilde{\mathbf{x}})$: for pair of test functions, $
\boldsymbol{\psi}(\tilde{\mathbf{x}}) \in \mathcal{C}^{\infty}_c(D)^3$ and $\varphi(\tilde{\mathbf{x}}) \in \mathcal{C}^{\infty}_c(D)$,
\begin{equation}\label{weaksolmodal::def}
\int_D  \begin{pmatrix}\boldsymbol{\psi} \\ \varphi \end{pmatrix} \, \cdot \mathcal{L}_m \begin{pmatrix} \bu_m \\ w_m \end{pmatrix} 
\eta \,d\tilde{\mathbf{x}} = \int_D   \begin{pmatrix} \mathbf{g}_m \\ h_m\end{pmatrix} \cdot\begin{pmatrix}\boldsymbol{\psi} \\ \varphi \end{pmatrix} \eta\, d\tilde{\mathbf{x}}\,.
\end{equation}

\subsection{Cylindrical expansion of the pressure perturbation Green's kernel}\label{cylinexpansion_GK::subsec}

We first note the equivalence of system \cref{eq:main-section-axisym:mode_pre} 
with $\mathbf{g}_m=0$, to a scalar second-order
PDE in $(\eta,z)$. For each mode $m$, introduce the 
differential operator $L_m:$
\begin{equation}
 L_m:= -[\nabla \cdot]_m \,A_m^{-1}\,  [\nabla]_m    -   V_m \cdot  [\nabla]_m  +  Q_m,
 \end{equation}
 with convection coefficient $V_m$ 
and potential $Q_m$ given by 
\begin{equation}V_m := A_m^{-t}[\boldsymbol{\beta}_2] +   A_m^{-1}[\boldsymbol{\beta}_1] , \hspace*{0.4cm} Q_m:= \varrho \, - \,[\boldsymbol{\beta}_2] \cdot A_m^{-1}\cdot  [\boldsymbol{\beta}_1] -[\nabla \cdot]_m (  A_m^{-1}[\boldsymbol{\beta}_1]).
\end{equation}

Following the same algebraic manipulation as in the proof of \cref{equiv2nd1st3d}, we have
\begin{equation}\label{cylLmdef}
\mathcal{L}_m \begin{pmatrix} \bu_m \\ w_m \end{pmatrix}
 =  \begin{pmatrix} \boldsymbol{0} \\ h_m\end{pmatrix} \\[0.3em]\quad 
 \Leftrightarrow \quad  L_m w_m \,=\, h_m.
 \end{equation}
Each side of \cref{cylLmdef} defines solution in the sense of \cref{weaksolmodal::def}, i.e
\begin{equation}\label{weaksolmodal_scalar::def} 
\begin{aligned}
\text{lhs of } \cref{cylLmdef} \hspace*{0.2cm}& \Leftrightarrow \hspace*{0.2cm}
 \int_D  \begin{pmatrix}\boldsymbol{\psi}(\tilde{\mathbf{x}}) \\\varphi(\tilde{\mathbf{x}})\end{pmatrix} \, \cdot \mathcal{L}_m \begin{pmatrix} \bu_m \\ w_m \end{pmatrix}
\eta \,d\tilde{\mathbf{x}} = \int_D   \begin{pmatrix}0 \\ h_m(\tilde{\mathbf{x}}) \varphi(\tilde{\mathbf{x}})\end{pmatrix}  \!\eta\, d\tilde{\mathbf{x}},\\
\text{rhs of } \cref{cylLmdef} \hspace*{0.2cm}& \Leftrightarrow \hspace*{0.2cm} \int_D \!\varphi(\tilde{\mathbf{x}}) L_m w(\tilde{\mathbf{x}})\, \eta d\tilde{\mathbf{x}} = \int_D\! h_m (\tilde{\mathbf{x}})\varphi(\tilde{\mathbf{x}})\, \eta\, d\tilde{\mathbf{x}}. 
 \end{aligned}
 \end{equation}

\begin{mdframed}
\begin{definition}[Azimuthal modal Green's kernel]\label{amodalGK::def} 
Consider the differential operators $L_m$, $\mathcal{L}_m$ 
defined in \cref{cylLmdef} and \cref{calLM::def}, respectively.
We denote by $G^m_p(\tilde{\mathbf{x}},\tilde{\mathbf{x}}')$ 
the fundamental solution to $L_m$ 
in the following sense: 
\begin{equation}\label{azithGP::def}
\int_D \varphi(\tilde{\mathbf{x}}) \,L_m G_{p}^m(\tilde{\mathbf{x}};\tilde{\mathbf{x}}') \, \eta \,d\tilde{\mathbf{x}} =  \varphi(\tilde{\mathbf{x}}') 
\hspace*{0.5cm}\Longleftrightarrow \hspace*{0.5cm} 
\int_D \! \begin{pmatrix}\boldsymbol{\psi} \\ w_m \end{pmatrix} \cdot \mathcal{L}^m\! \begin{pmatrix} \bu^m_p \\ G^m_p \end{pmatrix}  
\eta\, d\tilde{\mathbf{x}} = \begin{pmatrix} \boldsymbol{0} \\  \varphi(\tilde{\mathbf{x}}')\end{pmatrix}\,,
\end{equation}
for all test functions, 
$\varphi(\tilde{\mathbf{x}}) \in \mathcal{C}^{\infty}_c(D)$, and 
$\boldsymbol{\psi}(\tilde{\mathbf{x}}) \in \mathcal{C}^{\infty}_c(D)^3$.
\end{definition} \end{mdframed}
  
 We write \cref{azithGP::def} as
\begin{equation}
 L_m G^m_p(\tilde{\mathbf{x}};\tilde{\mathbf{x}}') = \dfrac{1}{\eta}\delta(\tilde{\mathbf{x}}-\tilde{\mathbf{x}}'), \hspace*{1cm} 
\hspace*{0.1cm}\Leftrightarrow \hspace*{0.1cm} \,\, \mathcal{L}_{m} \begin{pmatrix} \bu^m_p \\ G^m_p \end{pmatrix} \,=\, \dfrac{1}{\eta}\begin{pmatrix}\boldsymbol{0} \\ \delta(\tilde{\mathbf{x}}-\tilde{\mathbf{x}}')\end{pmatrix} \,.
 \end{equation}
For a regular source $h$, the solution to \begin{equation}
\mathcal{L}_m \begin{pmatrix} \bu_m \\ w_m \end{pmatrix} = \begin{pmatrix} 0 \\ h_m(\tilde{\mathbf{x}})\end{pmatrix} 
\quad \text{ or equivalently }  \quad L_m w_m = h_m \, ,
\end{equation}
in the sense of \cref{weaksolmodal_scalar::def} is given by 
\begin{equation}\label{modalweaksolwG}
w_m (\tilde{\mathbf{x}})\!= \!\int_{D} \!G^m_p(\tilde{\mathbf{x}},\tilde{\mathbf{x}}') \, h(\tilde{\mathbf{x}}')\, \eta' d\tilde{\mathbf{x}}' \,.
\end{equation}

\begin{identity}\label{cylexpanGK::identity} 
The Green's kernel $G_p$ \cref{scalarppGKgen::def} 
has a cylindrical expansion with coefficients 
given by $G^m_p$ \cref{azithGP::def}:
\begin{equation}
 G_p(\mathbf{x},\mathbf{s}) = \dfrac{1}{2\pi} \sum_{m\in \mathbb{Z}} G^m_p(\tilde{\mathbf{x}},\tilde{\mathbf{x}}')\, e^{\ii m (\phi-\phi')}.
\end{equation}
\end{identity}
\begin{proof}
Consider regular functions 
$w(\mathbf{x})$ and $h(\mathbf{x})$, 
with coefficients in the cylindrical 
expansion denoted by $w_m$ and $h_m$, i.e.,
\begin{equation}
\begin{aligned}
w = \sum_{m\in \mathbb{Z}} w_m(\tilde{\mathbf{x}})\, e^{\ii m \phi}, \quad  w_m = \dfrac{1}{2\pi}\int_0^{2\pi} w(\tilde{\mathbf{x}},\phi) \, e^{-\ii m \phi} \, d\phi,\\
h= \sum_{m\in \mathbb{Z}}h_m(\tilde{\mathbf{x}})\, e^{\ii m \phi}, \quad h_m  = \dfrac{1}{2\pi}\int_0^{2\pi} h(\tilde{\mathbf{x}},\phi) \, e^{-\ii m \phi} \, d\phi \,. \\
 \end{aligned}\end{equation}
 From the derivation of the azimuthal modal problem \cref{eq:main-section-axisym:mode_pre}, we have: 
\begin{equation}
\mathcal{L} \begin{pmatrix} \bu \\ w \end{pmatrix}
  = \begin{pmatrix} 0 \\ h\end{pmatrix} \hspace*{0.2cm}\Leftrightarrow \hspace*{0.2cm}\mathcal{L}_m \begin{pmatrix} \bu_m \\ w_m \end{pmatrix}
   = \begin{pmatrix} 0 \\ h_m\end{pmatrix} , \,\, 
 \forall \,m\in \mathbb{Z} \hspace*{0.2cm}\Leftrightarrow \hspace*{0.2cm} L_m w_m = h_m, \,\, 
 \forall \,m\in \mathbb{Z} \,. 
\end{equation}

From \cref{modalweaksolwG} and the definition of 
$G_p$ of \cref{scalarppGKgen_sol}, we have,
\begin{equation}
w_m(\tilde{\mathbf{x}}) = \int_D G^m_p(\tilde{\mathbf{x}},\tilde{\mathbf{x}}')\, h_m(\tilde{\mathbf{x}}') \eta\, d\tilde{\mathbf{x}}', \qquad 
w = \int_{\mathbb{B}_{\odot}} G_p(\mathbf{x},\mathbf{s}) \, h(\mathbf{s})\, d\mathbf{s} \,.
\end{equation}
Replacing the first equality in the expansion of $w$, we obtain,
\begin{equation}
\begin{aligned}
w &= \sum_{m\in \mathbb{Z}} w_m(\tilde{\mathbf{x}})\, e^{\ii m \phi} = \sum_{m\in \mathbb{Z}} \left(\int_D G^m_p(\tilde{\mathbf{x}},\tilde{\mathbf{x}}')\, h_m(\tilde{\mathbf{x}}') \eta\, d\tilde{\mathbf{x}}'\right) e^{\ii m \phi}\\
&= \sum_{m\in \mathbb{Z}} \int_D G^m_p(\tilde{\mathbf{x}},\tilde{\mathbf{x}}')\,e^{\ii m \phi}\, \left(\dfrac{1}{2\pi}\int_0^{2\pi} h(\tilde{\mathbf{x}}',\phi') \, e^{-\ii m \phi'} \, d\phi' \right)\eta'\, d\tilde{\mathbf{x}}'\\
&= \dfrac{1}{2\pi}\int_0^{2\pi} \int_D\left( \sum_{m\in \mathbb{Z}}  G^m_p(\tilde{\mathbf{x}},\tilde{\mathbf{x}}')\, e^{\ii m (\phi-\phi')} \right) h(\tilde{\mathbf{x}}',\phi')  \eta'\, d\tilde{\mathbf{x}}'d\phi' \,.
\end{aligned}
\end{equation}
We thus obtain the stated expansion for $G_p$.

\end{proof}

\section{HDG formulation of the cylindrical modal equations}
\label{section:axi-symmetric-hdg}

In \cref{section:axi-symmetric}, we have derived the modal boundary value 
problem \cref{eq:main-section-axisym:mode_pre,bc:main-section-axisym:mode_pre} 
for each azimuthal mode $m$ on the meridional half-disk $D$, \cref{figure:sketch-domain-2.5d}. 
For the remaining of the section, the subscript $m$ is dropped 
from the unknowns and sources, such that we consider the system,
\begin{equation}\label{eq:main-section-axisym:mode}
\begin{dcases}
  \boldsymbol{A}_m \bu
  + \boldsymbol{\beta}_1 w
  + [\nabla]_m w = \mathbf{g} ,\\ 
          [\nabla\cdot]_m \bu
    + \boldsymbol{\beta}_2 \cdot\bu +\varrho\, w = h.
\end{dcases}   \text{ on } D,\hspace*{0.1cm} \text{ with boundary conditions }  \begin{dcases}  \bu \cdot \n= 0,\hspace*{3.9em} \text{on } \partial D^a ,\\ w +  \alpha_{\mathrm{bc}} \,\bu \cdot \n  = 0, \,\,      \text{on } \partial D^b. \end{dcases}
\end{equation}
Recall that $\bu(\eta,z) = (u_\eta(\eta,z),u_\phi(\eta,z),u_z(\eta,z))^t$ 
and $\n$ is the normal vector defined from $\tilde{\n}$ 
of $\partial D= \partial D^a \cup \partial D^b$, cf. \cref{eq:normals-axisym}. 

We solve this problem with the Hybridizable Discontinuous Galerkin 
method (HDG, \cite{Cockburn2009}), following the perspective of 
\cite{Pham2024stabilization} for notations.
In the HDG formulation, one distinguishes between the value of 
an unknown on the union of the interior of mesh cells and its 
trace on the skeleton of the mesh, here corresponding to a set 
of edges. 
Specifically, in addition to the volume unknowns 
$(\bu,w)$ approximating the value (of the corresponding quantities) 
in the interior of the cells, another unknown that we referred to as the 
HDG numerical trace $\lambda$ is introduced and defined on the set of 
edges, to represent the Dirichlet trace of $w$ there.  
With this variable, the global system \cref{eq:main-section-axisym:mode_pre} 
on $D$ is written as a union of boundary value problems (local BVP) 
each defined on a mesh cell with boundary data $\lambda$.

These local BVPs are decoupled from one another, i.e., independent per cell,
forming block-diagonal structures in the global linear system 
(in terms of unknowns $(\bu,w,\lambda)$).
The link between cells is carried out by the constraints on the 
boundary data $\lambda$, and imposes a certain degree of regularity 
of the solution across the interfaces.
The aforementioned block-diagonal structure allows to employ 
static condensation to yield a linear system solely in terms 
of $\lambda$, which forms the essence of the HDG method in 
the DG family. 
Another specificity of a HDG method is the choice of stabilization 
of the error in the Neumann trace, which in implementation provides 
an expression for the numerical trace $\yhwidehat{\bu \cdot \n}$
(which approximates the trace of $\bu\cdot \n = \tilde{\bu}\cdot\tilde{\n}$, 
see below), in terms of $\lambda $ and the volume unknowns. 

\subsection{Domain discretization}

We discretize $D$ with a non-overlapping 
conforming mesh denoted by $\mesh$, 
which consists of $\ncell$ elements $K$. 
The skeleton of the mesh is denoted by $\setface$ and 
is composed of $\nface$ faces, which are divided into
$\setface^i$ the set of interfaces, $\setface^b$ 
discretizing $\partial D^b$ and $\setface^a$ 
discretizing $\partial D^a$. We have
\begin{equation}
  \mesh = \bigcup_{e=1}^{\ncell} K_e \,, \qquad \qquad 
  \setface = \bigcup_{e=1}^{\ncell} \partial K_e  = \setface^b \cup \setface^a \cup  \setface^i \,=\, \bigcup_{k=1}^{\nface} \face_k \,.
\end{equation}

\paragraph{Normal vectors and jump notation}
For a cell $K$, we write its outward-pointing normal vector 
in the meridional plane as $\tilde{\n}$ defined on $\partial K$; 
we also introduce  its extension $\n$ as in \cref{eq:normals-axisym}, 
\begin{equation}\label{norvecK}
\tilde{\n} := (n_\eta, n_z)^t, \qquad \n := (n_\eta,0,n_z)^t\,.
\end{equation}
We employ $\n$ to limit the use of $\tilde{\cdot}$, particularly in expressions with numerical traces, by using interchangeably the left-hand side 
and right-hand side of the following identity: 
\begin{equation}\label{cmpt_notation::rmk}
\tilde{\n}\cdot \tilde{\mathbf{v}} = \n \cdot \mathbf{v}, \quad \text{ for } \mathbf{v}=(v_\eta,v_\phi,v_z)^t,  \,\,\,\,\tilde{\mathbf{v}} = (v_\eta,v_z)^t.
\end{equation}
The normal jump of a vector along an interior face $\face \in \setface^i$
between cell $K^+$ and $K^-$, $\face = \overline{K^+} \cap \overline{K^-}$ 
(following \cref{cmpt_notation::rmk}), is denoted by $\llbracket\cdot\rrbracket$:
\begin{equation}\label{jumpvector}
 \llbracket \bu \cdot \n \rrbracket
   \,=\, \bu^+ \cdot \n^+ \,+\, \bu^- \cdot \n^-.
\end{equation}
Here, the vectors $\n^{\pm}$ are defined from the outward pointing 
normal vectors $\tilde{\n}^{\pm}$ of $K^{\pm}$ by \cref{norvecK}, 
and $\bu^{\pm}$ are the interior Dirichlet traces of $\bu|_{K^{\pm}}$ 
on $\partial K^{\pm}$.

\paragraph{Local and global indexes of a face} 
A face $\face\in \setface$ bears two systems of indexing, first as a member of $\setface$, and secondly as a member of the boundary of an element $K\in \mesh$,
\begin{equation}\label{global_local}
\begin{array}{c c c}
\text{Face Global index:}\hspace*{4em} & \face_k, & 1\leq k \leq \nface\,,  \\[0.50em]
\text{Local index as a face of $K_e$:}
&  \face^{(e,\ell)} , & 1\leq e\leq \ncell, \,\, 1\leq \ell \leq \nface^e \,,
\end{array} 
\hspace*{1em} \text{ such that } \face_k = \face^{(e,\ell)}  \text{ (as a set).}
\end{equation}
We denote by $\nfaceloc$ the number of faces in a cell $K$,
\begin{equation}
\partial K_e  = \bigcup_{\ell=1}^{\nfaceloc} \face^{(e,\ell)} , 
\qquad \text{for simplex cell in 2D (triangle), $\nfaceloc = 3$.}
\end{equation}

\subsection{Statement of the HDG formulation}

The weak problem is stated in broken Lagrange finite elements, denoted by $\mathbb{P}_{r}$ indicating polynomials of order $r$:
\begin{equation}\label{global_fem_spaces} \left. \qquad \begin{aligned}
 W_h &= \left\lbrace  w_h \in L^2(\domain): \hspace*{6.0em} w_h\!\mid_{K_e} \in \mathbb{P}_{r_e}(K_e), \hspace*{0.2cm} \forall K_e \in \mesh  \right\rbrace ;\\[0.3em]
 \boldsymbol{W}_h= [W_h]^3 &= \left\lbrace  \boldsymbol{w}_h =\begin{psmallmatrix} w_{\eta h} \\[0.3em] w_{\phi h}\\[0.3em] w_{z h}\end{psmallmatrix} \in L^2(\domain)^3 : 
                     \hspace*{1em}\boldsymbol{w}_h \!\mid_{K_e} \in [\mathbb{P}_{r_e}(K_e)]^3, \hspace*{0.2cm} 
                     \forall K_e \in \mesh  \right\rbrace; \\[0.3em]
 V_h &= \left\lbrace  v_h \in L^2(\setface)\,:\hspace*{6em} 
                   v_h \mid_{\face_k} \in \mathbb{P}_{\tilde{r}_k}(\face_k) , \hspace*{0.2cm}\face_k \in \setface \right\rbrace.
\end{aligned} \right. \end{equation}
Note that the definitions allow for \emph{$p$-adaptivity}, i.e. different orders of approximation between cells.

\begin{mdframed}

\paragraph{HDG formulation of \cref{eq:main-section-axisym:mode}:} find 
$(w_h,\bu_h, \lambda_h)\in W_h\times \boldsymbol{W}_h\times V_h $ such that: 
\begin{enumerate}
\item for each cell $K \in \mathcal{T}_h$, 
with prescribed boundary value $\lambda_h \in L^2(\partial K)$, $(w_h,\bu_h)$ satisfies 
\begin{subequations}\label{eq:main-section-axisym:mode:hdg}
\begin{empheq}[left={\empheqlbrace}]{align}
  \boldsymbol{A}_m \, \bu_h
  \,+\, [\boldsymbol{\beta}_1 ]\, w_h  
  \,+\, [\nabla]_m w_h \,=\, \mathbf{g},  \quad &\text{ on } K \label{eq:main-section-axisym:mode1}\\
        [\nabla\cdot]_m \bu_h
    \,+\, [\boldsymbol{\beta}_2] \,\cdot\,\bu_h \,+\,\varrho\, w_h \,=\, h, \quad & \text{ on } K \label{eq:main-section-axisym:mode2}\\
    w_h = \lambda_h  , \quad & \text{ on } \partial K.
\end{empheq}\end{subequations}
\item the trace $\lambda_h$ satisfies problems defined on 
the set of faces $\Sigma_h$, which consists of the 
continuity in the normal jump of $\bu_h$ along the interfaces $\setface^i $, 
\begin{equation}\label{jumpstrong}
  \llbracket  \yhwidehat{  \bu_h \cdot \n} \rrbracket = 0 \, \quad \text{ on }  \setface^i, 
\end{equation}
and boundary conditions \cref{bc:main-section-axisym:mode_pre} on exterior faces, $\setface^b\cup \setface^a$, 
\begin{equation}\label{bcstrong}
  \yhwidehat{\bu_h\cdot \n} = 0\,,\quad \text{on } \setface^a \,; \hspace*{3em}
 \dfrac{1}{\alpha_{\mathrm{bc}}  }\lambda_h\, + \, \yhwidehat{ \bu_h\cdot \n} \,=\, 0\,,\quad      \text{on } \setface^b \,.
\end{equation}
\end{enumerate}
The above problems are supplemented with 
the definition of the numerical Neumann 
trace, with $\tau$ the stabilization term:
\begin{equation}\label{numHdgNeutrace}
\yhwidehat{ \bu_h \cdot \n}  =  \bu_h\cdot \mathbf{n} \,-\, \tau \Big( w_h - \lambda_h \Big), \quad \text{ on } \partial K \text{ for each $K$ } \in \mathcal{T}_h\,.
\end{equation}

\end{mdframed}

\begin{remark}\label{summarynumtrace::rmk}
The definitions of the numerical traces can be summarized as follows: on boundary edges, 
\begin{equation}\label{numHdgNeutrace_boundary}
\widehat{w_h} = \lambda_h , \hspace*{0.3cm} \text{and} \hspace*{0.3cm} \yhwidehat{ \bu_h\cdot \n} = \bu_h\cdot \mathbf{n} \,-\, \tau \Big( w_h - \lambda_h \Big), \hspace*{0.5cm}  \text{ on } \setface \setminus \setface^i;
\end{equation}
for an interior face $\face \in \setface^i$, with $\face = \overline{K^+} \cap \overline{K^-}$, we have
\begin{equation}\label{numHdgNeutrace_inter}
\widehat{w_h}^{\pm} = \lambda_h,  \hspace*{0.3cm} \text{and} \hspace*{0.3cm} \yhwidehat{ \bu_h \cdot \mathbf{n}}^{\pm}  =  \bu_h^{\pm}\cdot \mathbf{n}^{\pm} \,-\, \tau^{\pm} \Big( w_h^{\pm} - \lambda_h \Big), \hspace*{0.5cm} \text{ on } \partial K^{\pm}\,.
\end{equation}
Here, the values associated with cells $K^{\pm}$ are distinguished by $\pm$.
\end{remark}

\begin{remark} In the second boundary condition of \cref{bcstrong}, we have employed that the numerical trace of $w_h$ is given by $\lambda_h$, i.e. $\yhwidehat{w_h} = \lambda_h$. By substituting the definition of $\yhwidehat{\bu_h\cdot \n}$ from \cref{numHdgNeutrace_boundary}, we can further rewrite the left-hand side 
of this condition as,
\begin{equation}\label{lhsLagbc}
\dfrac{1}{  \alpha_{\mathrm{bc}} }\lambda_h  +  \yhwidehat{\bu_h\cdot \n}   =\dfrac{1}{  \alpha_{\mathrm{bc}} }\lambda_h + \bu_h\cdot \mathbf{n} \,-\, \tau \Big( w_h - \lambda_h \Big)
 = \bu_h\cdot \mathbf{n}  - \tau w_h  + \left(\tau + \dfrac{1}{ \alpha_{\mathrm{bc}} }\right) \lambda_h .
\end{equation}

\end{remark}
\subsection{Statement of the HDG problem in weak form}

To state the weak formulation, we will need the following notations:
\begin{itemize}[leftmargin = *]
\item  We have already introduced the $\tilde{\cdot}$ notation which 
       restricts to the $\eta$ and $z$ components 
\begin{equation} \bu = (u_\eta,u_\phi,u_z)^t, \quad \tilde{\bu} \,=\, (u_\eta, u_z)^\mathrm{t}, \quad \tilde{\bu} \cdot \tilde{\n} = \bu\cdot \n \,.   \end{equation}

\item We extend this to the gradient and divergence operator, in introducing, 
\begin{equation}\label{tildegrad}
\widetilde{[\nabla\cdot]} (v_1, v_3)^t:= (\partial_\eta + \dfrac{1}{\eta} ) v_1  + \partial_z v_3  , \quad \widetilde{[\nabla]} w = (\partial_\eta  w, \partial_z w)^t.
\end{equation}
Relating to $[\nabla\cdot]_m$ and $[\nabla]_m$ defined in \cref{modalgraddiv}, we have 
\begin{equation}\label{tildegrad::id}
[\nabla\cdot]_m \bu=  \widetilde{[\nabla\cdot]}\, \tilde{\bu} \, +\, \ii \dfrac{m}{\eta} u_{\phi}, \hspace*{0.5cm}
\boldsymbol{\psi} \cdot [\nabla]_m w= \tilde{\boldsymbol{\psi}} \cdot \widetilde{[\nabla]} w  + \ii\dfrac{m}{\eta} \psi_\phi w\,.
\end{equation}

\item
With $\tilde{\mathbf{x}}:= (\eta,z)$, we denote by $d\tilde{\mathbf{x}}$  the Cartesian volume form in $\mathbb{R}_{\eta}\times \mathbb{R}_z$, i.e. $d\tilde{\mathbf{x}} = d\eta dz$,
 and by $d s_{\tilde{\mathbf{x}}}$ its restriction to a curve. We have 
 the identity for integration by parts\footnote{It suffices to consider the integration by parts in variable $\eta$, also recall $n_\eta$ is the component along $\mathbf{e}_\eta$ of $\tilde{\mathbf{n}} = (n_\eta,n_z)$, 
\begin{equation}\int_K g (\partial_\eta f + \dfrac{1}{\eta} f) \eta \,d\eta = \int_K  (-f \partial_\eta (\eta g)   + gf  ) \,d \eta + \int_{\partial K} n_\eta f g \,ds_{\eta} = -\int_K f (\partial_\eta g) \eta d\eta + \int_{\partial K} n_\eta f g \,ds_{\eta}    .\end{equation}
},
\begin{equation}\label{IPP_etaz}
\int_K  \varphi  \left(\widetilde{[\nabla\cdot]} \,\tilde{\mathbf{v}} \right)\eta \,d\tilde{\mathbf{x}}
 = - \int_K    \tilde{\mathbf{v}} \cdot \left(\widetilde{[\nabla]}\,\varphi  \right)  \eta\, d \tilde{\mathbf{x}}
 + \int_{\partial K}  \varphi \,  \mathbf{v} \cdot \mathbf{n}  \, \eta \,d s_{\tilde{\mathbf{x}} }  \,.
\end{equation}

\end{itemize}

To obtain the variational formation of the HDG problem, we 
integrate \cref{eq:main-section-axisym:mode} against test 
functions $ (\bpsi,\varphi) \in \boldsymbol{W}_h\times W_h$, 
with volume form $\eta d\tilde{\mathbf{x}}$.  
After substituting the definition of the Neumann trace \cref{numHdgNeutrace}, 
we carry out a reverse integration by parts in the second equation. 
The details of this derivation is listed below. For the problem 
defined on the edges, we integrate against test functions $\zeta\in V_h$ 
with surface form $d s_{\tilde{\mathbf{x}}}$, and substitute in the 
relation for the numerical trace \cref{numHdgNeutrace}; in particular,
\begin{subequations}
\begin{align}
\cref{bcstrong} ,\cref{lhsLagbc} \hspace*{0.5cm} \Rightarrow \hspace*{0.5cm}  
& \bu_h\cdot \mathbf{n}  - \tau w_h  
+ \left(\tau + 1/ \alpha_{\mathrm{bc}} \right) \lambda_h  = 0 ,  \quad \text{ on } \setface^b, \\
\cref{bcstrong} ,\cref{numHdgNeutrace_boundary} \hspace*{0.5cm} \Rightarrow \hspace*{0.5cm}  
&  \bu_h\cdot \mathbf{n} -  \tau ( w_h - \lambda_h ) = 0   ,  \hspace*{2.1cm} \text{ on } \setface^a,\\
\cref{jumpstrong},  \cref{numHdgNeutrace_inter}  \hspace*{0.5cm} \Rightarrow \hspace*{0.5cm}  &  \bu_h^+\cdot \mathbf{n}^+ - \tau^+ ( w_h^+ - \lambda_h )+\bu_h^{-}\cdot \mathbf{n}^{-} - \tau^{-} ( w_h^{-} - \lambda_h )=0, \quad \text{ on } \setface^i.
\end{align}
\end{subequations}

\begin{mdframed}

\textbf{Weak formulation of 
\cref{eq:main-section-axisym:mode}--\cref{bcstrong}}:
Find $(w_h,\bu_h,\lambda_h ) \in W_h\times \boldsymbol{W}_h\times V_h$, that solve
\begin{enumerate}[leftmargin = *]
\item the local problems on each element $K^e \in \mathcal{T}_h$: 
      for all test functions $ (\bpsi,\varphi) \in W_h\times \boldsymbol{W}_h$,  
\begin{subequations} \label{eq:hdg-weak1}
\begin{empheq}[left={\hspace*{-2em} \empheqlbrace}]{align}
& \int_K \left( \boldsymbol{A}_m  \bu_h \cdot \bpsi 
+w_h\,[\boldsymbol{\beta}_{1}]    \cdot  \bpsi 
- w_h\, \widetilde{[\nabla\cdot]}\, \widetilde{\bpsi} \right) \!\eta \, d\tilde{\mathbf{x}}  \nonumber \\
 &\hspace*{2.55cm}  + \int_K  \ii  m \, w_h  \psi_\phi \,d\tilde{\mathbf{x}}
  + \int_{\partial K}  \lambda_h \,  \bpsi \cdot \n \, \eta  ds_{\tilde{\mathbf{x}}}
  = \int_K  \mathbf{g} \cdot \bpsi\, \eta \,d\tilde{\mathbf{x}}, \label{eq:hdg-weak1a}\\[1em]
&   \int_K \!\!\left( \varphi\, \widetilde{[\nabla\cdot]}\, \widetilde{\bu}_h  
  +  \varphi \,[\boldsymbol{\beta}_{2}] \cdot\bu_h 
 +  \varrho\, w_h \, \varphi\right)  \!\eta \, d\tilde{\mathbf{x}}
  + \int_K\!\! \ii m  \, u_{h\,\phi} \, \varphi \,d\tilde{\mathbf{x}}
  - \int_{\partial K} \!\! \tau ( w_h - \lambda_h )  \varphi \, \eta \, ds_{\tilde{\mathbf{x}}} = 0. \label{eq:hdg-weak1b}
\end{empheq}\end{subequations}

\item On interior edges $\setface^i$: for $\face \in \setface^i$ 
(following notation in \cref{numHdgNeutrace_inter}), 
for test functions $\zeta \in V_h$, 
\begin{equation}\label{eq::hdg_int}
  \int_{\face} \left( \bu_h^+ \cdot \n^+  -
 \tau^+ (w_h^+ - \lambda_h)  \right) \zeta \,ds_{\tilde{\mathbf{x}}}
+ \int_{\face} \left(\bu_h^- \cdot \n^-  -
     \tau^- (w_h^- - \lambda_h) \right) \zeta\,ds_{\tilde{\mathbf{x}}}= 0 .
\end{equation}
\item On boundary edges: for test functions $\zeta \in V_h$, 
      on edges along rotation axis $\setface^a$, we have
\begin{equation}\label{eq::hdg_bdya}
  \int_{\face} \left(\bu_h \cdot \n    -
\tau (w_h - \lambda_h)\right) \zeta \,ds_{\tilde{\mathbf{x}}} = 0 , \quad \forall\,\face \in \setface^a,
\end{equation}
and vanishing Lagrangian pressure perturbation condition along $\setface^b$, 
\begin{equation}\label{eq::hdg_weaklag}
  \int_{\face} \left( \bu_h\cdot \mathbf{n}  - \tau w_h  + \left(\tau + \dfrac{1}{ \alpha_{\mathrm{bc}} }\right) \lambda_h \right)\,\zeta ds_{\tilde{\mathbf{x}}} \,=\, 0 , \quad  \forall \face \in \setface^b\,.
\end{equation}

\end{enumerate}

\end{mdframed}

\noindent \emph{Derivation of weak HDG local problems \cref{eq:hdg-weak1}}: 
Employing notation $\widetilde{[\nabla]}$ 
and $\widetilde{[\nabla\cdot]}$ introduced 
in \cref{tildegrad} and identity \cref{tildegrad::id}, 
in integrating \cref{eq:main-section-axisym:mode} against test functions $\boldsymbol{\psi} = (\psi_\eta,\psi_\phi,\psi_z)^t$ and $\varphi$, we have
\begin{subequations} \label{eq:hdg-weak0}
\begin{empheq}[left={\hspace*{-2em} \empheqlbrace}]{align}
& \int_K \! \Big( \boldsymbol{A}_m  \bu_h \cdot \bpsi 
+ w_h\,[\boldsymbol{\beta}_{1} ]   \cdot  \bpsi  +  \widetilde{\bpsi}\cdot \widetilde{[\nabla]} w_h \Big)\eta\,d\tilde{\mathbf{x}}
  + \int_K  \ii  m \, w_h \, \psi_\phi\, d\tilde{\mathbf{x}}
  = \int_K \eta \, \mathbf{g}\cdot \bpsi \, d\tilde{\mathbf{x}}, \\[1em]
&   \int_K \!\Big( \varphi\, \widetilde{[\nabla\cdot ]} \widetilde{\bu_h} + \varphi \, [\boldsymbol{\beta}_{2}] \cdot\bu_h 
+  \varrho \, w_h \, \varphi\Big) \eta\,d\tilde{\mathbf{x}}
  + \int_K \ii m  \, u_{\phi h} \, \varphi\,d\tilde{\mathbf{x}} = 0.
\end{empheq}\end{subequations}
Next we employ the integration by parts \cref{IPP_etaz} and substitute in 
the definition of the numerical traces on $\partial K$, cf. \cref{summarynumtrace::rmk},
\begin{equation}
\widehat{w}_h = \lambda_h \, , 
\hspace*{0.5cm} \text{and} \hspace*{0.5cm} 
\yhwidehat{ \bu_h\cdot \n} = \bu_h\cdot \mathbf{n} \,-\, \tau \Big( w_h - \lambda_h \Big),
\end{equation}
we make appear the role of $\lambda_h$:
\begin{subequations}
 \begin{align}
\int_K \widetilde{\bpsi}\cdot (\widetilde{[\nabla]} w_h )\eta\,d\tilde{\mathbf{x}}
& = -\int_K w_h (\widetilde{[\nabla\cdot ]}\widetilde{\bpsi}) \,\eta \,d\tilde{\mathbf{x}}+
 \int_{\partial K} \widehat{w_h} \n \cdot \bpsi \, \eta \,d s_{\tilde{\mathbf{x}}} \nonumber\\
 &= -\int_K w_h (\widetilde{[\nabla\cdot ]}\widetilde{\bpsi})\, \eta \,d\tilde{\mathbf{x}}+
 \int_{\partial K} \lambda_h \,\n \cdot \bpsi \, \eta\, d s_{\tilde{\mathbf{x}}} ,\label{derivingweakprob_1}\\
\text{and } \hspace*{0.1cm} \int_K \!  \varphi\, (\widetilde{[\nabla\cdot ]} \widetilde{\bu_h}) \eta  \,d\tilde{\mathbf{x}}
&= - \int_K\!   \bu_h (\widetilde{[\nabla ]} \varphi ) \,\eta  \,d\tilde{\mathbf{x}}
+ \int_{\partial K}  \yhwidehat{\bu_h \cdot \n}  \varphi \, \eta d s_{\tilde{\mathbf{x}}} \nonumber\\
& \hspace*{-3em} =- \int_K\!   \bu_h \,(\widetilde{[\nabla ]} \varphi ) \,\eta  \,d\tilde{\mathbf{x}}
+ \int_{\partial K} \Big(\bu_h\cdot \n - \tau(w_h - \lambda_h) \Big)  \varphi \, \eta \,d s_{\tilde{\mathbf{x}}} \nonumber \\
& \hspace*{-3em} = \left(- \int_K\!   \bu_h \,(\widetilde{[\nabla ]}\, \varphi ) \,\eta  \,d\tilde{\mathbf{x}} + \int_{\partial K} \bu_h\cdot \n \,\varphi \, \eta\, d s_{\tilde{\mathbf{x}}} \right) - \int_{\partial K} \tau(w_h - \lambda_h )  \varphi \, \eta \,d s_{\tilde{\mathbf{x}}} \,. \label{derivingweakprob_2temp}
\end{align}
\end{subequations}
In carrying out a reverse integration by parts, 
the term in parentheses of \cref{derivingweakprob_2temp} simplifies,
with that the derivative falling back to $\tilde{\bu}_h$, i.e.,
\begin{equation}\label{derivingweakprob_2}
 \int_K \!  \varphi\, (\widetilde{[\nabla\cdot ]} \widetilde{\bu_h}) \,\eta  \,d\tilde{\mathbf{x}}
  = \int_K \!  \varphi\, (\widetilde{[\nabla\cdot ]} \widetilde{\bu_h}) \,\eta  \,d\tilde{\mathbf{x}}- \int_{\partial K} \tau(w_h - \lambda_h)  \varphi \, \eta \,d s_{\tilde{\mathbf{x}}}\,.
\end{equation} 
Substituting \cref{derivingweakprob_2,derivingweakprob_1} 
in \cref{eq:hdg-weak0} to replace the terms with derivatives, 
we arrive at the problem stated \cref{eq:hdg-weak1}.

\subsection{Matrix description of the discrete problems}

\paragraph{Local basis functions} 
We introduce the notation for the basis elements of the 
finite elements mentioned in \cref{global_fem_spaces},
\begin{equation}\label{localbasisfcn}
\begin{array}{ c| c |c} \text{Local finite element space} &  \text{Basis functions}  & \text{Dimension}  \\[0.2em]\hline
 \mathbb{P}_{r_e}(K_e) ,  \,\, 1\leq e\leq \ncell &   \varphi^e_j \,, \,\, 1\leq j\leq \ndof  \hspace*{0.3cm}& \ndof^e  \\[0.5em] 
\mathbb{P}_{\tilde{r}_k}(\face_k),  \,\,  1\leq k\leq \nface \hspace*{0.2cm} & \hspace*{0.2cm}  \zeta^k_j \,, \,\, 1\leq j\leq  \ndofface \hspace*{0.3cm} & \hspace*{0.2cm}  n_{\mathrm{dof}}^{\face_k}    \end{array}
 \end{equation}
We also denote the total number of face degrees of freedom, 
\begin{equation} N_{\mathrm{dof}}
 = \sum_{k  =1}^{\nface}  n_{\mathrm{dof}}^{\face_k} .
\end{equation}
In the rest of the discussion, we simply write  $\ndof$ and $\ndofface$ while bearing in mind these values are cell-dependent.

\paragraph{Volume discrete unknowns}
Recall that the components of unknown $\mathbf{u}=(u_\eta,u_\phi,u_z)^t$ and $w$ 
are approximated on each cell $K\in \mesh$, by $\mathbf{u}_h = ( u_{\eta h},u_{\phi h},u_{z h})^t $ and $w_h$.
We denote the coefficients of $(\mathbf{u}_h,w_h)$ in the basis functions $\varphi^e_j$ in \cref{localbasisfcn} by $\mathrm{p}_j^{[\bullet]}$, with $\bullet $ indicating the corresponding unknown,
\begin{equation}
  w_h\!\mid_{K_e} \,=\, \sum_{j=1}^{\ndof} \mathrm{p}^{[w]}_j \, \varphi_j^e \,; \qquad 
  u_{\eta h}\!\mid_{K_e} \,=\, \sum_{j=1}^{\ndof} \mathrm{p}_j^{[u_\eta]} \, \varphi_j^e \,; \qquad 
  \text{similarly for } u_{\phi h}\!\mid_{K} \text{ and } u_{z h}\!\mid_{K}.
\end{equation}
We gather the coefficients $\mathrm{p}_j^{[\bullet]}$ associated with a cell $K_e$ into the vector $\mathsf{U}^e$, containing four blocks,
\begin{equation}
\mathsf{U}^e \,  =\, 
\begin{pmatrix}  \mathbb{p}_{[u_\eta]}  &  \mathbb{p}_{[u_\phi]} &   \mathbb{p}_{[u_z]} &  \mathbb{p}_{[w]} \end{pmatrix}^t, \hspace*{4em} 
\text{ with } \hspace*{0.1cm}   \mathbb{p}_{[\bullet]} 
\, =\, \left(  \mathrm{p}^{[\bullet]}_j \right)_{j=1}^{\ndof}. 
\end{equation}
%

\paragraph{Edge discrete unknowns}
The HDG trace unknown $\lambda_h$
is represented on each face $\face_k \in \setface$ in the basis functions $\zeta_j^k$ \cref{localbasisfcn} with coefficients $\mathrm{q}_j^k$ (also listed in local indexing \cref{global_local}),
\begin{equation}
 \lambda_h\!\mid_{\face_k} \,=\, \sum_{j=1}^{\ndofface}\, \mathrm{q}_j^k \, \zeta^k_j, \quad \text{with }
 \mathrm{q}_j^k = \mathrm{q}_j^{(e,\ell)} \hspace*{0.1cm} \text{ where }  \hspace*{0.1cm}\face_k = \face^{(e,\ell)}.
\end{equation}
We gather $\mathrm{q}_j^k$ into the global vector $\Lambda$ of length $N_{\mathrm{dof}}$ which is the unknown of the global HDG problem,
\begin{equation}
\Lambda = \left( \mathbb{q}^k\right)_{k=1}^{\nface} , \hspace*{0.2cm} \text{ with } \hspace*{0.2cm}  \mathbb{q}^k = ( \mathrm{q}_j^k )_{j=1}^{\ndofface}.
\end{equation}
Its restriction to an element $K_e$ is given by $\mathbb{R}_e$, which gathers the 
coefficients from each face $\face^{(e,\ell)} \in K_e$, $1\leq \ell\leq \nface$ of 
an element $K_e$,
\begin{equation}
   \Lambda^e := \hdgR_e \Lambda\,=\, \begin{pmatrix}\mathbb{q}^{(e,\ell)} \end{pmatrix}_{\ell=1}^{\nface}, 
   \quad \text{ with } \quad \mathbb{q}^{(e,\ell)}  = \left( \mathrm{q}_j^{(e,\ell)}\right)_{j=1}^{\ndofface}.
\end{equation}
We also note that $\hdgR_e^t$ assigns contribution 
defined on an element $K_e$ to the global vector, 
\cite[Equation (3.39)]{Pham2024stabilization}.

\subsubsection{Local problems}
The local problem  \cref{eq:hdg-weak1} is discretized and 
written in matrix form such as, for each $K_e$, $1\leq e\leq \ncell$:
\begin{equation}
  \hdgA^e \, \mathsf{U}^e \,+\, \hdgC^e \, \hdgR_e \Lambda \,=\, \hdgG^e \,.
\end{equation}
The matrix $\hdgA^e$ has a $4\times 4$ block structure, 
while the matrix $\hdgC^e $ has a $4\times \nfaceloc$ block structure,
\begin{subequations}
\begin{align}
 \hdgA^e  &= \left( \hdgA^e[I,J] \right)_{I,J=1,\ldots, 4} = \begin{pmatrix}  \hdgA^e[1,1]  & \ldots & \hdgA^e[1,4] \\
  \vdots  & \ddots & \vdots\\[0.3em]
  \hdgA^e[4,1]& \ldots & \hdgA^e[4,4]
  \end{pmatrix},  \label{hdgAblock}\\
    \hdgC^e & = \left( \hdgC^e[I,\ell] \right)_{\substack{I=1,\ldots, 4;\\[0.3em]  \ell=1,\ldots,\nfaceloc}} = \begin{pmatrix}  \hdgC^e[1,1]  & \ldots & \hdgC^e[1,\nfaceloc] \\
  \vdots  & \ddots & \vdots\\
  \hdgC^e[4,1]& \ldots & \hdgC^e[4,\nfaceloc]  \label{hdgCblock}
  \end{pmatrix}.\end{align}
\end{subequations}
The first three row-blocks, $\hdgA^e[1:3,:]$ and $\hdgC^e[1:3,:]$, 
correspond to the volume integrals in \cref{eq:hdg-weak1a} and 
with test functions containing  non-zero components 
only in $\eta$, $\phi$ and $z$ respectively. 
The fourth row-blocks, $\hdgA^e[4,:] $ and $\hdgC^e[4,:]$ correspond 
to the volume integrals in \cref{eq:hdg-weak1b}.

The vector source has a length $4 \times \ndof$,
we list its components for the case of a Dirac source 
on the scalar equation at position $\tilde{\mathbf{s}}=(s_\eta,s_z)$:
\begin{equation}
\hdgG^e = (\hdgG^e[I])_{I=1,\ldots,4}, \text{ with } 
\left\{
\begin{array}{l l}
\tilde{\mathbf{s}} \notin K^e: & \hspace*{0.3cm} \hdgG^e[:] = \boldsymbol{0}_{\ndof\times 1}, \\[0.3em]
\tilde{\mathbf{s}} \in K^e:    & \hspace*{0.3cm} 
\hdgG^e[1:3] = \boldsymbol{0}_{\ndof\times 1}\,; \quad
\hdgG^e[4] = \big(\varphi_i(\tilde{\mathbf{s}})\big)_{i=1}^{\ndof} \,.
\end{array}\right.
\end{equation}

\paragraph{Components of $\hdgA^e$}
Since each component of $\mathbf{u}_h$ and $w_h$ is discretized 
with the same basis functions $\{\varphi_j^e\}$ \cref{localbasisfcn}, 
we describe all together the elements $(i,j)$ of each 
sub-block listed in \cref{hdgAblock}. 
For this, we introduce the vector-valued differential operator  $\mathcal{D}$
acting on a scalar function $\varphi$,
\begin{equation}
\mathcal{D}_m \varphi:= \left(\partial_{\eta} \varphi + \dfrac{1}{\eta} \varphi\,,\,\, \ii \dfrac{m}{\eta} \varphi\,,\, \partial_z \varphi\right)^t , \quad \overline{\mathcal{D}_m} \varphi:= \left(\partial_{\eta} \varphi + \dfrac{1}{\eta} \varphi\,,\, - \ii \dfrac{m}{\eta} \varphi \,, \,\partial_z \varphi\right)^t  .\end{equation}
The first three row-blocks of $\hdgA^e$ 
are listed as $3\times 4$ matrices: for $i,j$, $1\leq i,j\leq \ndof$, 
\begin{equation}
\begin{aligned}
\Big(   \hdgA^e[I,J](i,j)  \Big)_{\substack{I=1,\ldots,3,\\ J=1,\ldots, 4}}
& = \int_K \eta \,\varphi_j \Big(\varphi_i\,\Amat_m  \hspace*{0.2cm} ,  \hspace*{0.2cm}\varphi_i\, [\boldsymbol{\beta}_1]  \, -\, \overline{\mathcal{D}_m}\, \varphi_i \,\Big)  \, d\tilde{\mathbf{x}}, \\[0.3em]
\Big(   \hdgA^e[I,J](i,j)  \Big)_{\substack{I=4,\\ J=1,\ldots 4}}  
&= \int_K \eta \,\varphi_i \Big( \varphi_j\, [\boldsymbol{\beta}_2]^t \,  +\,  \mathcal{D}_m^t\,\varphi_j \hspace*{0.2cm},\hspace*{0.2cm}  \rho \Big) \, d\tilde{\mathbf{x}}
  - \int_{\partial K} \eta \,  \Big( \boldsymbol{0} \,, \,   \tau \Big) \varphi_j\, ds_{\tilde{\mathbf{x}}}.
\end{aligned} \end{equation}
Here the integration is carried out component-wise.
Recall that the matrix-valued function $\Amat_m$ has size $3\times 3$, 
while the vector-valued functions $[\boldsymbol{\beta}_1]$ and
$[\boldsymbol{\beta}_2]$ are of size $3\times 1$; 
$\boldsymbol{0}$ is the zero row vector of length 3.

\paragraph{Components of $\hdgC^e $} For $ \ell=1,\ldots, \nfaceloc$, we describe the components for each column-block $\ell$, 
within which
the $(i,j)$-th element of the four blocks are listed as a column vector of length 4, 
\begin{equation}
  \Big(\,\hdgC^e[I,\ell](i,j) \,\Big)_{I=1}^4
 = \int_{\face^{(e,\ell)}}    \varphi^e_i \,  \zeta_j^{(e,\ell)} \,\ \big(n_\eta\,,\, 0 \,,\, n_z\,,\, \tau \big)^t\, \eta\, d s_{\tilde{\mathbf{x}}}.
\end{equation}
Recall that $(n_\eta,n_z)$ is the outward-pointing normal vector along $\partial K^e$.


\subsubsection{Problems on edges}

The discretization of the edge problems \cref{eq::hdg_int}--\cref{eq::hdg_weaklag} 
on $\setface$ takes the following form,
\begin{equation}
\sum_{e=1}^{\ncell} \hdgR_e^t \left( \hdgB^e \, \mathsf{U}^e \,+\, \hdgL^e \, \hdgR_e \, \Lambda\right)   = \boldsymbol{0},
\end{equation}
where the matrix  $\hdgL^e $ associated with the 
trace unknown $\Lambda$ is block-diagonal containing $\nface^e$ blocks,  
 \begin{equation}
 \hdgL^e = \mathrm{Diag} \big( \hdgL^e[1]  ,  \ldots  , \hdgL^e[\nface^e] \big) \,.
 \end{equation}
 Each sub-block $\hdgL^e[\ell]$, $1\leq \ell \leq \nface^e$, is of  size 
 $\ndofface \times \ndofface $ representing contribution from face $(e,\ell)$.
 The matrix $\hdgB^e$ has a block structure $ \nface^e \times 4$,
 \begin{equation}
 \hdgB^e = \left( \hdgB^e[\ell,J] \right)_{\substack{J=1,\ldots, 4;\\[0.3em]  \ell=1,\ldots,\nfaceloc}}= \begin{pmatrix}   \hdgB^e[1,1] & \ldots &  \hdgB^e[1,4] \\
              \vdots      &  \ddots  &  \vdots \\   \hdgB^e[\nface^e,1] & \ldots &  \hdgB^e[\nface^e,4]    \end{pmatrix}, \end{equation}
with each column-block associated with the volume unknown $\mathsf{U}^e$, 
and each row-block representing contribution (of all volume unknowns) 
from the face $(e,\ell)$.
           
We next describe the components of the sub-blocks in $\ell$-th column block:  $\ell=1,\ldots, \nface^e$,
\begin{itemize}[leftmargin = *]
\item  For matrix $\hdgL^e$, the $(i,j)$-th element 
       of the $\ell$-th block: for $ i,j=1,\ldots, \ndofface$, is, 
\begin{equation}\label{matbbL_u}
\mathbb{L}^e[\ell](i,j )\hspace*{0.em}= \hspace*{0.1em}
\begin{dcases} \int_{\face^{(e,\ell)}} \tau \,  \zeta^{(e,\ell)}_i\,  \zeta^{(e,\ell)}_j    \,  \mathrm{d}s_{\tilde{\mathbf{x}}}, &\hspace*{0.2cm} \face^{(e,\ell)} \in \setface^i \cup \setface^a,\\[-0.1em]
 \int_{\face^{(e,\ell)}} \zeta^{(e,\ell)}_i \,  \zeta_j^{(e,\ell)} \left(\tau + \dfrac{1}{\alpha_{\mathrm{bc}}} \right) \mathrm{d} s_{\tilde{\mathbf{x} }}
 \,,  &\hspace*{0.2cm} \face^{(e,\ell)} \in \setface^b.
\end{dcases}
 \end{equation}         
 
 \item 
       For matrix $\hdgB^e$,  we group the $(i,j)$-th element from the four sub-blocks in this row-block: for $i=1, \ldots, \ndof^f$, and $j=1,\ldots , \ndof$,
 \begin{equation}
 \Big(\hdgB^e[I,\ell](i,j) \Big)_{I=1}^4 =  \int_{\face^{(e,\ell)} } \zeta^{(e,\ell)}_i \, \varphi^e_j \,\, \big( n_{\eta}, 0 , n_z,-\tau\big) \, \eta\, d s_{\tilde{\mathbf{x}}} .
 \end{equation}                 
     \end{itemize}                     
                                                                 


\subsubsection{Summary}

We can now put the discretized local and global problems together,
taking the generic form of the HDG method showing the two-level characteristics.
The description follows notation of \cite[Section 3.3.4]{Pham2024stabilization}
(where the HDG method is used for elastic anisotropic waves),
\begin{subequations}\label{dis_forwardHDG_v0}
\begin{empheq}[left={ \empheqlbrace\,}]{align}
&\hspace*{0.7cm}  \hdgA^e \, \mathsf{U}^e 
  \,\, + \,\,\hdgC^e\, \hdgR_e \, \Lambda   = \, \mathsf{G}^e \,\,, 
\hspace*{1.cm} \forall \,\,e =1\, \ldots, \ncell\,, \label{forward_hdg_local_prob} \\[0.em]
&  \displaystyle\sum_{e=1}^{\ncell}\,\, \hdgR^t_e\,   \Big(\,    \hdgB^e \, \mathsf{U}^e
\, \, + \,\,  \hdgL^e \, \hdgR_e \, \Lambda \, \Big) \,  = \, \boldsymbol{0}_{N_{\mathrm{dof}}\times 1}. \label{forward_hdg_globalprob}
\end{empheq}
\end{subequations}
With the local problems \cref{forward_hdg_local_prob} decoupled from one another, 
and assuming the invertibility of $\hdgA^e$, one can solve for $\mathsf{U}^e$ in terms of $\Lambda$,
\begin{equation}\label{eq:hdg-U} \mathsf{U}^e \, =\,   \left(\hdgA^e\right)^{-1} \left( \mathsf{G}^e \, -\, \hdgC^e\, \hdgR_e \, \Lambda    \right) \,, \hspace*{1cm} \forall e =1, \ldots, \ncell.\end{equation}
Inserting this into \cref{forward_hdg_globalprob}, we obtain a problem only in terms of $\Lambda$, called the global problem,
\begin{equation} \label{eq:hdg-global-linear-system}  
\mathbb{K} \,  \Lambda \, = \, \mathsf{S} \hspace*{0.5cm}\text{where } \left\{\begin{array}{l}
  \mathsf{S} := -\displaystyle \sum_{e=1}^{\ncell} \hdgR^t_e   \,   \mathbb{B}^e \, (\hdgA^e)^{-1} \, \mathsf{G}^e ,\\[1.5em]
  \mathbb{K}:=\hspace*{0.3cm} \displaystyle \sum_{e=1}^{\ncell}\,\hdgR^t_e\,\mathbb{K}^e\,  \hdgR_e
\,, \hspace*{0.3cm} \text{with} \hspace*{0.2cm}\mathbb{K}^e:=  \hdgL^e\, -\,    \hdgB^e \,
(\hdgA^e)^{-1}\, \hdgC^e  .
\end{array}\right.
\end{equation}

%

\begin{remark} 

 Implementation is carried out in the open-source software \texttt{Hawen}, 
 \cite{Hawen2021} (\url{https://ffaucher.gitlab.io/hawen-website/})
 which uses \texttt{MPI} and \texttt{OpenMP} parallelism. 
 Once the global sparse linear system \cref{eq:hdg-global-linear-system} 
 is assembled, it is solved with the direct solver MUMPS, \cite{Amestoy2001,Amestoy2019},
 allowing for the fast resolution with multiple right-hand sides.
 Once the discrete solution $\Lambda$ is obtained, the local system
 \cref{eq:hdg-U} are solved. This step is embarrassingly parallel on each cell.

\end{remark}

\subsection{Stabilization operator}
\label{subsection:hdg-stabilization}

We devise a strategy to select the stabilization operator $\tau$ 
appearing in the numerical trace of $\bu\cdot\n$ \cref{numHdgNeutrace},
\begin{equation}\label{numHdgNeutrace_rep}
\yhwidehat{ \bu\cdot \n}  =  \bu_h\cdot \mathbf{n} \,-\, \tau \Big( w_h - \lambda \Big), \quad \text{ on } \partial K \text{ for each $K$ } \in \mathcal{T}_h\,.
\end{equation}
 Recall this relation was employed for the problem defined on edges $\setface$ as well as in the local volume problem, \cref{eq:hdg-weak1}--\cref{eq::hdg_weaklag}. 
With our system \cref{eq:main-section-axisym:mode} bearing similarities to that of the linear convection-diffusion equation considered in \cite{nguyen2009implicit,rouxelin2021mixed}, we adapt the strategy of \cite{nguyen2009implicit} to select $\tau$.

\medskip

\paragraph{Derivation of \cref{numHdgNeutrace}}
We first show that the general form  \cref{numHdgNeutrace} 
is obtained containing a more specific decomposition of $\tau$.
Consider an edge $\face \in \setface$, where
we assume that $\boldsymbol{A}_m$ is invertible, and that the 
support of source $\mathbf{g}_m$ does not intersect with $\face$.  
Let us introduce the following quantity,
\begin{equation}
\eta_{\boldsymbol{\beta}}:=\n\cdot\boldsymbol{A}_m^{-1}\cdot\boldsymbol{\beta}_1 , 
\qquad \eta_{\n} := \n\cdot\boldsymbol{A}_m^{-1}\cdot\n. 
\end{equation}
Their meanings are investigated for radially symmetric with rotation backgrounds in \cref{stab_radsym::rmk}.

\begin{itemize}[leftmargin=*]
\item Under the current assumption, from relation \cref{eq:main-section-axisym:mode1}, we have
\begin{equation}\label{buhn_expression}
\bu_h = - \boldsymbol{A}_m^{-1}\left( \boldsymbol{\beta}_1 \, w_h
  \,+\, [\nabla]_m w_h \right) \quad \Rightarrow \quad \bu_h\cdot \n= - \eta_{\boldsymbol{\beta}} w_h - \n\cdot\boldsymbol{A}_m^{-1}\cdot[\nabla]_m w_h.
\end{equation}
From this relation, the numerical trace is defined employing notation $\eta_\beta$, 
\begin{equation}\label{bunnumtrace::def}
\yhwidehat{ \bu_h \cdot\n} := - \yhwidehat{\eta_{\boldsymbol{\beta}}\, w_h}  \,\,-\,\,   \yhwidehat{\n\cdot\boldsymbol{A}_m^{-1} [\nabla]_m w_h} .
\end{equation}

\item Next, following the approach in \cite[Section 3.6.3]{nguyen2009implicit}, 
which distinguishes the term of convection type, $\yhwidehat{\eta_\beta\, w_h} $, 
from one of diffusion type,  $\yhwidehat{\n\cdot\boldsymbol{A}_m^{-1} [\nabla]_m w_h}$, 
we define
\begin{subequations}\label{convdiffnumtrace::def}
\begin{align}
\yhwidehat{\eta_{\boldsymbol{\beta}} w_h } 
&  :=  \hspace*{0.2cm}\eta_{\boldsymbol{\beta}}  \lambda  \hspace*{0.3cm}  \textcolor{black}{+}\,\, \tau_c (  w_h - \lambda  ),\\
\yhwidehat{\mathbf{n}\cdot \Amat_m^{-1} \cdot [\nabla]_m w_h  }
  &:= \mathbf{n}\cdot A_m^{-1} \cdot [\nabla]_m w_h  \,\,  \textcolor{black}{+} \,\, \tau_d (w_h - \lambda).
\end{align}\end{subequations}
In substituting the definitions \cref{convdiffnumtrace::def} in \cref{bunnumtrace::def}, we obtain\footnote{We also add and subtract  $\eta_\beta w_h$, to make appear $\bu_h\cdot \n$ in employing expression \cref{buhn_expression} 
\begin{equation}
 \yhwidehat{\bu_h \cdot\mathbf{n}}   = -\eta_{\beta} \lambda \,\, \textcolor{black}{-}\,\,  \tau_c (  w_h - \lambda  ) - \mathbf{n}\cdot \Amat^{-1} \cdot \nabla w_h  \textcolor{black}{-} \tau_d (w_h - \lambda)
  = \bu_h\cdot \n + \eta_\beta(w_h-\lambda) - (\tau_c + \tau_d) (w_h-\lambda). 
\end{equation}   
 }
   \begin{equation}\label{defNeutrace_spec}
\boxed{\yhwidehat{\bu\cdot\n}   = \bu_h\cdot\n \,\, \textcolor{black}{-}\,\,  ( \tau_c \textcolor{black}{-} \eta_\beta \textcolor{black}{+}  \,\,\tau_d) (w_h - \lambda)}
    \end{equation}
\end{itemize}

\paragraph{Choice of stabilization} We choose $\tau_c$ and $\tau_d$ in \cref{defNeutrace_spec} as follows:
\begin{equation}\label{stabtauc}
\tau_c:= \lvert \eta_{\boldsymbol{\beta}}\rvert + \eta_{\boldsymbol{\beta}}, \hspace*{1cm} \text{ i.e. } \,\,\tau_c \textcolor{black}{-} \eta_{\boldsymbol{\beta}}   = \lvert \eta_{\boldsymbol{\beta}}\rvert,
\end{equation}
We further allow for some tuning in the stabilization with a scaling parameter $\alpha_{\text{tune}}$ such that,
\begin{equation}\label{stabtaud}
\tau_d :=  \ii \omega \, \alpha_{\text{tune}} \, \lvert \eta_{\n}\rvert \,.
\end{equation}

\begin{remark}[Interpretation of HDG stabilization in radially symmetric backgrounds]\label{stab_radsym::rmk}
Here, we provide some interpretation for stabilization \cref{stabtauc,stabtaud} 
for radially symmetric backgrounds.
We first recall that without rotation and 
with radially symmetric background, from \cref{radsym::background}, we have
\begin{equation}
\eta_{\boldsymbol{\beta}} = \dfrac{n_r \beta_1  }{N^2-\sigma^2}, \qquad \eta_{\n} = \dfrac{c_0^2}{N^2-\sigma^2} n_r^2 - \dfrac{c_0^2}{\sigma^2} n_{\parallel}^2 \, .
\end{equation}
We also recall from \cref{ellhyp::def} the distinction between the elliptic region and the hyperbolic region, defined respectively as, 
\begin{equation}
\mathbb{B}_{\mathrm{ell}}:=\{ \omega^2 > N^2(r)\}  \hspace*{0.3cm} \text{(Elliptic region)}, \quad \mathbb{B}_{\mathrm{hyp}}:=\{\omega^2 < N^2(r)\} \hspace*{0.3cm}\text{(Hyperbolic region)}.
\end{equation}

We introduce some working notions in the hyperbolic region. Compared to a classical wave operator $\partial_t^2 - \Delta_x$, the radial direction here is analogous to the time direction, while the tangential direction to the spatial direction. Faces are named based on the sign of $\eta_{\n}$ at zero attenuation, i.e. with $\sigma^2=\omega^2 > 0$,
\begin{equation}\face= \begin{dcases} \text{ is space-like}, & \quad \eta_{\n} >0, \\ \text{ is time-like}, & \quad \eta_{\n} <0 \end{dcases}, \quad \text{for } \face \in  \mathbb{B}_{\mathrm{hyp}}. \end{equation}
In particular when $n_r = 0$ on $\face$, $ \eta_{\n} < 0$ and $\face$ is space-like. We also define speed $\frak{c}$ as,
\begin{equation}
\frak{c} = \sqrt{\dfrac{ N^2-\omega^2}{\omega^2}} >0, \quad \text{ on } \mathbb{B}_{\mathrm{hyp}}.\end{equation}
We can write $\eta_{\n}$ as (at zero attenuation),
\begin{equation}\eta_{\n} =  \dfrac{c_0^2}{N^2-\omega^2} \left( n_r^2 - \frak{c}^2  n_{\parallel}^2\right), \quad \text{ on } \mathbb{B}_{\mathrm{hyp}}.\end{equation}

We next observe the different nature of $\tau_d$ (diffusion-type stabilization) and $\tau_c$ (convection-type stabilization) at zero attenuation. The definition of $\tau_c$ which depends on $\eta_{\n}$ reflects the nature of the PDE on $\face$, while $\tau_d$ which depends on $\eta_{\boldsymbol{\beta}}$ only signals the direction of $\tilde{\mathbf{n}}$.
\begin{itemize}[leftmargin=*]
\item Diffusion-type: with the sign of $\eta_{\n}$ given by
\begin{equation}
\eta_{\n} =  \begin{dcases} <0, & \quad \text{ elliptic or hyperbolic time-like}, \\ >0,  &  \quad \text{  hyperbolic space-like}, \end{dcases}
\end{equation}
the stabilization $\tau_d$ takes the following form,
\begin{equation}
\tau_d = \ii \omega \, \alpha_{\text{tune}} \,  \eta_{\n}\times  \begin{dcases} - 1, & \quad \text{ elliptic or hyperbolic time-like face}, \\ 1,  &  \quad \text{  hyperbolic space-like}\,. \quad   \end{dcases}
\end{equation}

\item Convection-type: Note that $\tau_c =0 $ when $\eta_{\boldsymbol{\beta} } < 0$ and 
$\tau_c = 2 \lvert \eta_{\boldsymbol{\beta}}\rvert $ when $\eta_{\boldsymbol{\beta} } > 0$. 
Both cases can happen on each region, depending on the sign of $n_r \beta_1$. 
The total contribution from the convective-type stabilization in \cref{defNeutrace_spec} is thus
\begin{equation}
\tau_c -  \eta_{\boldsymbol{\beta}}=
\lvert \eta_{\boldsymbol{\beta}}\rvert = \eta_{\boldsymbol{\beta}} \times  \begin{dcases} -\mathrm{sgn}\, (n_r\beta_1) , & \quad \text{ elliptic}, \\ \mathrm{sgn}\, (n_r\beta_1),  &  \quad \text{  hyperbolic }.   \end{dcases}
\end{equation}

\end{itemize}

\end{remark}

\begin{remark}
We note that solar gravity affects differently the numerical resolution in 1.5D 
than in 2.5D  although within the same perspective of discretization (HDG) method.
For the same radially symmetric background, the current working equations lead to a different 1.5D problem having as variables, radial displacement and pressure perturbation,
in contrast to the 1.5D system considered in 
 \cite{Pham2024stabilization} which works  with radial displacement and its derivative. 
 The interior singularity (in the context of scalar ODE theory) encountered in \cite{Pham2024stabilization} as analyzed in 
 \cite{Pham2021Galbrun} is related to the Lamb frequencies, 
cf. discussion in \cite{Pham2024stabilization,Pham2021Galbrun}. Secondly, 
the 1.5D problem can employ mesh refinement more freely than in 2.5D, 
which allows for larger leeway with HDG stabilization. 
We note that \cite{Pham2024assembling} studies effects of gravity however in the context of devising absorbing boundary conditions.
\end{remark}

\section{Qualitative wavefield comparisons with solar background} 
\label{sect:numerical}

In this section we conduct a qualitative investigation of the 
behaviour of the axisymmetric HDG solver using the solar 
background parameters. 
The solar background presents a challenge due to the 
decrease of the parameters near-surface, and the non-zero 
buoyancy frequency, cf. \cref{subsection:numerical-solar-backgrounds}.
In \cref{subsection:numerical-comparison-formulation}, we compare
the variant formulations introduced in \cref{Liouville::subsec}
in terms of amplitudes of the numerical solutions. 
Subsequently, we investigate the choice of HDG stabilization in 
\cref{numerical-hdg:qualitative}.  
In particular, we emphasize that for frequencies below the cut-off 
and a source position in the region $\Bhyp$ (where $N^2 > \omega^2$)
a propagation of singularity occurs and the choice of stabilization
plays an important role to maintain the accuracy. 
Note that such a source position corresponds to the actual 
situation for creating helioseismic observables, as detailed in
\cref{section:numerical-helio}. The numerical experiments are carried out with the 
open-source software \texttt{Hawen}, 
\cite{Hawen2021} (\url{https://ffaucher.gitlab.io/hawen-website/})
employing direct solver MUMPS, \cite{Amestoy2019} for linear 
system resolution. The setups to reproduce the numerical experiments
are provided in the dedicated Zenodo repository, 
\url{https://doi.org/10.5281/zenodo.21297984}.

\begin{remark}[Rescaling]
  The model parameters are adimensionalized with respect to a 
  reference length which is taken to be the solar radius $\rsun$.
  The scaled background quantities are then plugged into the 
  equations, \cite{AtmoI2020}, \cite[Section 2]{Pham2021Galbrun}. 
  By denoting with subscript $\mathrm{S}$ the unscaled solar 
  background models, we have the scaled quantities as follow, 
  \cite{AtmoI2020,Pham2021Galbrun} 
  \begin{subequations}\begin{align}
   \rho_0 \,:=\, \rho_0^{\mathrm{S}} \,; \hspace*{2em}    
    c_0 \,:=\, c_0^{\mathrm{S}} / \rsun \,; \hspace*{2em}
    \Phi_0 \,=\, \Phi_0^\mathrm{S} / \rsun^2\,; \hspace*{2em}
    p_0 \,=\, p_0^\mathrm{S} / \rsun^2\,.
  \end{align}\end{subequations}
  Other quantities such as rotation and frequency remain
  in their original units. Therefore, in the numerical 
  experiments the scaled position $r=1$ corresponds to the 
  actual solar position $\rsun=\num{696000}\si{\km}$.

\end{remark}

\subsection{Solar background radial model}
\label{subsection:numerical-solar-backgrounds}

\begin{figure}[ht!]\centering
  \includegraphics[scale=1]{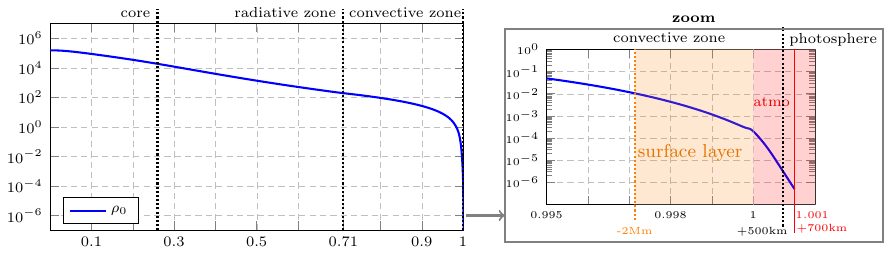}
  
  \includegraphics[]{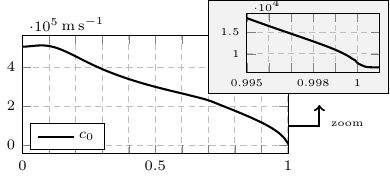} \hspace*{1em}
  \includegraphics[]{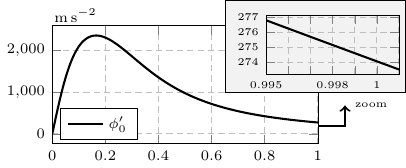}
\caption{Radial solar background models for the density $\rho_0$ (\textbf{top}), 
         wave-speed $c_0$ (\textbf{bottom left}) and 
         gravity potential $\phi_0'$ (\textbf{bottom right}) 
         corresponding to the standard \texttt{model-S}, 
         \cite{christensen1996current,AtmoI2020} (all units in SI).
         The plot of the density also highlights the different
         layers of the Sun.
         The position $r=1$ corresponds to the solar 
         radius $\rsun$, numerical experiments are
         carried out up to scaled radius $\rmax=\num{1.001}$
         which corresponds to 700\si{\km} above solar radius.}
\label{figure:solar-models}
\end{figure}

\begin{figure}[ht!]\centering
\subfloat[Buoyancy and Lamb frequencies.]
         {\includegraphics[]{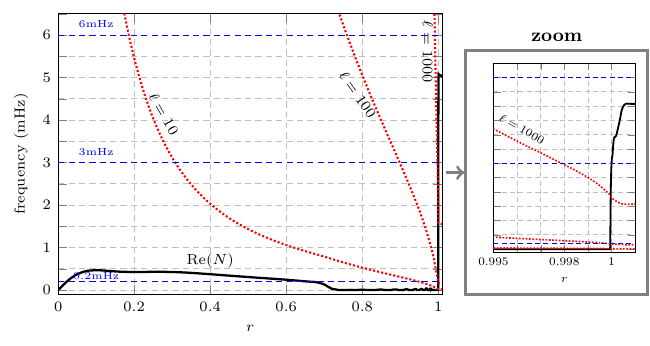}
          \label{figure:solar-models:N2Sl2:N}} \hfill
\subfloat[$N^2$ crossing with $\omega^2$ at 0.2 and 3\si{\milli\Hz} near surface.]
         {\includegraphics[]{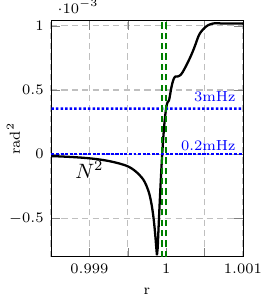}
          \label{figure:solar-models:N2Sl2:N2}}
\caption{The buoyancy frequency (or Brunt--V\"ais\"al\"a frequency) is represented
         with a solid line, Lamb frequencies $S_\ell$ are represented 
         with dotted line for different harmonic degree $\ell$. 
         The horizontal dashed lines indicate the three representative
         frequencies that we use in the numerical experiments, \num{0.2},
         \num{3}, and \num{6}\si{\milli\Hz}.
         The right panel shows $N^2$ crossing with $\omega^2$ near the
         surface at 0.2\si{\milli\Hz} (in $r=$\num{0.999949}) and 
         at 3\si{\milli\Hz} (in $r=$\num{0.9999995}).
         }
\label{figure:solar-models:N2Sl2}
\end{figure}

In \cref{figure:solar-models}, we show the radial solar 
background models that correspond to the standard 
\texttt{model-S} in the interior, \cite{christensen1996current}, and extension 
in the atmosphere, \cite{AtmoI2020}. 
The density $\rho_0$ is further plotted on a logarithmic axis 
which highlights its drastic decrease, with about twelve 
orders of magnitude, and the different Sun's layer are indicated. 
In \cref{figure:solar-models:N2Sl2}, the 
buoyancy (or Brunt-V\"ais\"al\"a) frequency 
\cref{Nfreq::def} is plotted together with 
Lamb frequencies for a few harmonic degrees $\ell$.
Using the radial backgrounds, the buoyancy 
frequency $N^2$ is readily used to define 
regions $\Bhyp$ and $\Bell$ of \cref{ellhyp::def}. 
Note that we picture the ordinary frequency
(i.e., $\mathrm{Re}(N)/(2\pi)$) in \cref{figure:solar-models:N2Sl2:N} 
and $N^2$ in \cref{figure:solar-models:N2Sl2:N2}. 

Regarding the solar buoyancy frequency, we see that: 
\vspace*{-0.50em}
\begin{itemize} \setlength{\itemsep}{-1pt}
  \item $N^2$ becomes negative close to the surface, as depicted 
        in \cref{figure:solar-models:N2Sl2:N2}, 
        leading to convective instabilities, a difficulty for time-domain simulations \cite{Schunker2011,Papini2014}.

  \item The buoyancy frequency is positive  in the interior, 
        up until a maximum of $\num{0.4664}\si{\milli\Hz}$,
        which corresponds to the maximal frequency at which internal 
        gravity waves can be excited, \cite{Pham2024assembling}.
\end{itemize}

In the experiments of this section, we will focus on three frequencies:
\num{0.2}, \num{3} and \num{6} \si{\milli\Hz}. These are representative
of the different configurations between $\omega^2$ and $N^2$:
\vspace*{-0.50em}
\begin{itemize} \setlength{\itemsep}{-1pt}
  \item For frequencies above the cut-off 
        ($\sim \num{5.2}\si{\milli\Hz}$, \cite{Pham2024assembling})
        $N^2 < \omega^2$ for all $r$. Therefore, in this case 
        $\Bell$ corresponds to the entire domain and $\Bhyp$ 
        is empty, \cref{ellhyp::def}.

  \item For frequencies below the cut-off, there is at least one 
        portion of the domain where $N^2 > \omega^2$, which means 
        that $\Bhyp$ is not empty, \cref{ellhyp::def}. 
        We further distinguish two situations:
        \vspace*{-0.60em}
        
        \begin{itemize}
            \item For frequencies above $\num{0.4664}\si{\milli\Hz}$
                  $\Bhyp$ corresponds only to an area stating in the 
                  near-surface/solar atmosphere. For instance at 
                  3\si{\milli\Hz}, $\Bhyp$ corresponds to the zone
                  $r > \num{0.9999995}$, \cref{figure:solar-models:N2Sl2:N2}.
                  This corresponds to about \num{350}\si{\meter} below 
                  the surface.

            \item For frequencies below $\num{0.4664}\si{\milli\Hz}$
                  $\Bhyp$ is also non-empty in the interior, with
                  at least three crossings between $N^2$ and $\omega^2$,
                  \cref{figure:solar-models:N2Sl2:N}.
                  For instance at 0.20\si{\milli\Hz}, 
                  $\Bhyp$ corresponds to
                  $\num{0.022606} < r < \num{0.659452}$ and 
                  $r > \num{0.999949}$ (\num{35.5}\si{\kilo\meter} below surface).
        \end{itemize}
\end{itemize}

Furthermore, a non-zero attenuation $\gamma_{\mathrm{att}}$
is always used to avoid the coefficient $\boldsymbol{A}_\bullet$ 
of \cref{eq:intro} to be zero at the position where $N^2 = \omega^2$.
Specifically, constant attenuation 
$\gamma_{\mathrm{att}} = 10\si{\micro\Hz}$ 
is used for experiments at 6 and 3\si{\milli\Hz} below,
while $\gamma_{\mathrm{att}} = 2\si{\micro\Hz}$ is used
at \num{0.2}\si{\milli\Hz}.
Despite having non-zero attenuation, in the following 
we still refer to $\Bhyp$ and $\Bell$ according to their definition \cref{ellhyp::def}.

For solar applications, the main source of excitation of acoustic waves is due to granulation and the source height is thus located a few hundred kms below the surface. The receiver height corresponds to the observation height which slightly depends on the position on the disk and the instrument but is located around 100-200 kilometers above the surface.
To generate helioseismic observables in \cref{section:numerical-helio},
we choose $r_{\mathrm{src}}=\num{0.99986}$
(about 97\si{\km} below surface) and 
$r_{\mathrm{rcv}}=\num{1.00021}$ (about 146\si{\km} above surface).

\paragraph{Discretization meshes}

The discretization meshes are created by first imposing the 
radial layers, using the wavelength computed from the solar 
background and potential associated to the spherical wave 
equation \cite{Pham2021Galbrun}.
Next, the tangential edge size is selected to be
similar to the layer size. Once the set of node is fixed, 
the software \texttt{Triangle}\footnote{\url{www.cs.cmu.edu/~quake/triangle.html}} \cite{shewchuk1996triangle} is used to assemble the mesh.
The meshes are thus more and more refined near the surface according 
to the variation of the background parameters and reduction of the wavelength. 
In the numerical experiments, two choices of maximal 
radius $\rmax$ are used to investigate the effect of 
the position of the outer boundary, with $\rmax=\num{1.001}$ 
and $\rmax=\num{1.01}$. These correspond to about 
\num{696} and \num{6960}\si{\km} above surface, respectively.
\begin{itemize}
  \item $\rmax=\num{1.001}$; three meshes used:
  \vspace*{-0.40em}
  \begin{itemize}
    \item \meshCa~with \num{24949} cells, used for the 
          low-frequency \num{0.2}\si{\milli\Hz}.
   \item \meshAa~with \num{175919} cells, it is 
         the main mesh used in simulations, created
         such that there are at least 10 points per 
         wavelength at frequency 6\si{\milli\Hz} 
         with polynomial order 4.
   \item \meshAb~with \num{594350} cells is a refined 
          version of \meshAa and is used to obtain reference solutions.
  \end{itemize}
  \item $\rmax=\num{1.01}$; the extended domain uses 
        \meshBa~that has \num{1183700} cells. 
        It is used to illustrate the effect of 
        moving the boundary further out, and it
        is based upon \meshAb~up until \num{1.001}. 
\end{itemize}

\paragraph{Relative difference} 
To evaluate the accuracy in the absence of analytic solution, 
we introduce the relative difference $\err$ between a reference 
solution $w_{\mathrm{ref}}$ and a simulation $w_{\tau}$. 
We further introduce quantities $\erra$ and $\errb$ that evaluate
the norm of $\err$ on different subdomains. We have,
\begin{equation}\label{eq:relative-difference}
\begin{aligned}
& \err(\eta,z) \,=\, \dfrac{\vert w_{\mathrm{ref}}(\eta,z) - w_{\tau}(\eta,z) \vert}
                        {\Vert w_{\mathrm{ref}} \Vert_2} \,, \\
& \erra = \Vert \err(\eta,z) \Vert\,, 
      \quad \text{$\eta \in [0:\num{e-3}:\rmax]$, $z \in [-\rmax:\num{e-3}:\rmax]$\hspace*{1.5em} with
                  $\sqrt{\eta^2 + z^2} \leq \rmax$} \, ; \\
& \errb = \Vert \err(\eta,z) \Vert\,, \,\,
      \text{$\eta \in [0.99:\num{e-5}:\rmax]$, $z \in [-0.005:\num{e-5}:0.02]$ with 
                  $\sqrt{\eta^2 + z^2} \leq \rmax$} \, .
\end{aligned}
\end{equation}
In \cref{eq:relative-difference}, $\erra$ corresponds to the norm of the 
relative difference on the entire computational interval, while $\errb$ 
focuses on a near-surface area close to the equator, and we use a structured
grid to evaluate the functions. 
As the experiments below consider the source on the equator, $\errb$ also 
emphasises the relative difference near the source position. However, for 
the computation of $\errb$, we exclude the actual position of the source 
in order to avoid instabilities at this position.
In the following, $w_{\mathrm{ref}}$ is typically computed with the 
refined mesh \meshAb, and we compare its difference with solutions 
computed with \meshAa.

\subsection{Comparison of formulations}
\label{subsection:numerical-comparison-formulation}

\begin{figure}[ht!] \centering
  \subfloat[Solution $w_{\mathrm{L0}}$ solving $\Lliouville$.]
           {\includegraphics[]{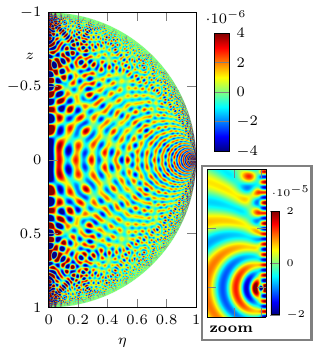}}   \hspace*{1em}
  \subfloat[Solution $w_{\mathrm{Lc}}$ solving $\Lliouvillec$.]
           {\includegraphics[]{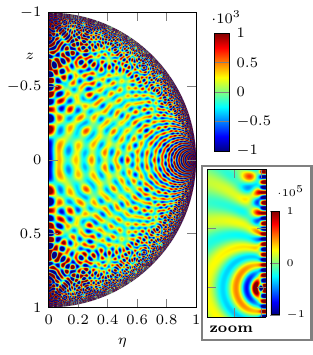}}  \hspace*{1em}
  \subfloat[Solution $w_{\mathrm{og}}$ solving $\Loriginaldiv$.]
           {\includegraphics[]{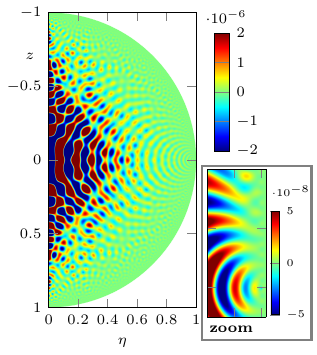} \label{figure:comparison-of-formulations_L0}} 

  \subfloat[Normalized solutions $\vert w_\bullet \vert/\max(\vert w_\bullet \vert)$ in $z=0$.]
           {\includegraphics[]{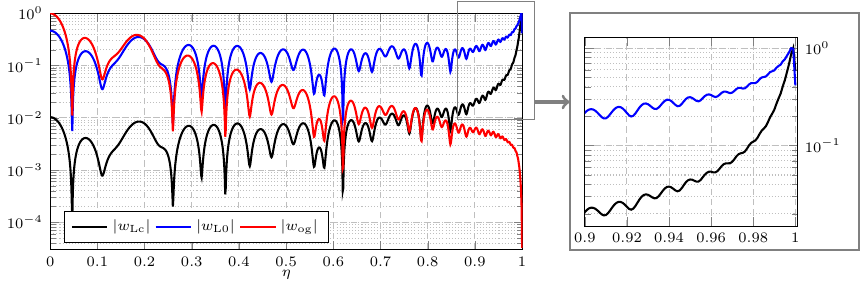}
            \label{figure:comparison-of-formulations:x-axis}} 

  \caption{Comparison of the solutions $\mathrm{Re}(w_\bullet)$ for the different 
           formulations for a source positioned in $(1,0)$, at mode 
           $m=0$ and frequency $6$\si{\milli\Hz}. 
           The zoom near the source corresponds to the zone 
           $(\num{0.99},\num{1.001})_\eta \times (\num{-0.005},\num{0.02})_z$,
           and employs a different scaling for better visualization.
           }
  \label{figure:comparison-of-formulations}
\end{figure}

We first compare the different formulations of the equations 
that have been introduced, namely $\Lliouville$, $\Lliouvillec$ 
and $\Loriginaldiv$, with their coefficients given in \cref{Liouville::subsec}. 
Their corresponding unknowns are denoted 
$(\bxi_{\mathrm{L0}},w_{\mathrm{L0}})$, 
$(\bxi_{\mathrm{Lc}},w_{\mathrm{Lc}})$, and 
$(\bxi_{\mathrm{og}},w_{\mathrm{og}})$ respectively,
with relations given in \cref{eq:liouville-unknowns}.
The solutions $w_{\bullet}$ are pictured in 
\cref{figure:comparison-of-formulations} for 
a source positioned in $(1,0)$, at azimuthal 
mode $m=0$ and frequency $6$\si{\milli\Hz}. 
Note that at this frequency, $\Bhyp$ is 
empty and $\Bell$ is the entire domain, 
i.e., $N^2 < \omega^2$, cf.~\cref{figure:solar-models:N2Sl2}.

The original formulation $\Loriginaldiv$ shows strong variation 
in amplitude between the center and the surface 
with a decrease of 4 orders in magnitude between
$r=0$ and $r=1$ in \cref{figure:comparison-of-formulations:x-axis}.
The discrepancy in scale renders difficult the simultaneous 
visualization the oscillation patterns in the interior and 
outer layers, as seen in \cref{figure:comparison-of-formulations_L0}
showing saturation in the interior but indistinguishable 
amplitude at higher height.
The latter affects in particular the near-surface layers 
where solar observables are generated,  and persists even 
in the zoom  of \cref{figure:comparison-of-formulations_L0}.
On the other hand, the Liouville variants 
$\Lliouville$ and $\Lliouvillec$ do not suffer from this 
contrast in amplitude between the interior and the surface, 
with only 2 orders of magnitude difference for $\Lliouvillec$, 
and about 1 for $\Lliouville$.
In view of the whole domain,  $\Lliouville$ yields solutions 
of more bounded variation compared to $\Lliouvillec$. 

To supplement these qualitative observations, the condition numbers 
of the global HDG matrices corresponding to the different formulations 
are given in \cref{table:condition-numbers} for frequencies between $1$ 
and $7$~\si{\milli\Hz}. The condition numbers are obtained with the direct 
sparse solver MUMPS also employed for resolution, cf. \cite{mumpsuserguide}.
We see that for both frequencies, the two Liouville variants 
yield similar condition numbers, both of which are approximately one order 
of magnitude lower than that of the original formulation.
They therefore  
improve the conditioning of the global matrix, 
and in this way enhance numerical robustness. 
On the other hand, we observe a general increase of the condition numbers 
with frequencies, except between 5 and 6 \si{\milli\Hz} where 
it decreases. This can be explained by the absence of $\Bhyp$ 
at frequencies above cut-off, cf. \cref{subsection:numerical-solar-backgrounds}.
In the rest of our investigation, we only present the results 
obtained with the variant $\Lliouvillec$.  

\begin{table}[ht!]
\begin{center} 
\caption{Conditioning of the global HDG matrix depending 
                on the formulation and frequency.}
\vspace*{-0.50em} 
\label{table:condition-numbers} 
\renewcommand{\arraystretch}{1.10} 


\begin{minipage}[t]{.48\linewidth}
\vspace*{0pt}
\begin{tabular}{|>{\arraybackslash}p{2.90em}|
                 >{\arraybackslash}p{4.0em}|
                 >{\arraybackslash}p{4.5em}|
                 >{\arraybackslash}p{4.5em}|}
\hline
Freq.            & $\Loriginaldiv$ & $\Lliouville$ & $\Lliouvillec$ \\ \hline
 1 \si{\milli\Hz} & \num{5.79e5} & \num{2.90e4} & \num{3.05e4} \\
 2 \si{\milli\Hz} & \num{6.75e5} & \num{4.51e4} & \num{4.75e4} \\
 3 \si{\milli\Hz} & \num{9.85e5} & \num{1.71e5} & \num{1.25e5} \\
 4 \si{\milli\Hz} & \num{3.22e6} & \num{2.84e5} & \num{3.09e5} \\
 \hline
\end{tabular}
\end{minipage}\begin{minipage}[t]{.48\linewidth}
\vspace*{0pt}

\begin{tabular}{|>{\arraybackslash}p{2.90em}|
                 >{\arraybackslash}p{4.0em}|
                 >{\arraybackslash}p{4.5em}|
                 >{\arraybackslash}p{4.5em}|}
\hline
Freq.            & $\Loriginaldiv$ & $\Lliouville$ & $\Lliouvillec$ \\ \hline
 5 \si{\milli\Hz} & \num{3.27e6} & \num{8.62e4} & \num{1.03e5} \\
 6 \si{\milli\Hz} & \num{3.30e5} & \num{5.73e4} & \num{3.62e4} \\
 7 \si{\milli\Hz} & \num{6.35e5} & \num{6.20e4} & \num{3.52e4} \\
 \hline
\end{tabular}
\end{minipage}

\end{center}
\end{table}

\subsection{Empirical study of the HDG stabilization} 
\label{numerical-hdg:qualitative}

The HDG discretization involves a stabilization parameter 
$\tau$, \cref{subsection:hdg-stabilization}, which is critical for
the accuracy of the method (e.g., \cite{Pham2024stabilization}).
Here we use the Liouville-c formulation of the equation 
and solves for unknowns $(\bxi_c,w_c)$, such that the 
relation for the numerical traces is written as 
\begin{equation} \label{eq:num:hdg-stabilization}
   \widehat{\bxi_{c\,m}} \,=\, \bxi_{c\,m} \,-\, \tau \big( w_{c\,m} \,-\, \lambda)\, \n \,,
   \qquad\quad \text{HDG Numerical trace relation.}
\end{equation}
Following the derivation carried out in 
\cref{subsection:hdg-stabilization}, the 
stabilization term $\tau$ is of the form
\begin{equation}\label{eq:num:hdg-tau-format}
  \tau \,=\, \vert \boldsymbol{A}^{-1} \boldsymbol{\beta}_1 \cdot \n \vert 
             \,+\, 
             \, \alpha_{\text{tune}} \, \vert  \n^\mathrm{t} \,\boldsymbol{A}^{-1}\,\n \vert \,,
\end{equation}
with $\alpha_{\text{tune}}$ a scaling factor. 
For the heterogeneous solar backgrounds considered, analytical solutions
do not exist, therefore two criteria are set up to decide of an acceptable 
stabilization:
(1) The absence of artifact in the solution,
and (2) Comparison with a reference solution based on a refined mesh.

We have performed a comparative study for values of $\alpha_{\text{tune}}$ 
in the complex plane, with amplitude $\pm \omega \num{e6}$. 
For the sake of conciseness, we do not provide all the choices 
of stabilization that have been experimented, and only work 
with the following illustrative choices:
\begin{subequations} \label{eq:list-if-tau}\begin{align}
 \taua & \,=\, 1 \, , \\
 \taub & \,=\, \vert \boldsymbol{A}^{-1} \boldsymbol{\beta}_1 \cdot \n \vert 
               \,-\, \num{e6} \ii \omega \, \vert  \n^\mathrm{t} \,\boldsymbol{A}^{-1}\,\n \vert
               \, , \\
 \tauc & \,=\, \vert \boldsymbol{A}^{-1} \boldsymbol{\beta}_1 \cdot \n \vert 
               \,+\, \num{e6} \omega \, \vert  \n^\mathrm{t} \,\boldsymbol{A}^{-1}\,\n \vert \, .
\end{align} \end{subequations}
In particular, we will see that the stabilization is critical mostly when 
the source is in or close to $\Bhyp$, i.e., when there are parts of the 
domain in which $N^2 > \omega^2$.

\subsubsection{Behaviour above cut-off frequency: $\Bhyp$ is empty}

\begin{figure}[ht!] \centering
  \subfloat[Simulations $\mathrm{Re}(w_c)$ 
            with different stabilizations. 
            \textbf{left}:   $\taua$, $\erra=6.\num{e-2}$, $\errb=8.\num{e-2}$ ; 
            \textbf{middle}: $\taub$, $\erra=5.\num{e-2}$, $\errb=4.\num{e-2}$ ;
            \textbf{right}:  $\tauc$, $\erra=4.\num{e-2}$, $\errb=7.\num{e-2}$.
            ]{
  \includegraphics[scale=1]{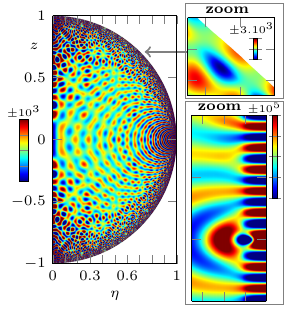}
  \hspace*{0.50em}
  \includegraphics[scale=1]{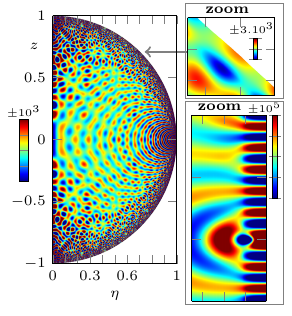}
  \hspace*{0.50em}
  \includegraphics[scale=1]{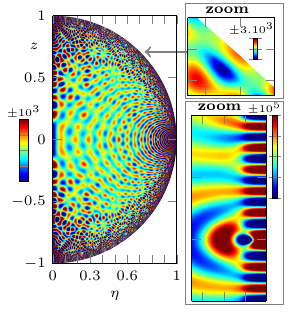}
  }

  \subfloat[Relative difference $\err$ \cref{eq:relative-difference}
            between a reference solution computed with \meshAb and 
            simulations using \meshAa~with 
            $\taua$ (\textbf{left}), $\taub$ (\textbf{middle})
            and $\tauc$ (\textbf{right}).
            ]{
  \includegraphics[scale=1]{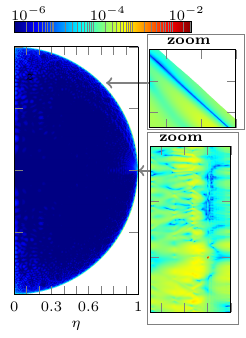}
  \hspace*{1.50em}
  \includegraphics[scale=1]{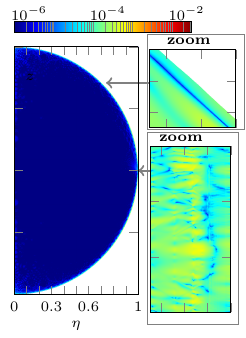}
  \hspace*{1.50em}
  \includegraphics[scale=1]{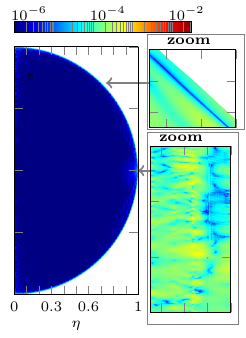}
  }  

  \caption{Simulations and relative difference $\err$ \cref{eq:relative-difference}
           for a Dirac source in $(1,0)$ at 6\si{\milli\Hz} 
           using \meshAa~for different choices of HDG stabilization 
           \cref{eq:list-if-tau}.
           The solutions on the entire domain are complemented with 
           a zoom near source position on interval 
           $(\num{0.9975},\num{1.001}) \times (-\num{0.005},\num{0.01})$, 
           and a zoom 
           in the area $(0.705,0.71)^2$. The attenuation is set to 10\si{\micro\Hz}.            
           }
  \label{fig:no-rot_2dwave_6mHz_m150k}
\end{figure}

In this experiment, the frequency is above 
cut-off, with $\omega/(2\pi)=6\si{\milli\Hz}$. 
The source is positioned in $(1,0)$ and simulations 
are performed at azimuthal mode $m=0$. 
For this frequency, $\Bell$ corresponds to the 
entire domain as $N^2 < \omega^2$ for 
all positions, see \cref{figure:solar-models:N2Sl2}.
We employ the mesh \meshAa~and polynomial order $4$, 
the results are shown in \cref{fig:no-rot_2dwave_6mHz_m150k}
for the different choices of stabilization \cref{eq:list-if-tau}. 
The solution on the axisymmetric domain is pictured,
as well as a zoom near the source position (close to the outer boundary) 
and also a zoom away from the source but still close 
to the boundary. 
In this case, the different choices of stabilization 
appear to give comparable results and we cannot distinguish 
any difference, this is confirmed by the relative 
difference computed with respect to simulation using 
the refined mesh \meshAb, which has the same pattern 
for each case. This relative difference is particularly 
low in the interior of the domain, and higher near the 
source position.

\subsubsection{Behaviour below cut-off frequency with $\Bhyp$ only near surface}

At frequency 3\si{\milli\Hz}, there is one crossing 
between $N^2$ and $\omega^2$; 
$\Bhyp$ corresponds to the part where the radius 
$r > \num{0.9999995}$, see \cref{figure:solar-models:N2Sl2}. 
We consider three sources to investigate the numerical 
solutions, all near the surface to be representative of 
helioseismology, \cite{Pham2024assembling}. 
The sources are positioned at $(0.999,0)$ (in $\Bell$, 
corresponding to about 696\si{\kilo\meter} below surface),
at $(0.99986,0)$ (in $\Bell$ but closer to the transition,
about 97\si{\kilo\meter} below the surface), 
which is the position used for helioseimic observables in 
\cref{section:numerical-helio}),
and a source in $(1,0)$ (in $\Bhyp$). 
The corresponding solutions for the different choices
of HDG stabilization 
are pictured in \cref{fig:no-rot_2dwave_3mHz_m150k}.
Here we also provide the relative error $\err_2$ and 
$\err_\infty$ of \cref{eq:relative-difference}, where
the reference solution is computed using \meshAb and $\tauc$.

\pgfmathsetmacro{\myscale}{1.0} 
\begin{figure}[ht!] \centering
   \subfloat[Simulations using $\taua$ for different sources, 
             \textbf{left}:  source in $(0.99,0)$, 
                             $\erra=1.7\cdot\num{e-2}$, $\errb=1.7\cdot\num{e-2}$;
             \textbf{middle}: source in $(0.99986,0)$, 
                             $\erra=1.8\cdot\num{e-2}$, $\errb=2.6\cdot\num{e-2}$;
             \textbf{right}:  source in $(1,0)$, 
                             $\erra=2.3\cdot\num{e-2}$, $\errb=7.7\cdot\num{e-2}$.]{
   \makebox[\linewidth][c]{
   \includegraphics[scale=\myscale]{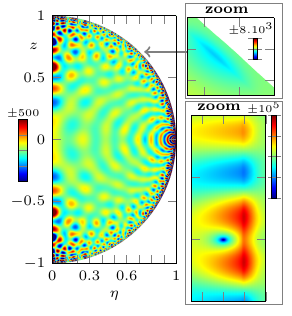}
   \hspace*{0em}
   \includegraphics[scale=\myscale]{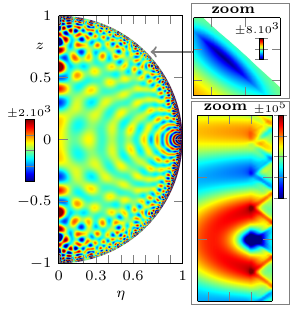}
   \hspace*{0em}
   \includegraphics[scale=\myscale]{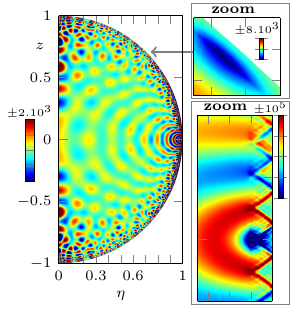}
   }}
   \vspace*{-0.50em} 

   \subfloat[Simulations using $\taub$ for different sources, 
             \textbf{left}:  source in $(0.99,0)$, 
                             $\erra=1.6\cdot\num{e-2}$, $\errb=1.8\cdot\num{e-2}$;
             \textbf{middle}: source in $(0.99986,0)$, 
                             $\erra=1.6\cdot\num{e-2}$, $\errb=2.6\cdot\num{e-2}$;
             \textbf{right}:  source in $(1,0)$, 
                             $\erra=1.7\cdot\num{e-2}$, $\errb=3.9\cdot\num{e-2}$.]{
   \makebox[\linewidth][c]{   
   \includegraphics[scale=\myscale]{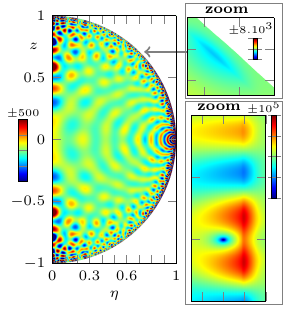}
   \hspace*{0em}
   \includegraphics[scale=\myscale]{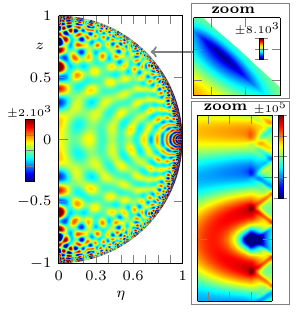}
   \hspace*{0em}
   \includegraphics[scale=\myscale]{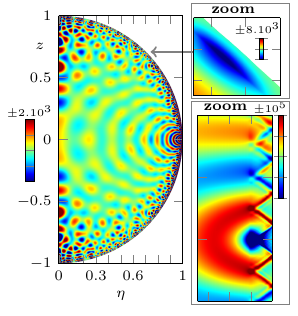}
   }}
   \vspace*{-0.50em} 

   \subfloat[Simulations using $\tauc$ for different sources, 
             \textbf{left}:  source in $(0.99,0)$, 
                             $\erra=2.3\cdot\num{e-2}$, $\errb=\num{e-2}$;
             \textbf{middle}: source in $(0.99986,0)$, 
                             $\erra=1.9\cdot\num{e-2}$, $\errb=1.9\cdot\num{e-2}$;
             \textbf{right}:  source in $(1,0)$, 
                             $\erra=2.2\cdot\num{e-2}$, $\errb=6.7\cdot\num{e-2}$.]{
   \makebox[\linewidth][c]{
   \includegraphics[scale=\myscale]{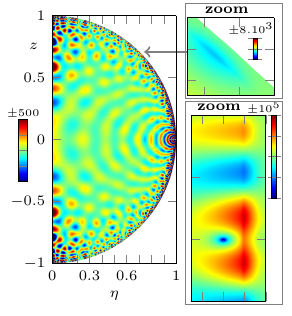}
   \hspace*{0em}
   \includegraphics[scale=\myscale]{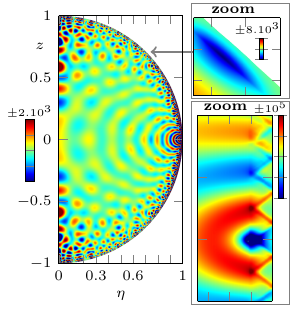}
   \hspace*{0em}
   \includegraphics[scale=\myscale]{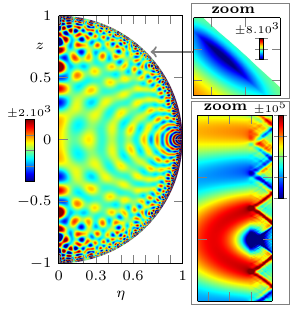}
   }}

  \caption{Simulations $\mathrm{Re}(w_c)$ at 3\si{\milli\Hz} using \meshAa for 
           a source positioned in 
           $(\num{0.999},0)$ (\textbf{left}), 
           $(\num{0.999876},0)$ (\textbf{middle}) 
           and $(\num{1},0)$ (\textbf{right}), 
           for different choices of stabilization 
           \cref{eq:list-if-tau}.
           At 3\si{\milli\Hz}, the transition between 
           $\Bhyp$ and $\Bell$ is in $r=\num{0.9999995}$.
           The solutions on the entire domain 
           are complemented with a zoom near 
           source position on interval 
           $(\num{0.9975},\num{1.001}) \times (-\num{0.005},\num{0.01})$, and a zoom 
           in the area $(0.705,0.71)^2$.
           The relative difference \cref{eq:relative-difference} 
           uses a reference solution computed with \meshAb and $\taub$.
           The attenuation is set to 10\si{\micro\Hz}.}
           
   \label{fig:no-rot_2dwave_3mHz_m150k}
\end{figure}

\begin{figure}[ht!] \centering
  \subfloat[The reference solution $\mathrm{Re}(w_c)$ is computed with refined mesh \meshAb.]{
  \includegraphics[scale=1]
                  {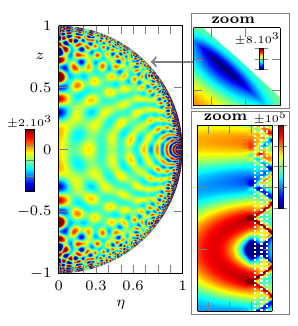}
                  \label{fig:no-rot_2dwave_3mHz_error_s1_ref}}
                   \hfill
  \subfloat[Relative difference $\err$ \cref{eq:relative-difference} 
            between the reference solution in \textbf{a)} and simulation
            using \meshAa~with $\taua$ (\textbf{left}),
            $\taub$ (\textbf{middle}) and $\tauc$ (\textbf{right}).]{
  \includegraphics[scale=0.90]{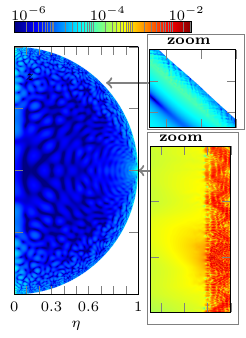}\hspace*{-0.50em}
  \includegraphics[scale=0.90]{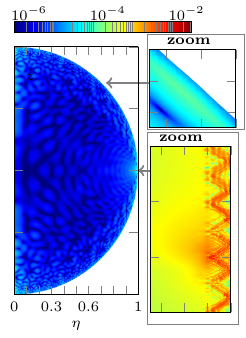}\hspace*{-0.50em}
  \includegraphics[scale=0.90]{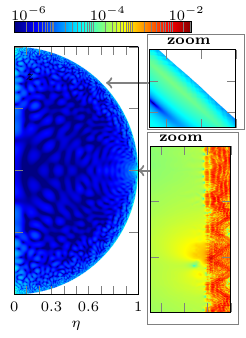}
  \label{fig:no-rot_2dwave_3mHz_error_s1_err}
  }
  
  \caption{Simulations are carried out at 3\si{\milli\Hz} for 
           a source positioned in $(\num{1},0)$, i.e., in $\Bhyp$.
           The solutions on the entire domain are complemented with 
           a zoom near source position on interval 
           $(\num{0.9975},\num{1.001}) \times (-\num{0.005},\num{0.01})$, and a zoom 
           in the area $(0.705,0.71)^2$.
           The attenuation is set to 10\si{\micro\Hz}.
           The white dotted lines on the zoom corresponds to height $r=\num{1.00021}$,
           \num{1.0005} and \num{1.00075}, 
           used to picture solutions in \cref{fig:no-rot_2dwave_3mHz_rcvplot_allheight}.}

  \label{fig:no-rot_2dwave_3mHz_error_s1}
\end{figure}

\begin{figure}[ht!] \centering
  \subfloat[Source at $(\num{0.99986},0)$, solutions
            $\mathrm{Re}(w_c)$ (\textbf{top}) and relative 
            difference $\err$ (\textbf{bottom}) at
            different heights.]{
  \includegraphics[scale=1]{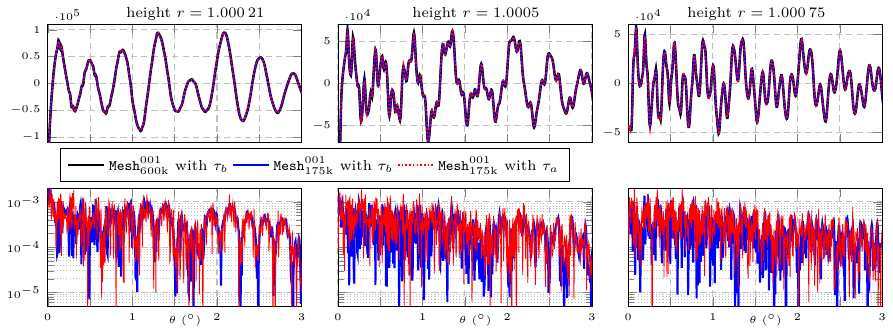}}

  \subfloat[Source at $(\num{1},0)$, solutions
            $\mathrm{Re}(w_c)$ (\textbf{top}) and relative 
            difference $\err$ (\textbf{bottom}) at
            different heights.]{
  \includegraphics[scale=1]{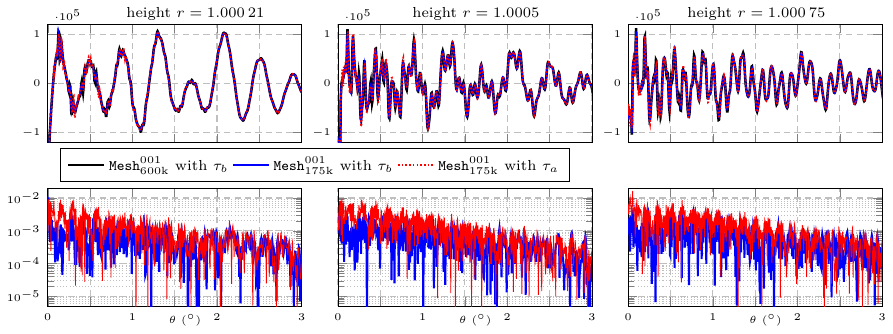}}
  \caption{Comparisons of the solutions at 3\si{\milli\Hz} for 
           fixed height $r$ and angle from 0 to 3\si{\degree}.
           The corresponding solutions on the entire domain
           are depicted in \cref{fig:no-rot_2dwave_3mHz_m150k,fig:no-rot_2dwave_3mHz_error_s1}.
           The relative difference $\err$ \cref{eq:relative-difference} is computed
           using the solution generated from the refined mesh \meshAb~for reference.}
  \label{fig:no-rot_2dwave_3mHz_rcvplot_allheight}
\end{figure}

With the source close to the surface, we see 
waves propagating in the interior, in a similar 
pattern for the three positions investigated in
\cref{fig:no-rot_2dwave_3mHz_m150k}. 
We notice the impact of the source position when 
we zoom near its position and in $\Bhyp$.
With the source in $\Bell$ (left of \cref{fig:no-rot_2dwave_3mHz_m150k}),
we observe that waves are evanescent in $\Bhyp$ and 
thus have small amplitudes. In this case all stabilization
gives comparable results visually, and have similar relative 
difference comparing with the refined solution using 
\meshAb. 
When the source is near or in $\Bhyp$ 
(middle and right of \cref{fig:no-rot_2dwave_3mHz_m150k}, respectively), 
we observe a propagation of singularity from the source position, which 
then reach the surface and is reflected in the layer $\Bhyp$.
It creates a zigzag pattern of waves reflecting back and forth 
in this layer.
We further notice that this pattern is located near the source, 
as solutions in the interior look similar for all three sources, 
and the zooms in the atmosphere away from the source always 
show a smooth pattern (cf. upper right zoom of each panel in 
\cref{fig:no-rot_2dwave_3mHz_m150k}).
Comparing the stabilizations $\tau_\bullet$, we see on 
the zoom near the source that the simulations using 
$\taua$ and $\tauc$ have artifacts in the region $\Bhyp$,
especially when the source is in this region as well
(right of \cref{fig:no-rot_2dwave_3mHz_m150k}), which
is the most challenging case. 
Regarding the relative differences, we see that $\erra$
remains relatively the same between the cases, as the 
solution in the interior of the domain remains accurate.
Regarding $\errb$ which focuses on the part close to the
source and surface, we observe the most contrast for the 
source in $(1,0)$, with almost a factor two between $\taua$ 
and $\taub$.
In \cref{fig:no-rot_2dwave_3mHz_error_s1}, the solution using 
the refined mesh \meshAb~and stabilization $\taub$ is pictured,
together with the full map for the relative error $\err$ 
\cref{eq:relative-difference} for the source in $(1,0)$.
In \cref{fig:no-rot_2dwave_3mHz_rcvplot_allheight}, we 
investigate different heights and different stabilizations.

The relative difference of~\cref{fig:no-rot_2dwave_3mHz_error_s1_err} 
confirms our observations: 
The solutions are accurate for all choices of stabilizations in 
the interior $\Bell$, as well as away from the source in $\Bhyp$. 
The difference is mostly located near the source area, more 
specifically restricted to the region $\Bhyp$. 
This zone of $\Bhyp$ is not accurate for $\taua$ and $\tauc$, and the
difference is only on the rays of the zigzag pattern for $\taub$.
Therefore, when the source is positioned close (or in) $\Bhyp$, 
waves are reflected back and forth in the solar atmosphere in a 
zigzag pattern; This effect is \emph{mostly localized in $\Bhyp$ and 
near the source position}: the solution in the interior or in the 
region further away from the source do not have this pattern as 
waves are rapidly attenuated from the source position. 
In addition, this effect of propagation of singularity is not 
triggered when the source is in $\Bell$.

The solutions at different heights in \cref{fig:no-rot_2dwave_3mHz_rcvplot_allheight}
show that the difference is mostly located near the source position, in the first few degrees. 
the difference increases with the height of the receivers, and is stronger for the source
in $(1,0)$ compared to the source in $(0.99986,0)$, as expected from the zoom provided 
in \cref{fig:no-rot_2dwave_3mHz_m150k}. 
Comparing the stabilization, we see that the difference with $\taub$ is 
smaller than $\tau$, especially for the source in $(1,0)$ and the highest
heights. Nonetheless, the difference is concentrated on the first few degrees
only, for positions below 2\si{\degree}.

\paragraph{Position of the boundary condition}

\begin{figure}[ht!] \centering
  \subfloat[Solution $\mathrm{Re}(w_c)$ increasing $\rmax$ to \num{1.01} with \meshBa.
            The white dash line on the zoom corresponds to height $r=\num{1.00021}$.]{
  {\raisebox{1.50em}{\includegraphics[scale=1]
                  {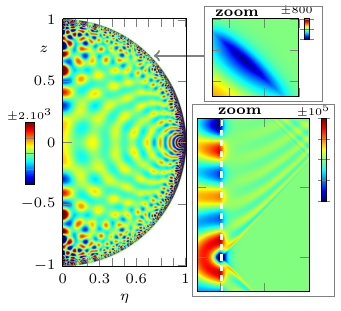}}} \label{fig:no-rot_2dwave_3mHz_s0.99986_rmax-rcv_wave}}
                   \hspace*{.20em}
  \subfloat[Comparison of the solutions $\mathrm{Re}(w_c)$ 
            at fixed height $r=\num{1.0002}$ depending 
            on the mesh (with different $\rmax$) (\textbf{top}) and relative difference
            $\err$ \cref{eq:relative-difference} using \meshBa~for the reference.
            The position $\theta=0$ corresponds to $(\num{1.00021},0)$.]{
  \begin{tikzpicture}
  \node[] (sol) at (0,0) {\includegraphics[scale=0.9]{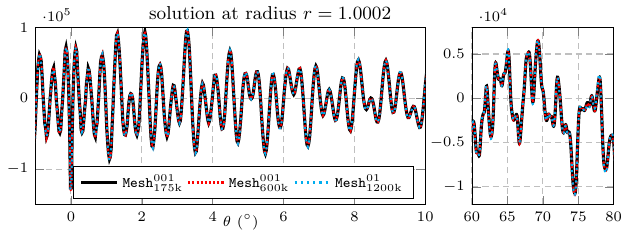}};
  \node[below = -0.4 of sol,xshift=1em] (err) {\includegraphics[scale=0.9]{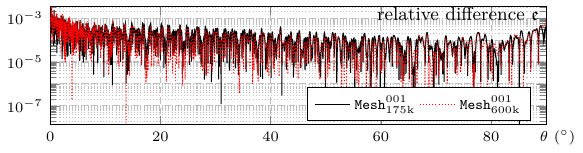}};
  \end{tikzpicture}
  \label{fig:no-rot_2dwave_3mHz_s0.99986_rmax-rcv_rcv}
  }

  \subfloat[Comparison of the solutions $\mathrm{Re}(w_c)$
            at different heights depending on the position
            of $\rmax$.]
           {\includegraphics[scale=1]{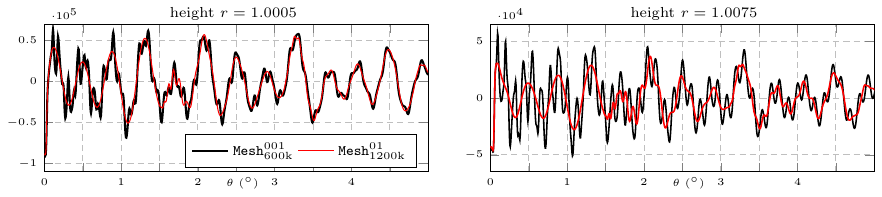}}

  \caption{Simulations are carried out at 3\si{\milli\Hz} for 
           a source positioned in $(\num{0.99986},0)$ using 
           stabilization $\taub$.
           The solutions on the entire domain are complemented with 
           a zoom near source position on interval 
           $(\num{0.995},\num{1.001}) \times (-\num{0.005},\num{0.01})$, and a zoom 
           in the area $(0.705,0.71)^2$.
           The attenuation is set to 10\si{\micro\Hz}.}
   \label{fig:no-rot_2dwave_3mHz_s0.99986_rmax-rcv}
\end{figure}

We now use the extended domain up to $\rmax=\num{1.01}$ 
(instead of \num{1.001}) with \meshBa, to evaluate the 
behaviour of the solutions with respect to the position
of the boundary condition for this frequency below cut-off.
We emphasize the configuration employed for the numerical 
helioseismic observables, where the source is slightly below 
the surface in $(\num{0.99986},0)$, and the solution at receivers 
slightly above, in $r=\num{1.00021}$, cf. \cref{section:numerical-helio}.
The wavefield at frequency $3$ \si{\milli\Hz} 
is shown in \cref{fig:no-rot_2dwave_3mHz_s0.99986_rmax-rcv_wave}. 
In \cref{fig:no-rot_2dwave_3mHz_s0.99986_rmax-rcv_rcv}, we compare
the solutions at $r=\num{1.00021}$ depending on $\rmax$ and the mesh,
and provide the relative difference $\err$ of \cref{eq:relative-difference}.

When the domain is extended, \cref{fig:no-rot_2dwave_3mHz_s0.99986_rmax-rcv_wave},
the singularity propagates further before reaching the surface, 
and the reflective pattern in $\Bhyp$ is milder as waves are attenuated.
We further observe that the solution in the interior does seem to change.
In fact, the solution remains similar even in the lower part of $\Bhyp$ where 
we cannot distinguish the difference between using $\rmax=\num{1.01}$ or 
$\rmax=\num{1.001}$ in the plot of the solutions at height $r=\num{1.00021}$
in \cref{fig:no-rot_2dwave_3mHz_s0.99986_rmax-rcv_rcv}.
This is particularly important as the helioseismic observables use 
this height for the data, see \cref{section:numerical-helio}. 
Consequently, we can retain \meshAa~ for the numerical experiments of 
solar observables. On the other hand, one must remind that if the solution
at higher height is used, it will be affected by the position 
of $\rmax$ for the computational domain.

\subsubsection{Behaviour at low frequency with $\Bhyp$ also in the interior}

At frequency \num{0.2}\si{\milli\Hz}, $\Bhyp$ 
corresponds to the interior of the domain in 
addition to the solar atmosphere, cf.~\cref{figure:solar-models:N2Sl2}. 
Namely, $\Bhyp$ comprises of the layer 
$\num{0.022606} < r < \num{0.659452}$ 
and the part where $r > \num{0.999949}$.
In this case we focus on sources near the interior 
region of $\Bhyp$, and show on \cref{fig:no-rot_2dwave_0.2mHz}
simulations for sources positioned in 
$(0.5,0)$ and $(0.7,0)$. Therefore, the sources 
are in the interior $\Bhyp$ and in $\Bell$, respectively.
As the frequency decreases, so does the wavelength and 
we can use a coarser mesh for the simulation with \meshCa.
The attenuation is reduced to 2\si{\micro\Hz} so that it 
is not too strong compared to the frequency simulated.

\pgfmathsetmacro{\myscale}{0.95}
\begin{figure}[ht!] \centering
  \includegraphics[scale=\myscale]{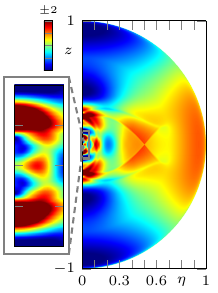}
  \hspace*{1em}
  \includegraphics[scale=\myscale]{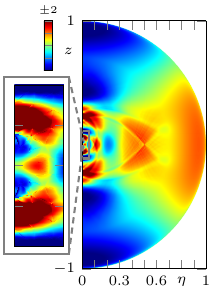} 
  \includegraphics[scale=\myscale]{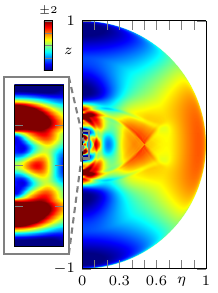}
  \includegraphics[scale=\myscale]{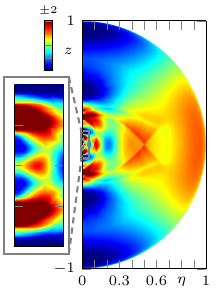}
  \vspace*{-1.50em}

  \subfloat[With source positioned in $(0.5,0)$, from left to right the relative difference 
           is for $\taua$ ($\erra=3.7\cdot\num{e-2}$), 
           $\taub$ ($\erra=3.7\cdot\num{e-2}$) 
           and $\tauc$ ($\erra=6.2\cdot\num{e-2}$).]{
  \hspace*{9.50em}
  \includegraphics[scale=\myscale]{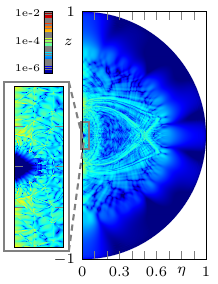} 
  \includegraphics[scale=\myscale]{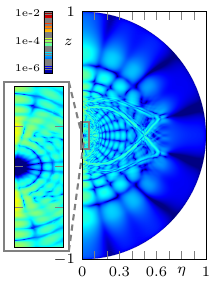}
  \includegraphics[scale=\myscale]{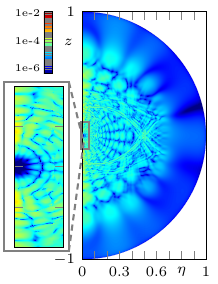} 
  \label{fig:no-rot_2dwave_0.2mHz_a}}
  \vspace*{0.00em}

  \includegraphics[scale=\myscale]{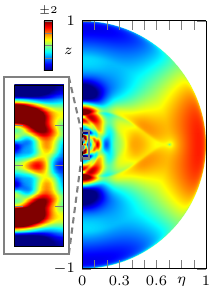}
  \hspace*{1em}
  \includegraphics[scale=\myscale]{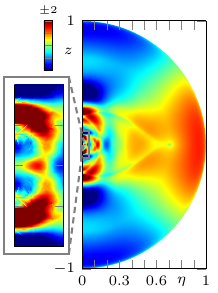} 
  \includegraphics[scale=\myscale]{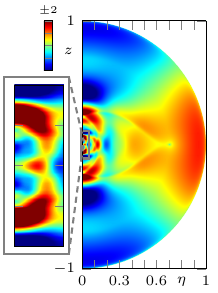}
  \includegraphics[scale=\myscale]{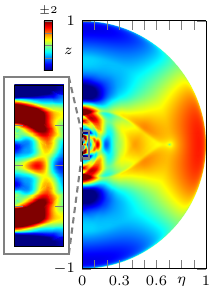}
  \vspace*{-1.50em}

  \subfloat[With source positioned in $(0.7,0)$, from left to right the relative difference 
           is for $\taua$ ($\erra=2.5\cdot\num{e-2}$), 
           $\taub$ ($\erra=2.7\cdot\num{e-2}$) 
           and $\tauc$ ($\erra=4.5\cdot\num{e-2}$).]{  \hspace*{9.50em}
  \includegraphics[scale=\myscale]{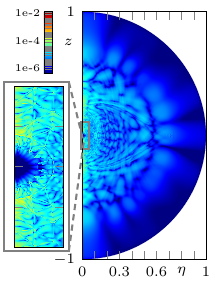} 
  \includegraphics[scale=\myscale]{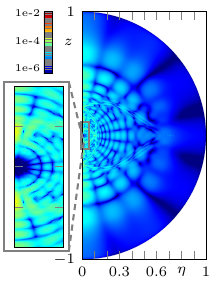}
  \includegraphics[scale=\myscale]{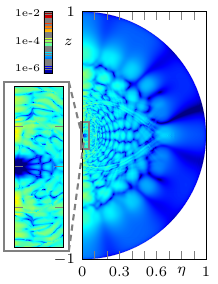} 
  \label{fig:no-rot_2dwave_0.2mHz_b}}
  
  \caption{Simulations at \num{0.2}\si{\milli\Hz}, the reference solution
           is on the left computed with \meshBa, then we show the solutions
           using \meshCa~for different stabilization 
           $\taua$, $\taub$ and $\tauc$ from left to right.
           In each subfigure, the first line shows the wavefield
           and the second one the relative difference
           with the reference solution \cref{eq:relative-difference}.
           The solutions on the entire domain are 
           complemented with a zoom near the origin
           on interval 
           $(\num{0}, 5\cdot\num{e-2}) \times (-\num{0.1},\num{0.1})$.
           The attenuation is set to 2\si{\micro\Hz}.
           }
  \label{fig:no-rot_2dwave_0.2mHz}
\end{figure}

\bigskip

In \cref{fig:no-rot_2dwave_0.2mHz}, we show both the reference 
solutions that use \meshBa, and the simulations using \meshCa~
for the three choices of the HDG stabilization. 
We also provide the picture of the relative difference $\err$ 
\cref{eq:relative-difference}.
On the wavefields, we see that the stabilizations $\taua$ 
and $\tauc$ give solutions that have artifacts near the 
origin, for the two sources.
When the source belongs in $\Bhyp$, \cref{fig:no-rot_2dwave_0.2mHz_a},
those artifacts are also visible near the source position
for stabilizations $\taua$ and $\tauc$. On the other hand, the 
stabilization $\taub$ gives solutions without artifact.
Comparing the relative difference, the numbers are relatively close
between each choice.
We also observe that the relative difference is higher near the 
$z$-axis, in particular near the origin. It is higher when the 
source belongs to $\Bhyp$ compared to when the source in is $\Bell$,
\cref{fig:no-rot_2dwave_0.2mHz_a} and \cref{fig:no-rot_2dwave_0.2mHz_b}
respectively.

\subsection{Ignoring gravity effects}

To ignore the effects of gravity in the problem, one can 
take the gradient of the gravity potential to zero ($\nabla\Phi_0 = 0$) 
in the equations. 
This assumption is reasonable  when one
focuses on the solar acoustic p-modes, cf.~\cite{Pham2020Siam}.
In this case, the buoyancy frequency $N^2$ defined in \cref{Nfreq::def}
is zero, hence $\Bhyp$ is empty.
Furthermore, the vacuum boundary condition \cref{vLpBC_gen} simplifies
to a Dirichlet boundary condition $\pressE=0$.
In \cref{fig:waves:no-gravity}, we picture the solutions with 
and without gravity effects, for frequencies \num{3} and \num{6}\si{\milli\Hz}. 
The source is positioned in $(\num{0.99986},0)$ and 
the attenuation is set to 10\si{\micro\Hz}. 
We also compare the solutions at the height $r=\num{1.00021}$.

\pgfmathsetmacro{\myscale}{0.78}
\begin{figure}[ht!] \centering
  \subfloat[3\si{\milli\Hz} solutions
            without (\textbf{left}) 
            and with (\textbf{right}) gravity.]{
  \includegraphics[scale=\myscale]{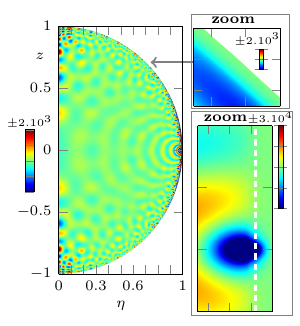} \hspace*{-0.75em}
  \includegraphics[scale=\myscale]{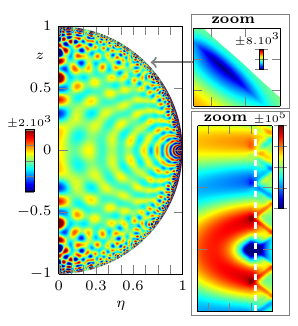}
  } \hspace*{0.5em}
  \subfloat[6\si{\milli\Hz} solutions
            without (\textbf{left}) 
            and with (\textbf{right}) gravity.]{\includegraphics[scale=\myscale]{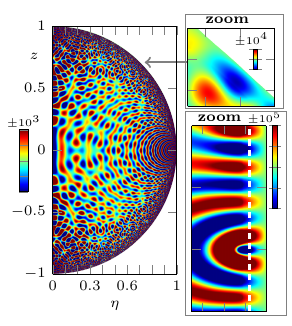} \hspace*{-0.75em}
  \includegraphics[scale=\myscale]{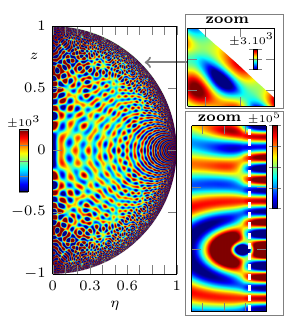}
  }

  \subfloat[Comparisons of the solutions at height 
            $r=\num{1.00021}$
            for the cases without and with gravity at 
            3\si{\milli\Hz} (\textbf{left}) and ;
            6\si{\milli\Hz} (\textbf{right}). 
            For 3\si{\milli\Hz}, we show from 0 to 15\si{\degree}, but 
            separate the case without gravity (\textbf{top left}) which
            has very small amplitude compared to the case with (\textbf{bottom left}).
            For 6\si{\milli\Hz}, we shown separately for angles 
            from 0 to 3.5\si{\degree} (\textbf{top right})
            and 5 to 9\si{\degree} (\textbf{bottom right}).
            ]{
  \includegraphics[scale=1]{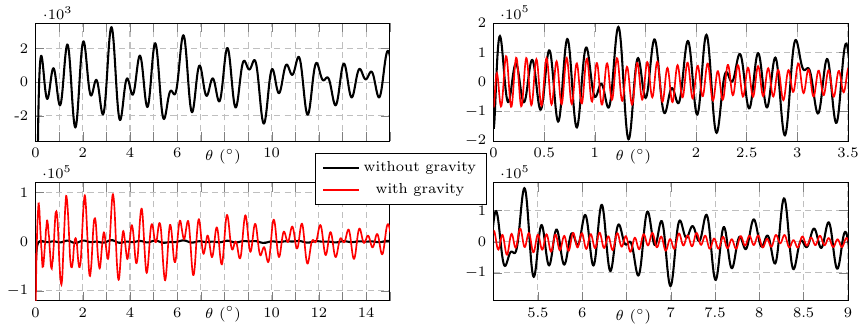}   
  \label{fig:waves:no-gravity:rcv}
  }

  \caption{Comparisons of the solutions with and without
           gravity effects for a source positioned in 
           $(\num{0.99986},0)$. 
           In the case without gravity, $N^2=0$ and $\Bhyp$ 
           is always empty.
           The solutions on the entire domain are 
           complemented with a zoom near source 
           position on interval 
           $(\num{0.9975},\num{1.001}) \times (-\num{0.005},\num{0.01})$, 
           and a zoom in the area $(0.705,0.71)^2$.
           The attenuation is set to 10\si{\micro\Hz}.
           The white dashed line on the zoom corresponds 
           to the height $r=1.00021$.}
  \label{fig:waves:no-gravity}
\end{figure}

\bigskip

We see that the solutions are very different between the cases
with and without gravity. For the frequency below cut-off, 3\si{\milli\Hz}, 
we do not have the multiple reflections back and forth as $\Bhyp$ is empty
in the case without gravity. For the frequency 6\si{\milli\Hz}, 
we also see a different pattern in the region 
close to the surface: ignoring the gravity, the waves decrease in 
amplitude rapidly while when considering the gravity, we see waves 
propagating in this region and particularly small wavelength.
We also notice that the amplitudes of the wavefields are different. 
Consequently, the signals are hardly comparable at 
the height \num{1.00021} (\cref{fig:waves:no-gravity:rcv})
with both amplitude and phase being different.
This experiment illustrates that incorporating the gravity induces drastic 
changes in the pattern of the solutions, in particular near the surface 
which is where the helioseismic observables are generated.


\subsection{Summary}

In this section, we have shown the behaviour of the wavefield 
with solar background for three representative frequencies. 
The difficulty of working with solar background is the non-zero
buoyancy frequency, which possibly leads to the existence of
the region $\Bhyp$. 
\begin{itemize}
  \item In the case where the frequency is above cut-off frequency, 
        $\Bhyp$ is empty and the solutions are stable with respect
        to the choice of the HDG stabilization, as depicted in \cref{fig:no-rot_2dwave_6mHz_m150k}.
  \item For frequencies below cut-off, $\Bhyp$ is not empty. 
        The behaviour of the solution is affected
        by the position of the source:
        if the source is in $\Bell$, the solutions are evanescent
        in $\Bhyp$, and the HDG stabilizations behave well 
        (left of \cref{fig:no-rot_2dwave_3mHz_m150k}).
        If the source is in $\Bhyp$ (or even close to), 
        propagation of singularity arises, with back and forth
        reflections in the region $\Bhyp$ in a zigzag pattern. 
        In this case the choice of stabilization plays a role 
        in avoiding artifacts near the source position 
        (right of \ref{fig:no-rot_2dwave_3mHz_m150k}).
  \item Ignoring the gravity effects, the buoyancy frequency 
        is zero in the entire domain, hence $\Bhyp$ is empty.
        This avoids the propagation of singularity for frequencies
        below cut-off. However, ignoring gravity drastically 
        changes the solutions, which are not comparable in terms
        of phase and amplitude, \cref{fig:waves:no-gravity}.
\end{itemize}
In the context of helioseismology, the sources and receivers
are taken near the surface and frequencies below cut-off are
studied. This corresponds precisely to the zone of $\Bhyp$. 
Note that here we have only shown the results for the azimuthal 
mode $m=0$, but the behaviour is the same for non-zero azimuthal 
modes, which are not included for the sake of conciseness.

%

\begin{remark}[Including rotation and differential flow]
  Due to the small values of the solar rotation, below \si{\micro\Hz},
  there are little visual differences in the solutions when incorporating
  the rotation for the frequencies used above. However, these differences 
  are important as they allow to infer the rotation,  
  cf.~\cref{section:numerical-helio,figure:model-rotation}.
  Differences may be expected when studying extremely 
  low frequencies with magnitude comparable to the solar 
  rotation, which is the case for inertial modes \cite{Gizon2021}.
  Nonetheless, this is out of the scope of the current paper 
  and it will be the focus of future works.
\end{remark} 

%
%
%
%
%

\section{Comparisons with helioseismic products}
\label{section:numerical-helio}

The objective of this section is to validate our 
axisymmetric solver by comparing with synthetic
helioseismic data. 
We first derive in \cref{syndata::subsec} 
expressions to compute synthetic 
helioseismic products, such as power spectra \cref{eq:power-spectrum}
and expected values of cross-covariance \cref{eq:mean-cc} 
from the azimuthal Green's kernels.
We illustrate with the synthetic power spectrum associated
with the standard solar model in \cref{heliossm::subsec}, 
and compare different models of solar rotation in \cref{heliorot::subsec}.
We will show that the synthetic power spectrum for standard \texttt{model-S}
contains peaks of energy that are in agreement with observations, and 
with the eigenfrequencies computed with software \texttt{Gyre} in \cref{heliossm::subsec}. 
The power spectrum computed from the solar background containing rotation
is shown to produce the observed effects on p-modes in \cref{heliorot::subsec},
namely a shift in the p-mode depending on the azimuthal order $m$.
Since the axisymmetric solver constructed in this work takes into account gravity 
effect and allows for uniform and differential rotation, they provide
more realistic simulations of the helioseismic products, compared to the 
1.5D solvers that ignore rotation in \cite{Pham2020Siam,Pham2024assembling} 
and the axisymmetric formulation that ignore gravity in \cite{Gizon2017}.

\subsection{Derivation of synthetic helioseismic data}
\label{syndata::subsec}

Helioseismic data correspond to line-of-sight Doppler 
velocities and are obtained from instruments on board of 
satellites (such as the Helioseismic and Magnetic Imager, HMI,  
of the Solar Dynamics Observatory mission), or from 
ground-based telescopes.
Here we show how to model helioseismic data with the azimuthal 
Green's kernel $G_p^m$ given \cref{amodalGK::def}, solved 
with the HDG discretization method in \cref{section:axi-symmetric-hdg}. 

\paragraph{Synthetic Doppler signal} 

In this work, in assuming single-height 
contribution\footnote{The observation height $\robs$ is commonly 
                      set $\geq 1$, we use $\robs=\num{1.00021}$ in our experiments.},
we model observed signal $\psi(\hat{\bx},\omega)$ 
at position $\hat{\bx}=(\theta,\phi)$ on the solar disk 
as the Eulerian pressure $\pressE(\robs,\hat{\bx},\omega)$;
secondly we assume
that each realization of the observation is 
caused by a scalar stochastic source $\frak{s}$.
The wave operator employed is from \cref{origdiv_form} with only scalar source,
\begin{equation}
\psi(\hat{\mathbf{x}}_{\rm obs};\omega) :=\pressE(\robs,\hat{\bx}_{\rm obs},\omega), \quad \text{where }
\Loriginaldiv \begin{pmatrix}\bxi \\[0.2em]\pressE \end{pmatrix} 
\,=\, \begin{pmatrix}0\\[0.2em] \frak{s} \end{pmatrix}.
\end{equation}
The synthetic signal, denoted by  $\psi(\hat{\mathbf{x}}_{\rm obs};\omega)$ is related to the scalar Green's 
function $G_p$ \cref{scalarppGKgen::def} as,
\begin{equation}\label{synobs::def}
 \psi(\hat{\mathbf{x}}_{\rm obs};\omega) :=\pressE(\robs,\hat{\bx}_{\rm obs},\omega)= \int_{\mathbb{B}_{\odot}} G_p(r_{\rm obs}, \hat{\mathbf{x}}_{\rm obs}, \mathbf{s};\omega) \, \frak{s}(\mathbf{s}) \mathrm{d}\mathbf{s}.
\end{equation}

The observables are then projected onto spherical harmonic 
functions $Y^m_\ell$, yielding coefficients denoted by
$ \psi^{\ell m}(\omega)$,
\begin{equation}
 \psi^{\ell m}(\omega) := \int_{\mathbb{S}^2} \psi(\hat{\mathbf{x}}_{\rm obs}; \omega) \,  \overline{Y_\ell^m(\hat{\mathbf{x}}_{\rm obs})}\, \mathrm{d} \hat{\mathbf{x}}_{\rm obs}= \int_{\mathbb{B}_{\odot}} \mathcal{K}_p^{lm}(r_{\rm obs},\mathbf{s};\omega) \, \frak{s}(\mathbf{s}) \mathrm{d}\mathbf{s}.
 \end{equation}
In the last expression, we have also introduced the kernel $\mathcal{K}_p^{\ell m}$ defined as,
\begin{equation}\label{calKlm::def}
\mathcal{K}_p^{\ell m}(r_{\rm obs},\mathbf{s};\omega) := \int_{\mathbb{S}^2} G_p( r_{\rm obs},\hat{\mathbf{x}}_{\rm obs},\mathbf{s};\omega) \, \overline{Y_\ell^m(\hat{\mathbf{x}}_{\rm obs})}\, \mathrm{d}\hat{\mathbf{x}}_{\rm obs}.
\end{equation}
Its expression is obtained by substituting the last expression 
of \cref{synobs::def} and interchanging the integration order between 
$\int_{\mathbb{B}_{\odot}}d\mathbf{r}$ and $\int_{\mathbb{S}} \mathrm{d}\hat{\mathbf{x}}_{\rm obs}$.

\paragraph{Azimuthal Green's kernel}
Recall \cref{cylexpanGK::identity} which states the cylindrical expansion of $G_p(\mathbf{r},\mathbf{s};\omega)$ 
with coefficients given by azimuthal Green's kernel $ G^m_p$, 
\begin{equation}
G_p(\mathbf{x},\mathbf{s};\omega) = \dfrac{1}{2\pi} \sum_{m\in \mathbb{Z}} G^m_p(\tilde{\mathbf{x}},\tilde{\mathbf{s}}';\omega)\, e^{\ii m (\phi-\phi')}, \quad \mathbf{s} \notin \text{ z-axis}.
\end{equation}
When the position of the Dirac source is on the axis, there is only the term associated with $m=0$ in the above summand,
\begin{equation} G_p(\mathbf{x},\mathbf{s};\omega) = \dfrac{1}{2\pi} G^0_p(\tilde{\mathbf{x}},0,z';\omega),  \hspace*{1cm} \mathbf{s} \in \text{ z-axis}.
 \end{equation}
In numerical resolution, our first choice is the  Liouville-c
variant $\Lliouvillec$, which yields $G_{c}^m$ related to $G^m_p$ by
\begin{equation}G_p^m (\eta,z,\eta',z')
 \,=\, G_{c}^m(\eta,z,\eta',z') \,\,   c_0(\eta,z)
       \sqrt{\rho_0(\eta,z)}
       \,\, c_0(\eta_{\mathrm{src}},z_{\mathrm{src}})
       \sqrt{\rho_0(\eta_{\mathrm{src}},z_{\mathrm{src}})}.
\end{equation}
The value of $G^m_{c}$ is obtained, in each resolution, for a frequency $\omega$, a point source position
$(\eta_{\mathrm{src}},z_{\mathrm{src}} )$, and mode $m$, as,
\begin{equation}\label{eq:local-lc}
\begin{aligned}
&G^m_{c} ( \eta,z; \eta_{\mathrm{src}}, z_{\mathrm{src}};\omega) := w_m(\eta,z), \\
&\hspace*{1cm} \text{ where } \quad\Lliouvillec^m \begin{pmatrix}
\bu^m(\eta,z))\\ 
\ w_m(\eta,z)\end{pmatrix} = \dfrac{1}{\eta}  \begin{pmatrix} \boldsymbol{0} \\ \delta(\eta-\eta_{\mathrm{src}}, z-z_{\mathrm{src}})\end{pmatrix} .
\end{aligned} \end{equation}

We next relate $\mathcal{K}_p^{lm}$ to the numerically computed 2.5D kernel $ G^m_{c}$.
We denote the representation of the source $\mathbf{s}$ in cylindrical and spherical coordinates by 
\begin{equation}
\mathbf{s} \leftrightarrow  (\eta',z',\phi') \leftrightarrow (s,\theta',\phi') \quad  \text{ where } \eta'=s \cos(\theta'), \quad z' = s\sin(\theta') \,.
\end{equation}
From the definitions of $\mathcal{K}_p^{lm}$ \cref{calKlm::def} and spherical harmonics 
$Y^m_\ell(\theta,\phi) = c(\ell,m) P^{\lvert m\rvert}_\ell(\cos \theta) e^{\ii m \phi}$,
we have,
\begin{align*}
& \mathcal{K}_p^{lm}(r_{\rm obs},\mathbf{s};\omega) =\mathcal{K}_p^{lm}(r_{\rm obs},;\omega)  =\mathcal{K}_p^{lm}(r_{\rm obs},s,\theta',\phi';\omega)\\
& = \int_0^{2\pi}\int_0^{\pi} G_p(r_{\rm obs},\theta_{\rm obs},\phi_{\rm obs}, \mathbf{s};\omega) \, \overline{Y_l^m(\theta_{\rm obs},\phi_{\rm obs)}}\, \sin\theta_{\rm obs}\mathrm{d}\theta_{\rm obs}  \mathrm{d} \phi_{\rm obs} \\
 &= \dfrac{1}{2\pi}\int_0^{2\pi}\hspace*{-0.2cm}\int_0^{\pi}\hspace*{-0.1cm} \left( \sum_{m'}  G^{m'}_p(r_{\rm obs}, \theta_{\rm obs} ,\eta',z'; \omega)   e^{\ii m' (\phi_{\rm obs} - \phi')}  \right) c(\ell,m) \overline{P^{\lvert m\rvert}_\ell(\cos \theta_{\rm obs})}  e^{-\ii m \phi_{\rm obs}} \, \mathrm{d} \phi_{\rm obs} \sin\theta_{\rm obs}\mathrm{d}\theta_{\rm obs}   \\
& =  c(\ell,m) e^{-\ii m \phi'} \int_0^{\pi} G^{-m}_p(r_{\rm obs},\theta_{\rm obs},\eta',z' ; \omega) \,   \overline{P^{\lvert m\rvert}_\ell(\cos \theta_{\rm obs})}  \sin\theta_{\rm obs}\, \mathrm{d}\theta_{\rm obs} \,.
\end{align*}
We thus obtain the following result.

\begin{identity}\label{obsGK::identity}
The Green's kernel corresponding to synthetic observables \cref{calKlm::def} 
is related to the azimuthal Green's kernel as
\begin{equation}  \label{eq:definition-Kplm}
 \mathcal{K}_p^{lm}( r_{\rm obs},s,  \theta', \phi'; \omega)  
 =  e^{-\ii m \phi}\,  \tilde{\mathcal{K}}_p^{lm}( r_{\rm obs},s,\theta'; \omega)  \,. 
\end{equation} where the quantity $\tilde{\mathcal{K}}_p^{lm}$   defined as
\begin{equation}\label{calKlm:25d}
\tilde{\mathcal{K}}_p^{lm}(   r_{\rm obs},s,\theta'; \omega) 
  = c(\ell,m) \int_0^{\pi} G^{-m}_p(r_{\rm obs},\theta_{\rm obs},\eta',z' ; \omega) \,\,    \, \overline{P^{\lvert m\rvert}_\ell(\cos \theta_{\rm obs})}  \sin\theta_{\rm obs}\, \mathrm{d}\theta_{\rm obs}
\end{equation}
is independent of $\phi'$.
We also have the independence of 
$\lvert\mathcal{K}_p^{lm}\rvert^2$ 
with respect to $\phi'$: 
\begin{equation}\label{calKlmsq:25d}
 \lvert\mathcal{K}_p^{lm}( r_{\rm obs},s,  \theta', \phi'; \omega) \rvert^2
 =  \lvert\tilde{\mathcal{K}}_p^{lm}( r_{\rm obs},s,\theta'; \omega) \rvert^2 .
\end{equation}
\end{identity}

\subsubsection{Synthetic power spectrum}
The power spectrum is defined with the expected 
value of the components $\psi^{lm}$ such that:
\begin{equation}
 \mathcal{P}_l^m(\omega) := \mathbb{E}[|\psi^{lm}(\omega)|^2].
\end{equation}
We further assume that the random (realization) sources 
are spatially uncorrelated and from a single depth $r_{\rm src}$ 
such that
\begin{equation}
 \mathbb{E}[s(\mathbf{r},\omega) \overline{s(\mathbf{r}',\omega)}] = S(\omega) \delta(\mathbf{r} - \mathbf{r}') \delta(r-\rsrc),
\end{equation}
where $S$ represents the power distribution of the sources with frequencies.
In this case, in terms of the quantities $\tilde{\mathcal{K}}_p^{lm}$ from \cref{calKlm:25d},  the power spectrum becomes
\begin{equation}\begin{aligned}\label{eq:power-spectrum_v0}
 \mathcal{P}_l^m(\omega) &= \int_{\mathbb{B}_{\odot}} \int_{\mathbb{B}_{\odot}} \mathcal{K}_p^{lm}(\robs,\mathbf{r},\omega) \overline{\mathcal{K}_p^{lm}(\robs,\mathbf{r}',\omega)} \mathbb{E}[s(\mathbf{r}) \overline{s(\mathbf{r}')}] \mathrm{d}\mathbf{r} \mathrm{d}\mathbf{r}' \\
 &= \int_{\mathbb{B}_{\odot}} |\mathcal{K}_p^{lm}(\robs,\mathbf{r},\omega)|^2  \mathbb{E}[s(\mathbf{r}) \overline{s(\mathbf{r})}] \mathrm{d}\mathbf{r} \\
 &= S(\omega) \int_{\mathbb{S}^2} |\mathcal{K}_p^{lm}(\robs, \rsrc, \hat{\mathbf{r}}, \omega)|^2 \mathrm{d}\hat{\mathbf{r}}\\
 &= S(\omega) \int_{0}^{2\pi} 
              \int_{0}^{\pi}  |\mathcal{K}_p^{lm}(\robs, \rsrc, \theta, \phi, \omega)|^2 \sin \theta \, \dd\theta \dd\phi.
              \end{aligned}\end{equation}
By applying  \cref{calKlmsq:25d}, we obtain
\begin{equation}\label{eq:power-spectrum}
\mathcal{P}_l^m(\omega) = 2\pi S(\omega) \int_0^\pi |\tilde{\mathcal{K}}_p^{lm}(\robs, \rsrc, \theta, \omega)|^2 \sin\theta\mathrm{d}\theta.
\end{equation}
Therefore, for every azimuthal mode $m$, the power spectrum
can be computed from the simulations for sources at fixed $\rsrc$ and at many $\theta$ locations to perform the integration. We will see in \cref{heliorot::subsec} that the dependency of the power spectrum on the azimuthal order $m$ is directly related to the rotation.


\subsubsection{Synthetic cross-covariance} 
In addition to the power spectrum \cref{eq:power-spectrum},
another useful helioseismic product is the cross-covariance 
between the wavefields at two positions $\hat{\bx}_1$ and 
$\hat{\bx}_2$ on the visible disk.
It is typically computed in the frequency domain as
\begin{equation}
  C(\hat{\bx}_1,\hat{\bx}_2,\omega) = \psi(\hat{\bx}_1,\omega) \overline{\psi(\hat{\bx}_2,\omega)}.
  \end{equation}
The expected value $\mathcal{C}$ of the cross-covariance is given by
\begin{align}
 \mathcal{C}(\hat{\bx}_1,\hat{\bx}_2,\omega) &= \mathbb{E}[\psi(\hat{\bx}_1,\omega) \overline{\psi(\hat{\bx}_2,\omega)}] \\
 &= \int_{\mathbb{B}_{\odot}} \int_{\mathbb{B}_{\odot}} G_p(\robs,\hat{\bx}_1,\mathbf{r},\omega) \overline{G_p(\robs,\hat{\bx}_2,\mathbf{r}',\omega)} \mathbb{E}[s(\mathbf{r}) \overline{s(\mathbf{r}')}] \mathrm{d}\mathbf{r} \mathrm{d}\mathbf{r}' \\
 &= \int_V G_p(\robs,\hat{\bx}_1,\mathbf{r},\omega) \overline{G_p(\robs,\hat{\bx}_2,\mathbf{r},\omega)} \mathbb{E}[s(\mathbf{r}) \overline{s(\mathbf{r})}] \mathrm{d}\mathbf{r}. \label{eq:covariance}
\end{align}
Writing the decomposition of $G_p$ from \cref{cylexpanGK::identity},
\begin{equation}
 G_p(\robs,\hat{\bx}_1,\mathbf{r},\omega) = \dfrac{1}{2\pi}\sum_{m} G_p^{m}(\robs,\theta_1,r,\theta,\omega) \textrm{e}^{\ii m (\phi_1 - \phi)},
\end{equation}
we obtain
\begin{equation} \label{eq:mean-cc}
 \mathcal{C}(\hat{\bx}_1,\hat{\bx}_2,\omega) = \dfrac{S(\omega)}{2\pi} \sum_{m} \textrm{e}^{\ii m (\phi_1 - \phi_2)} \int_0^\pi G_p^{m}(\robs,\theta_1,\rsrc,\theta,\omega) \overline{G_p^{m}(\robs,\theta_2,\rsrc,\theta,\omega)} \sin\theta \mathrm{d}\theta.
\end{equation}

\subsubsection{Simplification for the case without rotation}

Without rotation, it is convenient to work in the spherical harmonic basis. In terms of the quantity introduced in \cref{calKlm::def},  the Green's function for a source at depth $r_{\rm src}$ and horizontal position $\hat{\bx}_s = (\theta_s, \phi_s)$ (or equivalently $(\eta_s,z_s)$) is written as, 
\begin{equation}
 G_p(r_{\rm obs},\hat{\bx}_1, r_{\rm src}, \hat{\bx}_s,\omega) = \sum_{\ell=0}^{\infty} \sum_{m=-\ell}^{\ell}\mathcal{K}_p^{\ell m}(r_{\rm obs}, r_{\rm src}, \hat{\bx}_s,\omega) Y_\ell^m(\theta_1, \phi_1). \label{eq:Glm0}
\end{equation}
If there is no background flow, then the Green's function for a source located at $\hat{\bx}_0 = (0,z_{\rm src})$ on the $z-$axis 
depends only on the mode $m=0$ and can be written as
\begin{equation}
 G_p(r_{\rm obs},\hat{\bx}_1, r_{\rm src}, \hat{\bx}_0,\omega) = \frac{1}{\sqrt{2\pi}}\sum_{\ell=0}^{\infty} \mathcal{K}_p^{\ell0}(r_{\rm obs}, r_{\rm src}, \hat{\bx}_0,\omega) P_\ell(\cos\theta_1),
\end{equation}
where $P_\ell$ represents the Legendre polynomial of degree $\ell$.

Using the rotation formula of spherical harmonics, $\mathcal{K}_p^{\ell m}(r_{\rm obs}, r_{\rm src}, \hat{\bx}_s,\omega)$ can be written as a function of $\mathcal{K}_p^{\ell0}$ such that
\begin{equation}
\mathcal{K}_p^{\ell m}(r_{\rm obs}, r_{\rm src}, \hat{\bx}_s,\omega) = \sqrt{\frac{4\pi}{2\ell +1}} \mathcal{K}_p^{\ell0}(r_{\rm obs}, r_{\rm src},\hat{\bx}_0,\omega) Y_\ell^m(\theta_s, \phi_s). \label{eq:Glmrot}
\end{equation}
Inserting \cref{eq:Glmrot} into \cref{eq:Glm0} leads to
\begin{equation}
  G_p(r_{\rm obs},\hat{\bx}_1, r_{\rm src}, \hat{\bx}_s,\omega)  = \sum_{l=0}^{\infty}\mathcal{K}_p^{l0}(r_{\rm obs}, r_{\rm src}, \hat{\bx}_0,\omega) \sqrt{\frac{4\pi}{2\ell +1}} \sum_{m=-\ell}^\ell Y_\ell^m(\theta_s, \phi_s) Y_\ell^m(\theta_1, \phi_1). \label{eq:Glm1}
\end{equation}
Using this expression in the expectation value 
of the cross-covariance \cref{eq:covariance} 
leads to
\begin{align}
  \overline{C}(\hat{\bx}_1,\hat{\bx}_2,\omega) &= S(\omega) \int_{\mathbb{S}^2} G_p(r_{\rm obs}, \hat{\bx}_1, r_{\rm src}, \hat{\bx}_s,\omega) \overline{ G_p(r_{\rm obs}, \hat{\bx}_2, r_{\rm src}, \hat{\bx}_s,\omega)} \sin\theta_s d\theta_s d\phi_s, \\
  &= S(\omega) \sum_\ell \frac{4\pi}{2\ell +1} |\mathcal{K}_p^{\ell 0}(r_{\rm obs}, r_{\rm src}, \hat{\bx}_0,\omega)|^2 \sum_{m=-\ell}^\ell Y_\ell^m(\theta_1,\phi_1) \overline{Y_\ell^m(\theta_2,\phi_2)}.
\end{align}
Employ the addition theorem of spherical harmonics to the last expression, 
we obtain the following identity.

\begin{identity}
For background without flow,  the (synthetic) cross-covariance depends only on the angular distance between 
the two observations points $\hat{\bx}_1$ and $\hat{\bx}_2$ and is given by
\begin{equation}\label{ccfnoflow}
 \overline{C}(\hat{\bx}_1,\hat{\bx}_2,\omega)  = S(\omega) \sum_{\ell=0}^{\infty} |\mathcal{K}_p^{\ell 0}(r_{\rm obs}, r_{\rm src}, \hat{\bx}_0,\omega)|^2 P_\ell(\hat{\bx}_1 \cdot \hat{\bx}_2).
\end{equation}
In this case, the (synthetic) power spectrum \cref{eq:power-spectrum} is also independent 
of $m$ and is given by
\begin{equation}\label{psnoflow}
\mathcal{P}_\ell^m = \mathcal{P}_\ell^0=  2\pi S(\omega) |\mathcal{K}_p^{\ell 0}(r_{\rm obs}, r_{\rm src}, \hat{\bx}_0,\omega)|^2.
\end{equation}

\end{identity}
\begin{remark}
These expressions \cref{ccfnoflow} and \cref{psnoflow} are computationally cheaper than the general expression 
of the power spectrum given by \cref{eq:power-spectrum} and \cref{eq:mean-cc} as they require
only one simulation for $m=0$ and for a single source (no integration over $\theta$ is required).
The computational steps for the case with and without rotation are further described in \cref{algorith:numerical-steps}.
\end{remark}

\begin{algorithm}[ht!]\centering
\caption{Computational steps for the simulation of helioseismic observables 
         using the axisymmetric formulation for the cases with and without 
         rotation. Considering rotation leads to increased computational cost 
         with the steps indicated in red.}

%
%
%
%

\begin{tikzpicture}[every node/.append style ={font=\small}]

\node[text width=23em,draw] (in) at (0,0) {\textbf{Inputs}: 
                             background models; frequency range for $\omega$; \\[0.10em]
                             \hspace*{3.90em} source height $r_{\mathrm{src}}$; 
                                              $\quad$ receiver height $r_{\mathrm{rcv}}$.
                                              }; 

\node[below = of in, yshift=2.50em,xshift=-11em,blue!70!white]{\textbf{without rotation}};
\node[below = of in, yshift=2.50em,xshift= 11em,blue!70!white]{\textbf{with rotation}};

\node[text width=17em, below right= of in,draw,xshift=-12em,yshift=1em] (r1a) {\textbf{Additional inputs}: \\[0.10em]
                                          $-$  rotation profile; \\[0.0em]
                                          $-$ number of sources $n_{\mathrm{src}}$; \\[0.0em]
                                          $-$ list of azimuthal mode $m$.
                                          };
\node[text width=17em, below left= of in,draw,xshift=12em,yshift=-4.3em] (r0a) 
     {\textbf{Solve} axisymmetric equation \cref{eq:local-lc} for\\[0.20em]
      $-$ one source on $z$-axis at $(0,r_{\mathrm{src}})$, \\[0.10em]
      $-$ one azimuthal mode $m=0$. \\[0.10em]
      };

\node[text width=17em, below = 0.2 of r1a,draw] (r1b) 
     {\textbf{Solve} axisymmetric equation \cref{eq:local-lc} for\\[0.20em]
      $-$ {\color{red!80!black}{$n_{\mathrm{src}}$ sources at height $r_{\mathrm{src}}$}}, \\[0.10em]
      $-$ {\color{red!80!black}{all azimuthal modes $m$ selected}}.
      };

\node[text width=0.90\linewidth, below = 4.5 of in,draw] (rrcv) 
     {Extract solutions at the receivers height 
      for all frequencies $\omega$, source $s$, and mode $m$.
      };

\node[text width=0.90\linewidth, below = 0.30 of rrcv,draw] (rintegrate) 
     {For each selected harmonic degrees $\ell$: \\[0.1em]
      \textbf{Integrate} over the receiver height with associated Legendre polynomials to 
      get $\mathcal{K}_p^{\ell m}$ \cref{eq:definition-Kplm,calKlm:25d,calKlmsq:25d}
      };

\node[below = of rintegrate, yshift=2.50em,xshift=-10em,blue!70!white]{\textbf{without rotation}};
\node[below = of rintegrate, yshift=2.50em,xshift= 10em,blue!70!white]{\textbf{with rotation}};

\node[text width=18.5em, below = 0.50 of rintegrate,draw,xshift=-10.70em] (r0end) 
     {Compute power spectrum \cref{psnoflow} and cross-covariance \cref{ccfnoflow}.};

\node[text width=18.5em, below = 0.50 of rintegrate,draw,xshift=10.70em] (r1end) 
     {{\color{red!80!black}{\textbf{Integrate} over source angle}}
      for power spectrum 
      \cref{eq:power-spectrum} and cross-covariance \cref{eq:mean-cc}.};

\node[text width=1em, below = 0.01 of r1end] (phantom) {} ;  

\draw[->,line width=1pt] ([xshift=-6em] in.south) -- ([yshift=-6.7em,xshift=-6em] in.south) ;
\draw[->,line width=1pt] ([xshift= 6em] in.south) -- ([yshift=-1.5em,xshift= 6em] in.south) ;
\draw[->,line width=1pt] (r1a.south) -- (r1b.north) ;
\draw[->,line width=1pt] (r1b.south) -- ([yshift=-0.90em] r1b.south) ;
\draw[->,line width=1pt] (r0a.south) -- ([yshift=-0.90em] r0a.south) ;
\draw[->,line width=1pt] (rrcv.south) -- (rintegrate.north) ;
\draw[->,line width=1pt] ([xshift=-5em] rintegrate.south) -- ([xshift=-5em,yshift=-1.2em] rintegrate.south) ;
\draw[->,line width=1pt] ([xshift= 5em] rintegrate.south) -- ([xshift= 5em,yshift=-1.2em] rintegrate.south) ;
      
\end{tikzpicture}%


\label{algorith:numerical-steps}
\end{algorithm}

\subsection{Validation of helioseismic products without rotation}
\label{heliossm::subsec}

\paragraph{Power spectrum}

\begin{figure}[ht!] \centering
  \subfloat[Spectrum computed with frequencies 
            from \num{0.5}\si{\micro\Hz} to 9\si{\milli\Hz}
            and harmonic degree $\ell$ from 0 to $200$.
            On the right, the ridges are labelled as
            f-mode and p-modes.
            ]{
  \includegraphics[scale=1]{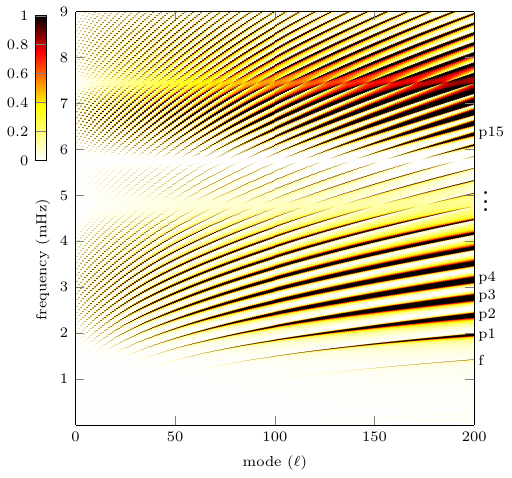}} \hspace*{0.50em}
  \subfloat[Zoom between frequencies 
            from 2.9 and 3.1\si{\milli\Hz}
            and harmonic degree $\ell$ from 
            0 to $100$ to compare with the eigenvalues
            obtained from \texttt{Gyre} (cyan).
            ]{
            \begin{tikzpicture}
\node[] (p) at (0,0) {\includegraphics[scale=1.0]{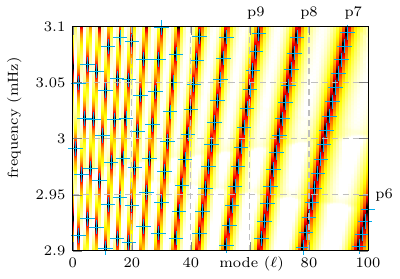}};
\node[below = -0.1 of p] {
\includegraphics[scale=0.9]{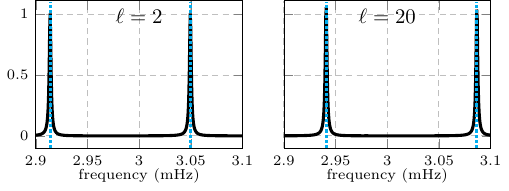}};
             \end{tikzpicture}}
  \caption{Power spectrum $\mathcal{P}_l^0(\omega)$ 
           \cref{psnoflow} without rotation (background color) 
           and comparison with eigenvalues (cyan crosses) computed 
           with the \texttt{Gyre} code.
           The bottom right panels show slices with frequencies at 
           fixed 
           $\ell$, with the \texttt{Gyre} eigenvalues indicated by the
           cyan vertical lines.}
  \label{fig:helioseismic-spectrum}
\end{figure}

We first consider the case without rotation and background flow.
A representation of the power spectrum $\mathcal{P}_l^0(\omega)$ 
of \cref{psnoflow} is given in \cref{fig:helioseismic-spectrum}. 
Without rotation, it requires the simulation for one source on 
the $z$-axis\footnote{To avoid having the source where a boundary 
                      condition is imposed on $\partial D^a$ and 
                      avoid numerical 
                      instabilities, the source is implemented 
                      slightly away from the $z$-axis in
                      the simulations, at $\epsilon=\num{e-12}$.
                      } 
and only for $m=0$, see \cref{algorith:numerical-steps}.
Here we use the source height $\rsrc = \num{0.99986}$, and 
observation height is set to $\robs = \num{1.00021}$. 
In this experiment, the axisymmetric system is solved for 
every frequency from 2\si{\micro\Hz} to 9\si{\milli\Hz} with 
step 2\si{\micro\Hz}. To compute the integral in \cref{obsGK::identity},
\num{18001} data-points are saved at the receiver height $\num{1.00021}$,
for equipartitioned angles between 0 and $\pi$.
The regions of strong power correspond to the eigenvalues of the forward operator,
\cite{Pham2020Siam,Pham2024assembling} and are in good agreement with the 
eigenvalues computed from the eigenvalue solver \texttt{Gyre} \cite{Townsend2013}. 
Nonetheless, the Green's function gives us a more complete
representation of the wavefield and contains for example information 
about the wave attenuation.

\paragraph{Cross-covariance}

The power spectrum is typically used in global helioseismology, 
for example to infer the solar differential rotation. 
In local helioseismology, the cross-covariance \cref{ccfnoflow} is 
the main observable. 
It is the basic ingredient to compute wave travel-times used 
in time-distance helioseismology or to compute holograms for 
helioseismic holography, \cite{Yang2023}. We compute it from the power spectrum 
using \cref{ccfnoflow} and performing a Fourier transform in time.

\begin{figure}[ht!]\centering
\subfloat[Time-distance diagram showing
          all angular positions for $\hat{\bx}_2$  
          between $0$ and $180$\si{\degree}.]
{\includegraphics[scale=1]{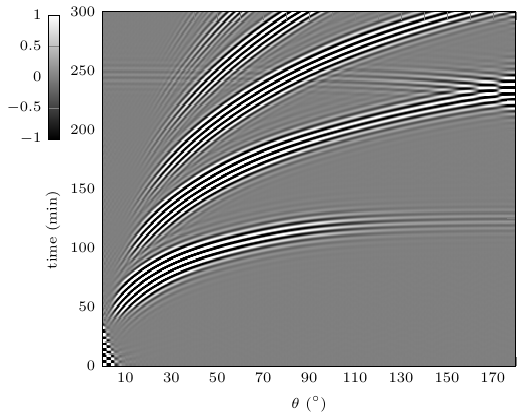}} \hspace*{1em}
\subfloat[Corresponding time-series at three different angular position.]
         {\includegraphics[scale=1]{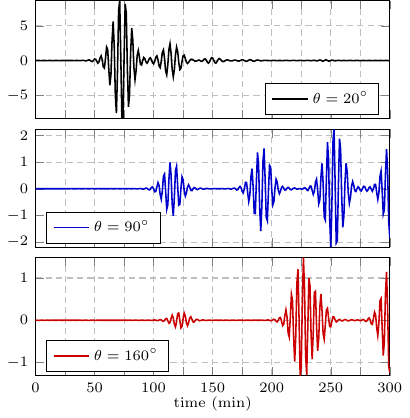}}
\caption{Time-distance diagram showing the cross-covariance \cref{ccfnoflow}
          for $\hat{\bx}_1$, $\hat{\bx}_2$ at height \num{1.00021}. 
          The source is positioned on the $z$-axis (i.e., $\hat{\bx}_1$). 
          }
\label{fig:time-distance-diagram}
\end{figure}

A classical representation used in helioseismology is the 
time-distance diagram that represents the cross-covariance 
$\overline{C}(\hat{\bx}_1,\hat{\bx}_2,t)$ as a function of 
separation distance between the points $\hat{\bx}_1$ and $\hat{\bx}_2$ and time. 
We use a maximum harmonic degree of 200 and, to avoid aliasing, 
we multiply the power at each $\ell$ by a filter $[1-\textrm{tanh}(0.03\ell-3)]/2$. 
The resulting time-distance diagram is shown in \cref{fig:time-distance-diagram}
where we can see the 
first skip which is the fastest wave that is travelling directly 
from $\hat{\bx}_1$ to $\hat{\bx}_2$. The second skip corresponds 
to waves that reflect once at the surface between $\hat{\bx}_1$ 
and $\hat{\bx}_2$ and so on. All these skips give information 
about different depths and are thus valuable inputs in the 
perspective of inversion.

\subsection{Validation of helioseismic products with rotation}
\label{heliorot::subsec}

In this section, we compute the
synthetic power spectrum obtained for various
rotation profiles listed in \cref{rotprofile::subsec}.
The most realistic one, called solar rotation, is 
inferred from global-mode helioseismology \cite{Larson2018}. 
With rotation incorporated, the power spectrum now depends on 
the azimuthal order $m$, \cref{eq:power-spectrum}, 
and so do the eigenvalues. 
Note that these eigenvalues correspond to the position of the 
peaks in the power spectrum, as depicted in \cref{fig:helioseismic-spectrum} 
without rotation.
In \cref{rotonPS::subsec}, we show how this effect is quantified,
in terms of the a-coefficients $a_j$, $j\in \mathbb{N}$,
which are coefficients of a specific polynomial expansions, 
cf. \cref{eq:a_coeff}. 
For the solar rotation,
we then compare the a-coefficients computed from our numerical power spectrum
 with those obtained from observed power spectrum and given in \cite{Schou1994}.
In this section, we choose a reference rotation frame 
with an angular velocity  $\rotzero/(2\pi)=\num{460}\si{\nano\Hz}$.
To compute the integrals, the simulations of these sections use
\num{361} source positions and the solution is saved at \num{1801} receivers.

\subsubsection{Rotation profiles for $\Omega$}\label{rotprofile::subsec}

The differential rotation $\Omega(r,\theta)$ that defines the
background flow in the rotating frame $\mathbf{v}_0$ 
(see \cref{purediffrot}) is often 
written on a basis function in the angular direction such that,
\begin{equation}
 \Omega(r,\theta) = \sum_{s=0}^{N_s} \Omega_s(r) \phi_s(\theta).
\end{equation}
As the variations are relatively smooth with latitude, good approximations can be obtained with a limited number of coefficients. Classical choices are $\phi_s(\theta) = \cos^s(\theta)$ or derivatives of Legendre polynomials. In this latter case, the first basis functions are
\begin{equation}
 \phi_0(x) = 1, \quad \phi_2(x) = -5x^2+1, \quad \phi_4(x) = 21x^4-14x^2+1.
\end{equation}
We will write $\Omega_i$ the coefficients in the basis of derivatives of Legendre polynomials and $\tilde{\Omega}_i$ the ones in the cosine expansion.

\begin{figure}[ht!] \centering
  \subfloat[Analytical rotation profile from \cite{Gizon2017}.]{\makebox[15em][c]{
  \includegraphics[scale=1]{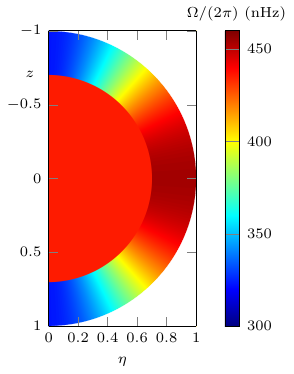}}
  \label{figure:model-rotation-v1}}
  \hspace*{2em}
  \subfloat[Rotation profile inferred from global-mode helioseismology \cite{Larson2018}.]{
  \makebox[15em][c]{
  \includegraphics[scale=1]{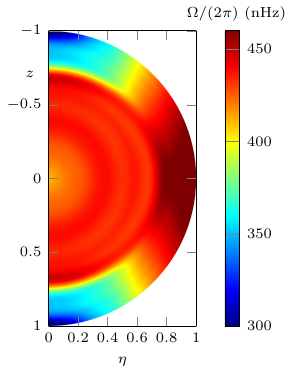}}
  \label{figure:model-rotation-v2}}
  
  \caption{Solar rotation $\Omega/(2\pi)$ in \si{\nano\Hz}.}
  \label{figure:model-rotation}
\end{figure}

\paragraph{Profiles under consideration}
For the numerical tests, we consider three different rotation profiles:
\begin{enumerate}[leftmargin = *]\setlength{\itemsep}{0pt}
 \item Non-rotating Sun $\Omega(r,\theta) = 0$
       which means that $\mathbf{v}_0=-r\sin\theta \rotzero \mathrm{e}_\phi$ (corresponding to the tracking rate).

 \item A \emph{simplified rotation} profile depicted in 
       \cref{figure:model-rotation-v1}, such that, 
       \cite{Gizon2017},
\begin{equation} \label{eq:rotation-v1}
 \Omega(r,\theta) = \tilde{\Omega}_0 + \tilde{\Omega}_2 \cos^2\theta + \tilde{\Omega}_4 \cos^4\theta, \quad \textrm{for } r > 0.7 \, \quad 
\text{and}\quad \tilde{\Omega}_c \, \text{ otherwise,}
\end{equation}
with
\begin{equation}
 \tilde{\Omega}_0 / 2\pi = 454 \textrm{nHz}, \quad \tilde{\Omega}_2 / 2\pi = -55 \textrm{nHz},  \quad \tilde{\Omega}_4 / 2\pi = -76 \textrm{nHz}, \quad \tilde{\Omega}_c / 2\pi = 435 \textrm{nHz}.
\end{equation}
It corresponds, in the basis of Legendre derivative functions, to
\begin{equation}
 \Omega_0 / 2\pi  = 436.5~\textrm{nHz}, \quad \Omega_2 / 2\pi = 21.1~\textrm{nHz}, \quad \Omega_4 / 2\pi = -3.6~\textrm{nHz}. \label{eq:omega-coeff_sun}
\end{equation}
\item The \emph{solar rotation profile} inferred from global-mode helioseismology
      \cite{Larson2018}, see \cref{figure:model-rotation-v2}. 
      It corresponds to an average of the inverted profiles for
      the first 6~years of HMI data.
\end{enumerate}
In \cref{figure:model-rotation}, we see that the two profiles of 
rotation are relatively close, although the one inferred from global
helioseismology \cite{Larson2018} has more variations with depth and contains a near-surface shear layer.


\begin{remark}
The derivatives of $\Omega$ with respect to the axisymmetric 
coordinates $\eta$ and $z$ are also needed in the numerical 
implementation, see \cref{appendix:rotation-axisym}. 
As $\Omega$ is usually known as a function of $r$ and $\theta$, 
we use,
\begin{equation}
 \frac{d \Omega}{d \eta} = \frac{d \Omega}{d \theta} \frac{d \theta}{d \eta} + \frac{d \Omega}{dr} \frac{dr}{d \eta} \qquad \textrm{and} \qquad \frac{d \Omega}{d z} = \frac{d \Omega}{d \theta} \frac{d \theta}{d z} + \frac{d \Omega}{dr} \frac{dr}{d z}.
\end{equation}
We have
\begin{equation}
 \eta = r \sin\theta \quad \textrm{and} \quad z = r\cos\theta,
\qquad \text{thus} \qquad
 r = \sqrt{\eta^2 + z^2} \quad \textrm{and} \quad  \theta = \textrm{arccos} \left( \dfrac{z}{\sqrt{\eta^2+z^2}} \right).
\end{equation}
We obtain the derivatives
\begin{align}
 \frac{d \theta}{dz} &= -\frac{1}{\sqrt{1 - \dfrac{z^2}{\eta^2+z^2}}} \frac{\sqrt{\eta^2+z^2} - \dfrac{2z^2}{2\sqrt{\eta^2+z^2}}}{\eta^2+z^2} = -\frac{\eta}{\eta^2 + z^2} = -\frac{\cos\theta}{r}, \\
 \frac{d \theta}{d\eta} &= \frac{1}{2} \frac{z}{\sqrt{1 - \dfrac{z^2}{\eta^2+z^2}}} (\eta^2+z^2)^{-3/2} 2\eta = \frac{z}{\eta^2 + z^2} = \frac{\sin\theta}{r}, \\
 \frac{dr}{dz} &= \frac{z}{\eta^2+z^2} = \frac{\sin\theta}{r}, \qquad \qquad \qquad
 \frac{dr}{d\eta} = \frac{\eta}{\eta^2+z^2} = \frac{\cos\theta}{r}.
\end{align}
In particular, the derivatives of the simplified rotation 
profile of \cref{figure:model-rotation-v1}, \cref{eq:rotation-v1}, 
are analytically given by
\begin{equation}
 \frac{d\Omega}{dz} =  \frac{2 \eta^2 z}{(\eta^2 + z^2)} \left[ \tilde{\Omega}_2 + 2 \tilde{\Omega}_4 \frac{z^2}{\eta^2+z^2} \right]; \qquad 
 \frac{d\Omega}{d\eta} = - \frac{2 \eta z^2}{(\eta^2 + z^2)} \left[ \tilde{\Omega}_2 + 2 \tilde{\Omega}_4 \frac{z^2}{\eta^2+z^2} \right].
\end{equation}
\end{remark}

\subsubsection{Effect of rotation on the power spectrum}
\label{rotonPS::subsec}

By incorporating rotation in the equations, 
the power spectrum depends on the azimuthal 
order $m$, \cref{eq:power-spectrum}, and so 
do the eigenvalues, that we label $\omega_{n\ell m}$ 
with their radial $n$, latitudinal $\ell$ and
longitudinal $m$ wavenumbers.
These eigenvalues
represent position of peaks in magnitude of the power spectrum. The extraction 
of their value from the power spectrum is discussed below.

\begin{definition} For a fixed $n$ and $\ell$, 
we define the frequency shift $\Delta\omega_{n\ell m}$ 
caused by rotation as the difference with
the eigenfrequency for $m=0$, such  that,
\begin{equation}\label{shift::def}
 \Delta\omega_{n\ell m} := \omega_{n\ell m} - \omega_{n \ell 0}, \quad -\ell \leq m\leq \ell.
\end{equation}
\end{definition}

\begin{figure}[ht!] \centering
  \subfloat[Using the approximate solar rotation profile of \cref{figure:model-rotation-v1}.
            \textbf{Left}: Cut for $m=11$ and associated Lorentzian fit to determine 
            the frequency of maximum power. 
            Middle: Central frequency for all $m$, and fit 
            using \cref{eq:a_coeff}. 
            \textbf{Right}: 2D power spectrum $\mathcal{P}_\ell^m(\omega)$ for $\ell=85$.]{
  \includegraphics[scale=1]{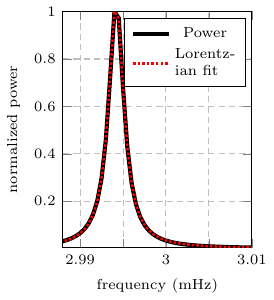}
  \includegraphics[scale=1]{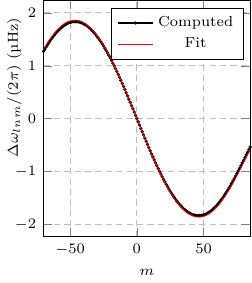}
  \includegraphics[scale=1]{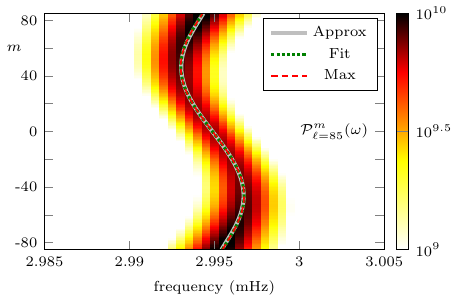}
  \label{fig:helioseismic-spectrum-rot-A}
  }

  \subfloat[Using $\rotzero/(2\pi)=460$\si{\nano\Hz} and $\Omega=0$.]{
  \includegraphics[scale=1]{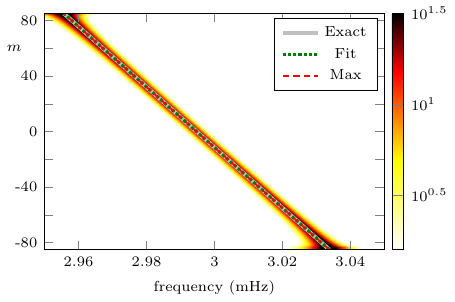}
  \label{fig:helioseismic-spectrum-rot-B}}
  \subfloat[Using the solar rotation profile of \cref{figure:model-rotation-v2}.]{
  \includegraphics[scale=1]{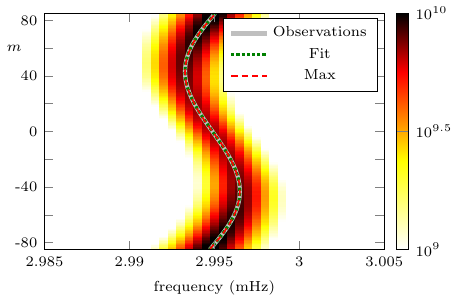}
  \label{fig:helioseismic-spectrum-rot-C}}

  \caption{Comparison of power spectra $\mathcal{P}_\ell^m(\omega)$ 
           for different rotation profiles. With rotation, the power
           spectrum changes with azimuthal order $m$, leading to a 
           shift in frequency in the position of maximal power, 
           shift that can be related to the rotation model.
           On the spectra, the red dashed lines correspond 
           to the maximum at each $m$  in the simulations, 
           and the green dotted lines correspond to the fit 
           according to \cref{eq:a_coeff}. 
           The gray lines are computed (in \textbf{a})
           from the coefficients of \cref{eq:a-coeff_sun}
           for the simplified rotation model;
           (in \textbf{b}) from the analytic expression for the case without rotation: 
           and (in \textbf{c}) from the observed coefficients \cref{eq:a-coeff_sun_measured} 
           for the rotation profile of \cite{Larson2018}.}
  \label{fig:helioseismic-spectrum-rot}
\end{figure}

\paragraph{Extraction of eigenfrequency from spectrum}
To compute the shift \cref{shift::def}, the eigenfrequency $\omega_{n\ell m}$ 
is extracted from the power spectrum $\mathcal{P}_l^m(\omega)$ 
by fitting a Lorentzian function, as shown in \cref{fig:helioseismic-spectrum-rot}. This approach is more accurate than using directly the maximum value of the power spectrum whose accuracy is limited by the frequency resolution

\paragraph{Shift quantification by a-coefficients}
To quantify the effect of rotation, the shift $\Delta\omega_{n\ell m}$ \cref{shift::def} is next represented as,
\begin{equation}
 \Delta\omega_{n\ell m} \,\, \textcolor{green!50!black}{\simeq}\,\,  \sum_{j=1}^{j_{\rm max}} a_j(n,\ell) \,\Pi_j^{(\ell)}(m) , \quad -\ell \leq m\leq \ell. \label{eq:a_coeff}
\end{equation}
At a fixed $\ell$, this fit is performed for all values of $m$ with $\lvert m\rvert \leq \ell$. Here $a_j(n,\ell)$ are the so-called a-coefficients and $\Pi_j^{(\ell)}$ is a polynomial of degree $j$. The maximum degree $j_{\rm max}$ depends on the complexity of the rotation profile and of the noise level in the case of observations. Standard helioseismic inversions use $j_{\rm max} = 6, 18,$ or 36 \cite{Larson2018}.
Here, we use $j_{\rm max} = 6$.

For the solar rotation \cref{figure:model-rotation-v2}, we will compare the numerical a-coefficients (obtained by fitting the synthetic power spectrum
\cref{eq:power-spectrum} with \cref{eq:a_coeff}) with the measured one given in \cite{Larson2018}:
\begin{equation}
 a_1 / 2\pi = 442.85\pm 0.05~\textrm{nHz}, \quad a_3 / 2\pi = 22.15 \pm 0.08~\textrm{nHz}, \quad a_5 / 2\pi = -3.21 \pm 0.10~\textrm{nHz}. \label{eq:a-coeff_sun_measured}
\end{equation}
For the case of simplified rotation \cref{eq:rotation-v1}, the computation 
of the a-coefficients is carried out with polynomials $\tilde{\Pi}_j^{(\ell)}$ which are equivalent to $\Pi_j^{(\ell)}$ at large $\ell$, cf. \cref{PIfnc::rmk}. In this case, the a-coefficients can be computed analytically as shown in \cref{acoeff_simprot::rmk}.

\begin{remark}[On the function $\Pi_j$ and their asymptotics]\label{PIfnc::rmk}
When measuring these coefficients 
on the Sun, the basis functions are chosen such that
\begin{equation}
  \Pi_i^{(\ell)}(\ell) = \ell, \quad \sum_{m=-\ell}^{\ell} \Pi_i^{(\ell)}(m)\,  \Pi_j^{(\ell)}(m) = 0 \quad \textrm{for } i \neq j.
\end{equation}
These polynomials are computed recursively starting from $\Pi_0^{(\ell)}(m) = \ell$ (see Appendix~A in \cite{Schou1994}). For large values of $\ell$, these polynomials are asymptotically equivalent to
\begin{equation}\label{tildePI::def}
 \tilde{\Pi}_j^{(\ell)}(m) = L P_j(m/L) \quad \textrm{with} \quad L = \sqrt{\ell(\ell+1)},
\end{equation}
where $P_j$ is the Legendre polynomial of order $j$.
\end{remark}

\begin{remark}[Computing a-coefficients for simplified rotation]\label{acoeff_simprot::rmk}
For simplified rotation \cref{eq:rotation-v1}, the fitting is carried out with the $\tilde{\Pi}_j^{(\ell)}(m)$
defined in \cref{tildePI::def}, i.e.
\begin{equation}
 \Delta\omega_{n\ell m} \,\, \textcolor{black}{\simeq}\,\,  \sum_{j=1}^{j_{\rm max}} a_j(n,\ell) \, \tilde{\Pi}_j^{(\ell)}(m) \,,
 \hspace*{0.5cm} 
 \text{ for simplified rotation \cref{eq:rotation-v1}.}
 \label{eq:a_coeff_simprot}
\end{equation}
In this basis, we can relate the a-coefficients 
to the rotation coefficients $\Omega_i$, \cite{Kosovichev1996},
\begin{equation}
a_k(n,\ell) = \Omega_{k-1} \left( 1 - \frac{1}{I_{n\ell}} \int_0^R \left[ 2 U_{n,\ell} V_{n,\ell} + V_{n,\ell}^2 \right] \rho_0 r^2 dr \right), \label{eq:ak_rot}
\end{equation}
where $U_{n\ell}$ and $V_{n\ell}$ are the radial and horizontal eigenfunctions of the mode at frequency $\omega_{n\ell}$ and $I_{n\ell}$ is the mode mass defined as 
\begin{equation}
I_{n\ell} =  \int_0^R \left[U_{n,\ell}^2 \, +\,  \ell (\ell+1) V_{n,\ell}^2 \right] \rho r^2 dr.
\end{equation}
Here, we have used that our simplified rotation profile does not depend on depth. The second term in \cref{eq:ak_rot} corresponds to the influence of the Coriolis force and is of order $1/L$ (\cite{Schou1994}), thus the $a_j$ coefficients give an approximation of $\Omega_{j-1}$. 

For the mode considered here, ($\ell=85$ and $n=7$), we compute the eigenfunctions $U_{n,\ell}$ and $V_{n,\ell}$ with \texttt{Gyre} \cite{Townsend2013} and we obtain that the second term in \cref{eq:ak_rot} is equal to $1.3 \times 10^{-3}$. We thus expect for the simplified rotation profile \eqref{eq:rotation-v1}
\begin{equation}
 a_1 / 2\pi  = 435.9~\textrm{nHz}, \quad a_3 / 2\pi = 21.1~\textrm{nHz}, \quad a_5/2\pi = -3.6~\textrm{nHz}, \label{eq:a-coeff_sun}
\end{equation}
which is obtained by multiplying the input 
coefficients $\Omega_j$ by $(1-1.3 \times 10^{-3}$)
following \cref{eq:ak_rot}.

\end{remark}

\paragraph{Numerical results} In \cref{fig:helioseismic-spectrum-rot}, we show the power spectra $\mathcal{P}_l^m(\omega)$ 
computed for $\ell=85$ in the range of frequency (2.9-3.1~mHz) which contains the mode with $n=7$. Three different rotation profiles listed in \cref{rotprofile::subsec} are considered: without rotation,  with a simplified model of solar rotation, and the profile inferred from global helioseismology.
We remind that in all the cases the tracking rate 
is fixed at $\Omega_c / 2\pi = 460$~nHz.
In each case, the location of maximum power on the spectra is different 
for each $m$ and the shift is compared to the reference frequency of 
$m=0$ to recover information about the differential rotation. For each $m$, we fit the power spectrum with a Lorentzian as shown in \cref{fig:helioseismic-spectrum-rot-A} in order to obtain the frequency of maximum power $\omega_{n\ell m}$.

We have the following observations:
\begin{itemize}[leftmargin = *]
  \item For the case without rotation, \cref{fig:helioseismic-spectrum-rot-B}, the shift in frequency depends linearly on $m$ with slope $\Omega_c$. We see that the maximal power in the simulations matches perfectly this slope.

  \item For the case with the simplified rotation \cref{figure:model-rotation-v1,eq:rotation-v1}, the theoretical a-coefficients can be computed analytically \cref{eq:ak_rot} and the location of maximal power matches very well the polynomial $\tilde{\Pi}^{(\ell)}$ obtained with these coefficients \cref{fig:helioseismic-spectrum-rot-A}.

  \item For the rotation inferred by helioseismology \cite{Larson2018}, the a-coefficients are the input of the inversion procedure. Using these coefficients in the function $\Pi^{(\ell)}$, we see that the maximal power in the simulations matches this polynomial \cref{fig:helioseismic-spectrum-rot-C}.
\end{itemize}

To better estimate the accuracy of our simulations, we fit the numerical power spectra to obtain a-coefficients and compare them to the observations for the solar profile and to the analytic solution for the simplified profile, see \cref{table:a-coeff}. In both cases, the accuracy is better than 0.03~nHz which is below the noise level in the measurements.
\begin{table} \begin{tabular}{c c || c| c|c }  \multicolumn{2}{c}{a-coeff \cref{eq:a_coeff} in (nHz)}              & $a_1 /(2\pi)$&  $a_3/(2\pi)$ & $a_5 /(2\pi)$\\[0.em]\hline\hline
                      \multirow{2}{*}{Solar profile} &    From simulation  &       442.86    & 22.17    & -3.22  \\[0.1em]
                      &  From observation \cite{Larson2018} &     442.85 $\pm$ 0.05   & 22.15 $\pm$ 0.08 &-3.21$\pm$ 0.10 \\ \hline
                      \multirow{2}{*}{Simplified profile} &    From simulation &       435.89    & 21.13   & -3.63  \\[0.1em]
                      &  \text{Theoretical} &     435.90   & 21.11 &-3.61
                           \end{tabular}
\caption{Comparison of the a-coefficients obtained by fitting the numerical simulations to the reference obtained analytically for the simplified rotation profile and from the observations for the profile infered by helioseismology.}
\label{table:a-coeff}
   \end{table}

\begin{remark}
 This quantitative comparison of the a-coefficients allows us to validate the implementation of rotation in our code and to estimate the numerical accuracy. It is stronger than the qualitative comparison presented in \cite{Gizon2017} where the measured solar a-coefficients were compared to the ones obtained from the simulations with a simplified rotation profile. A consistent comparison is also performed in \cite{Stejko2021} between their numerical simulation and the eigenfrequencies computed using first-order perturbation but only a visual comparison is presented.
\end{remark}

\section{Conclusion}

We have developed a forward solver for the equations of 
stellar oscillations with background flow  and gravity,
and have validated it in the context of helioseismology.
This is realized with the Hybridizable Discontinuous Galerkin 
method, and used to solve the equations for realistic solar backgrounds.
These backgrounds are challenging due to their drastic
variations and the non-zero buoyancy frequency in the case
with gravity, which leads to 
a change in the nature of the problem from ellipticity
to hyperbolicity, at zero attenuation and for frequencies below acoustic cut-off.
Despite encoding attenuation, appropriate choices of HDG stabilization and meshing
are required in numerical implementations to maintain accuracy.
It is not only due to the existence of these regions,  
but also to the fact that
the formation and observation heights employed in helioseimology
are positioned in the transition zone between the two behaviors.

From the Green's kernels, we derive helioseismic products such as 
the power spectrum. The positions of high power match the
eigenfrequencies obtained independently with the GYRE eigensolver.
Including rotation, the power spectrum depends on the azimuthal 
order $m$ and the change in the eigenfrequency can be determined 
analytically for simple rotation profiles. 
The validation presented in this paper has been conducted in the solar context, 
but our assumption of azimuthal symmetry allows for the extension of our solver 
to other objects such as gas giants and stars, including fast rotators like 
Altair, which has recently been found to exhibit non-radial pulsations \cite{Rieutord2023}.

The developed solver is a valuable tool for generating synthetic observables, 
enabling the study of differential rotation using various helioseismic techniques, 
such as time-distance \cite{Duvall1993} and holography \cite{Lindsey1997}. 
Moreover, we readily get access to the 3D displacement field as the HDG 
method solves the mixed system. 
This will allow for more accurate modeling of observables, taking into 
account projection and radiative transfer effects \cite{Fournier2025}. 
These experiments allow us 
to validate our code without and with rotation.
This efficient forward solver also provides a crucial step towards nonlinear 
inversions (see, e.g., \cite{Mueller2024} for iterative holography).
In addition to the acoustic modes that are traditionally used in 
helioseismology, the inclusion of rotation further gives access 
to inertial modes such as Rossby modes that have recently been 
observed on the Sun \cite{Loeptien2018,Gizon2021}. These modes have 
sensitivity deep inside the Sun and have a potential for better 
constraining the base of the convection zone \cite{Gizon2024}.


\section*{Acknowledgments}

  This work was supported by the ANR-DFG project 
  \textsc{Butterfly}, grant numbers ANR-23-CE46-0009-01 and DFG-530101854. 
  This work was partially supported by the EXAMA 
  (Methods and Algorithms at Exascale) project under 
  grant ANR--22--EXNU--0002.
  FF acknowledges funding by the European Union 
  with ERC Project \textsc{Incorwave} -- grant 101116288.
  DF and LG acknowledge support from ERC Synergy grant WHOLESUN 810218.
  Views and opinions expressed are however those of the authors 
  only and do not necessarily reflect those 
  of the European Union or the European Research Council 
  Executive Agency (ERCEA). Neither the European Union nor the
  granting authority can be held responsible for them.

\appendix
\section{Cylindrical expansion of $\Amat^0_\Omega$ and its coefficient $\Amat_m$}
\label{appendix:rotation-axisym}
In this appendix 
we compute the coefficients $\Amat_m$ appearing in the system of boundary 
value problems \cref{eq:main-section-axisym:mode_pre}
defined on mode $m$.
These are also the coefficients in the cylindrical expansion \cref{actionAmat::def} of
matrix-valued differential operator $\Amat^0_\Omega$ \cref{Amat::def},
\begin{equation}
[\Amat \bu](\eta,\phi,z)  = \sum_{m \in \mathbb{Z}}
\Amat_m(\eta,z)  \bu_m(\eta,z)  \,e^{\ii\, m \, \phi}\,.
\end{equation}
Recall from \cref{section:axi-symmetric} that the components of a tensor in basis $\{\mathbf{e}_\eta, \mathbf{e}_\phi , \mathbf{e}_z\}$ are denoted by $[\cdot]$, e.g. for a vector $\bu$, and its vector product with matrix $M$,
\begin{equation*}
\bu = u_\eta \mathbf{e}_\eta  +  u_{\phi} \mathbf{e}_\phi  +  u_z  \mathbf{e}_z
\hspace*{0.1cm} \Rightarrow \hspace*{0.1cm} [ \bu ]  = (u_\eta ,  u_{\phi} ,  u_z)^t;\hspace*{0.5cm}
\left[ M \bu \right] =\left[ M \right] \left[\bu \right].
\end{equation*}
The cylindrical expansion of a vector $\bu$ is given by,
\begin{equation}\label{cylexpanV_rep}[\bu](\eta,\phi,z)\, = \,  
\sum_{m\in \mathbb{Z}} \bu_m(\eta,z)\,  e^{\mathrm{i}m\phi} 
 \,, \quad \text{where} \,\,  \bu_m(\eta,z) = (u_{\eta m}, u_{\phi m}, u_{z m} )^t\,.
 \end{equation}

We recall the definition of $\Amat_0^{\Omega}$ from
\cref{Amat::def}
\begin{equation}\Amat_0^{\Omega} : = \Bmat_0^{\Omega} \,-\,2\ii\omega\gamma_{\mathrm{att}}\,+\, \phifull \otimes
      \Big(\boldsymbol{\alpha}_\rho \,-\, \dfrac{\boldsymbol{\alpha}_{p}}{\Gamma_1}\Big),
      \end{equation}
where the matrix-valued differential operator $ \Bmat_0^{\Omega}$ is, \cref{Bmat::def},
      \begin{equation}\Bmat_0^{\Omega} :=    \big(-\ii \omega  + \mathbf{v}_0\cdot\nabla\big)^2
    \,+\, 2 \Rotzero \times 
              (-\ii \omega + \mathbf{v}_0 \cdot \nabla)
      \, -\, 2 \Rotzero \times \nabla \mathbf{v}_0 
     \,  -\, \nabla \left( (\mathbf{v}_0\cdot \nabla) \mathbf{v}_0\right).
      \end{equation}
      The zero-th order term containing $\phifull$, and the inverse scale height vectors are,
      \cref{scaleheight::def},
\begin{equation}
  \phifull   =   \nabla \phi_0  + \ \rotvec\times \rotvec\times \mathbf{x}  \overset{\cref{phifullexp}}{=} 
\nabla \phi_0  - \Upomega^2 \eta \mathbf{e}_\eta , \quad 
\boldsymbol{\alpha}_p = -\dfrac{\nabla p}{p}, \quad \boldsymbol{\alpha}_{\rho} = - \dfrac{\nabla \rho}{\rho}\,.
\end{equation}

The computation is carried out for differential rotation flow background, cf. \cref{inderotrate::identity},
\begin{equation}\label{v0phi0_app}
\mathbf{v}_0 = (\rotvec - \Rotzero) \times \mathbf{x}= \eta \,( \rot - \rotzero) \mathbf{e}_\phi, 
\end{equation}
with the observation frame rotation rate $\rotzero$, full rotation rate $\rot$ and the difference between the two rates $\tilde{\rot}$,
\begin{equation} \Rotzero = \rotzero(\eta,z) \,\mathbf{e}_z, \,\, \text{ with } \rotzero \text{ constant} ,  \hspace*{0.5cm}  \rotvec = \rot(\eta,z)\, \mathbf{e}_z, \hspace*{0.5cm} \tilde{\rot} (\eta,z):= \rot(\eta,z) - \rotzero. \end{equation}
The second expression of $\mathbf{v}_0$ follows from the identities: $\mathbf{e}_z \times \mathbf{e}_{\phi} = - \mathbf{e}_\eta$ , $ \mathbf{e}_z \times \mathbf{e}_\eta = \mathbf{e}_\phi$, we have
\begin{equation}\label{ezcurlid}\mathbf{e}_z =\dfrac{\eta}{r} \mathbf{e}_\eta + \dfrac{z}{r}\mathbf{e}_z , \quad \mathbf{x}= r \mathbf{e}_r \hspace*{0.3cm}
\Rightarrow  \hspace*{0.3cm}
\mathbf{e}_z \times \mathbf{x}  = \dfrac{\eta}{r} \mathbf{e}_\phi, \quad \mathbf{e}_z \times \mathbf{e}_z \times \mathbf{x}
= - \dfrac{\eta}{r} \mathbf{e}_\eta.
\end{equation}
We thus have the following representation in cylindrical basis are,
\begin{equation}\label{v0cyl}
\left[\Rotzero\right] = \begin{pmatrix} 0\\0\\\rotzero \end{pmatrix}, \quad\left[\mathbf{v}_0\right]=  \begin{pmatrix} 0 \\ \eta\,\tilde{\rot}\\ 0 \end{pmatrix} , \quad  [\Rotzero \times ] = \begin{pmatrix}
          0 & -\rotzero & 0 \\
          \rotzero & 0 & 0  \\
          0 & 0 & 0
          \end{pmatrix} .
\end{equation}
The last identity is the representation of the cross product with $\Rotzero$, which also gives, 
\begin{equation}\label{rotmatrix} 
 [\Rotzero \times \,\mathbf{v}_0 ] = [\Rotzero \times ] [\mathbf{v}_0] = \begin{pmatrix} -\eta \,\tilde{\rot}\, \rotzero\\0 \\0 \end{pmatrix}, \hspace*{0.6cm} \text{ i.e}  \hspace*{0.6cm}
          \Rotzero\times\mathbf{v}_0 = - \eta \,\tilde{\rot}\,\rotzero \,\mathbf{e}_\eta.
\end{equation}

To compute $\Amat_m $, we compute $\Bmat_m$ (corresponding coefficient of $\Bmat_0^{\Omega}$ in the cylindrical expansion). We also provide the expressions for $
[\nabla \phi_0]$, $[\boldsymbol{\alpha}_p]$, $[\boldsymbol{\alpha}_{\rho}]$ for radially symmetric backgrounds, in \cref{radsymfcn_cyl}. 

\begin{identity}
For $\Amat_m$ appearing in \cref{actionAmat::def} and \cref{eq:main-section-axisym:mode_pre}, we have
\begin{equation}
\Amat_m = \Bmat_m    \,-\,2\ii\omega\gamma_{\mathrm{att}} \, + \,  \left[ \phifull \right] \otimes
      \Big(\left[\boldsymbol{\alpha}_\rho\right] \,-\, \left[ \dfrac{\boldsymbol{\alpha}_{p}}{\Gamma_1} \right]\Big),
\end{equation}
with $\left[ \phifull \right] $ given in \cref{phifullcyl} and
 \begin{equation}
\Bmat_m   :=     \mathsf{R}_{m}^2 +  2\mathsf{R}_{\Omega\,m} 
         \, -\, 2 \left[ \Rotzero \times \nabla \mathbf{v}_0 \right]
     \,  -\,\left[ \nabla  (\mathbf{v}_0\cdot\nabla)\mathbf{v}_0 \right], \end{equation}
where the matrices $\mathsf{R}_{m}$ and $\mathsf{R}_{\Omega\,m}$
are defined in \cref{RmROm::def}, $\left[ \Rotzero \times \nabla \mathbf{v}_0 \right]$
in \cref{rotgradv_cyl}, and $\left[ \nabla  (\mathbf{v}_0\cdot\nabla)\mathbf{v}_0 \right]$ in \cref{gradmatDv_cyl}.

\end{identity}

\begin{proof}
$-$ We first obtain the expression for $\phifull$. From \cref{ezcurlid}, we also have
\begin{equation}
 \left[ \rotvec\times \rotvec\times \mathbf{x}\right] \,\, \overset{\cref{phifullexp}}{=}   \begin{pmatrix} -\rot^2\, \eta\\ 0\\0\end{pmatrix} \\
\Rightarrow \hspace*{0.2cm} \left[ \phifull \right] = \left[ \nabla \phi_0 \right] +  \begin{pmatrix} -\rot^2\, \eta\\ 0\\0\end{pmatrix}\label{phifullcyl}.
\end{equation}

$-$ We consider the terms in $\Bmat_0^{\Omega}$ containing material derivative $\mathbf{v}_0\cdot \nabla$. Define matrices, $ \mathsf{R}_{m} $ and $\mathsf{R}_{\Omega\,m}$
 \begin{equation}\label{RmROm::def}
 \mathsf{R}_{m} \,:=\, \begin{pmatrix}
                 -\ii\omega + \ii m \widetilde{\Omega} & -\widetilde{\Omega} & 0 \\
                 \widetilde{\Omega} & -\ii\omega + \ii m \widetilde{\Omega} & 0 \\
                 0 & 0 & -\ii\omega + \ii m \widetilde{\Omega} 
                 \end{pmatrix}, \quad \mathsf{R}_{\Omega\,m} \,:=\,[\Rotzero \times ] \,\mathsf{R}_m ,
\end{equation}
where the second equation 
is given by a product of the matrix $\mathsf{R}_{m}$
with matrix $[\Rotzero \times ]$, cf. \cref{rotmatrix}.
With $\mathbf{v}_0$ containing only nonzero components in $\mathbf{e}_\phi$, the material derivative $\mathbf{v}_0\cdot \nabla$ acting vectors $\mathbf{u}$ decomposes to matrix multiplication operation in its cylindrical expansion,
\begin{equation}\big[(-\ii \omega + \mathbf{v}_0 \cdot \nabla) \bu\big]
                 \, =\, \sum_{m\in \mathbb{Z}} \mathsf{R}_{m} \, \bu_m(\eta,z)\,  e^{\mathrm{i}m\phi}, \quad \bu_m \text{ defined in } \cref{cylexpanV_rep}.
\end{equation}
Thus first two terms in $\Bmat_0^{\Omega}$ decomposes to multiplication by $ \mathsf{R}_{m}$ and $ \mathsf{R}_{\Omega\,m}$, on each $m$,
\begin{subequations}
\begin{align}
\left[  (-\ii \omega + \mathbf{v}_0 \cdot \nabla)^2 \bu\right] \,&=\, \sum_{m\in \mathbb{Z}} \mathsf{R}_{m}^2 \, \bu_m(\eta,z)\,  e^{\mathrm{i}m\phi} \,; \\[.80em]
\left[  \Rotzero \times\,  (-\ii \omega + \mathbf{v}_0 \cdot \nabla) \bu\right] 
       \,& =\, \sum_{m\in \mathbb{Z}} \mathsf{R}_{\Omega m} \, \bu_m(\eta,z)\,  e^{\mathrm{i}m\phi} .
 \end{align}\end{subequations}

\noindent $-$ For the third term $\Bmat_0^{\Omega}$, we first list the gradient of $\mathbf{v}_0$ in cylindrical basis,
\begin{equation}\label{nablav0_cyl}\left[ \nabla\mathbf{v}_0\right]  \,=\, \begin{pmatrix}
                                                    0 & -\widetilde{\Omega} & 0\\
                                               \partial_\eta(\eta\widetilde{\Omega}) & 0 & \partial_z(\eta\widetilde{\Omega}) \\
                                                    0 & 0 & 0 
                                                  \end{pmatrix}.\end{equation}
 From this, we obtain
\begin{equation}\label{rotgradv_cyl}
   \left[\Rotzero \times \nabla \mathbf{v}_0 \right]  = \left[ \Rotzero \times\right]  \left[\nabla \mathbf{v}_0\right] 
      \,=\,\begin{pmatrix}
            -\rotzero \, \partial_\eta(\eta\, \widetilde{\Omega}) & 0 & -\rotzero \,\partial_z(\eta\, \widetilde{\Omega}) \\
            0 & - \rotzero \,\widetilde{\Omega} & 0 \\
            0 & 0 & 0
         \end{pmatrix}. \end{equation}
 
\noindent $-$ For the last term in $\Bmat_0^{\Omega}$, we rewrite the material derivative $(\mathbf{v}_0\cdot\nabla)$ acting on $\mathbf{v}_0$ as a product of matrix $\nabla\mathbf{v}_0$ and $\mathbf{v}_0$, $(\mathbf{v}_0\cdot\nabla)\mathbf{v}_0=(\nabla \mathbf{v}_0) \mathbf{v}_0  $ and employ this to compute its components in cylindrical basis,
\begin{equation}\label{matder_cyl}
\left[(\mathbf{v}_0\cdot\nabla)\mathbf{v}_0 \right] = \left[ \nabla \mathbf{v}_0\right] \, \left[ \mathbf{v}_0\right] 
\overset{  \cref{nablav0_cyl} ,\cref{v0cyl}}{=} \begin{pmatrix} - \eta\,\,\widetilde{\Omega}^2 \\ 0 \\ 0 \end{pmatrix} , \quad \text{i.e.} \quad
(\mathbf{v}_0\cdot\nabla)\mathbf{v}_0 
\,=\, - \eta\,\,\widetilde{\Omega}^2 \, \mathbf{e}_{\eta}  .
\end{equation}
The gradient of the above vector  $(\mathbf{v}_0\cdot\nabla)\mathbf{v}_0$ is
\begin{equation}\label{gradmatDv_cyl}
\left[ \nabla  (\mathbf{v}_0\cdot\nabla)\mathbf{v}_0 \right] \,=\, 
                       \,-\, \begin{pmatrix}
                         \partial_\eta(\eta \widetilde{\Omega}^2) & 0 & \partial_z(\eta \widetilde{\Omega}^2) \\
                         0 & \widetilde{\Omega}^2 & 0 \\
                         0 & 0 & 0 
                       \end{pmatrix} .
\end{equation}
The final expression for $\left[\Bmat_0^{\Omega}\right]$ is obtained by putting together expressions from \cref{RmROm::def,rotgradv_cyl,gradmatDv_cyl}. 
\end{proof}

\paragraph{Gradient of radially symmetric function in cylindrical basis}
We will employ radially symmetric parameters $\phi_0$, $\rho_0$, and $p_0$ even with background
containing rotation, cf. \cref{rolefghost::rmk}. It is useful to write their components in this basis.
We have
\begin{gather}\label{radsymfcn_cyl}
\mathbf{e}_r = \dfrac{\eta}{r} \mathbf{e}_\eta + \dfrac{z}{r} \mathbf{e}_z,\quad
\boldsymbol{\alpha}_g = \alpha_g(r) \mathbf{e}_r 
\hspace*{0.2cm}\Rightarrow \hspace*{0.2cm}\left[\boldsymbol{\alpha}_g \right] = \alpha_g(r)\begin{pmatrix} \eta/r \\ 0 \\ z/r\end{pmatrix},\\
\Longrightarrow  \hspace*{0.3cm}\nabla \phi_0 = \partial_r \phi_0 \begin{pmatrix} \eta/r \\ 0 \\ z/r\end{pmatrix}, \quad 
\left[ \boldsymbol{\alpha}_p + \dfrac{\boldsymbol{\alpha}_{\rho}}{\Gamma_1}\right]= 
\left(\alpha_p  + \dfrac{\alpha_p}{\Gamma_1}\right)\begin{pmatrix} \eta/r \\ 0 \\ z/r\end{pmatrix}.
\end{gather}

\newcommand{\wex} {w_{\mathrm{ex}}}
\newcommand{\fex} {h_{\mathrm{ex}}}
\newcommand{\buex}{\bu_{\mathrm{ex}}}
\newcommand{\uex} {u_{\mathrm{ex}}}
\newcommand{\hmesh}{\mathfrak{h}}
\section{Investigations with synthetic background parameters}
\label{appendix:synthetic-background}

In this appendix, we consider synthetic backgrounds to
analyse the numerical behaviour of the solutions when 
there are both $\Bell$ and $\Bhyp$ within the computational 
domain.
We first perform a convergence study of the solver 
using a manufactured solution in \cref{appendix:convergence},
and investigate the choice of scaling factor for the HDG
stabilization.
Then, in \cref{appendix:singularity}, we consider localized sources
and highlight the phenomenon arising when the
source is close to the interface between $\Bell$ and $\Bhyp$, 
called the sonic line\footnote{This terminology follows \cite{payne1996interior}.}.
In \cref{appendix:attenuation}, we further investigate the 
numerical behaviour as the attenuation is reduced, although 
non-zero attenuation is a key requirement for ensuring well-posedness,
as discussed in the paper.

We consider the wave system \cref{uniformulation}, with radial 
coefficients given by
\begin{equation} \label{eq:manufactured-parameter}
\begin{aligned}
& \Amat(r) \,=\, -\sigma^2\mathbb{Id} + N^2(r) \,\mathbf{e}_r \otimes \mathbf{e}_r, 
           \qquad \text{with}\qquad \sigma^2:= \omega^2 +2\ii\omega\gamma_{\mathrm{att}}, \\
& \varrho(r) = 1/c_0^2, \text{ with sound speed $c_0$} \,, \hspace*{7em} 
\boldsymbol{\beta}_1 \,=\, \boldsymbol{\beta}_2 \,=\,0 \, .
\end{aligned} \end{equation}
In particular, in this case we have, 
\begin{equation}
   \Amat^{-1}(r) \,=\, 
   - \dfrac{1}{\sigma^2} \mathbb{Id} + \left(\dfrac{1}{N^2(r)-\sigma^2} 
                                     + \dfrac{1}{\sigma^2}\right) \mathbf{e}_r\otimes \mathbf{e}_r\,.
\end{equation}


\paragraph{Variation profile for $N^2$}

The distinction between the regions $\Bhyp$ and $\Bell$ is determined 
by the sign of $(N^2-\omega^2)$.
To ensure their co-existence, we construct $N^2$ as
a smooth function 
so that $(N^2-\omega^2)$ ranges between $\pm\omega^2$.
Furthermore, $N^2$ is parametrized by the scalar parameters
$a$ and $b$, whose difference controls the width of the
transition region. The sonic line is located at $r=(a+b)/2$, where 
$N^2=\omega^2$. To satisfy these condition, we choose $N^2$ defined on
$(0, r_{\max})$ with $0 < a < b < r_{\max}$, as
\begin{equation}\label{eq:appendix-N2-var}
N^2(r)  \,=\, 2\omega^2 \chi(r)\,, \qquad\text{where}\quad 
  \chi(r) \, =\, \dfrac{1}{2} \left(1 + 
  \tanh\left(   
                \dfrac{r - (a+b)/2}{(b-a)/6}
       \right)
  \right) \,.
\end{equation}

\subsection{Convergence test with a manufactured solution}
\label{appendix:convergence}

Given a manufactured solution $\wex$, the corresponding right-hand 
side $\fex$ associated with \cref{uniformulation} and the background 
parameters defined in \cref{eq:manufactured-parameter} is radial 
and given by
\begin{equation}\label{eq:manufactured-rhs}
  \fex(r) \,=\, \varrho \wex(r)       
     \,+\, \uex'(r)  + 2\dfrac{\uex(r)}{r} \,, 
\end{equation}
with
\begin{equation}
\uex(r) \,=\, -\dfrac{\wex'(r)}{N^2(r)-\sigma^2} \,, \quad\text{and}\quad
\uex'(r) \,=\, \dfrac{(N^2(r))'}{\left( N^2(r)-\sigma^2 \right)^2} \wex'(r)
         \,-\, \dfrac{\wex''(r)}{N^2(r)-\sigma^2} \,. 
\end{equation}

The numerical experiments are carried out on the
meridional half-disk of radius $r_{\max}=1$, and we consider
an oscillatory solution defined 
in terms of the Bessel function $J_0$:
\begin{equation} \label{eq:appendix-wex-bessel}
  \wex(r) \,=\, J_0(k\,r) \,\, (1-r^2)^2 \, , \qquad \text{with } k=\omega/c_0 \,, \qquad
  \text{for $r\in(0,1)$}.
\end{equation}
From $\wex$, the right-hand side $\fex$ is computed 
from \cref{eq:manufactured-rhs} and is used in the 
numerical solver. 
We employ the following background parameters:
\begin{equation} \label{eq:appendix:background-manufactured-a}
  \text{$N^2$ from \cref{eq:appendix-N2-var} with $a=0.1$ and $b=0.9$}, \qquad
  c_0 \,=\, 1 \,, \qquad 
  \omega \,=\, 8\,\pi\,, \qquad
  \gamma_{\mathrm{att}} \,=\, 2\,.
\end{equation}

\subsubsection{Scaling factor of the HDG stabilization}

\begin{figure}[ht!]\centering
  \includegraphics[scale=0.75]{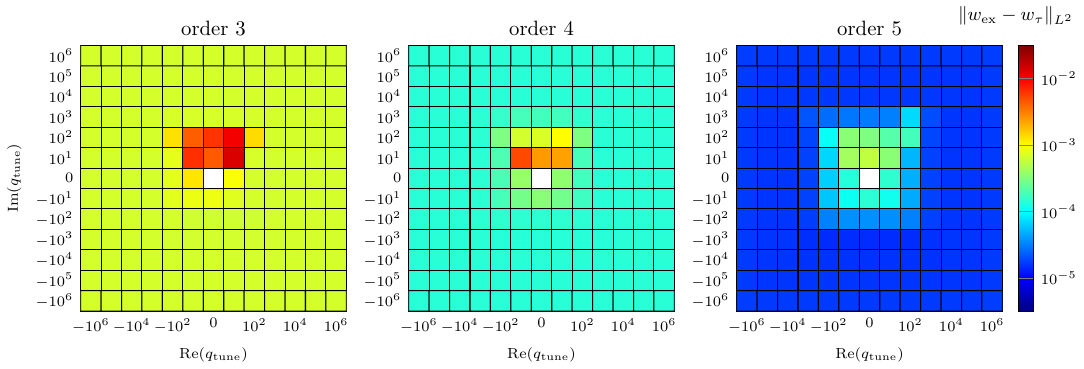}
  \caption{\modifadd{Evolution of the $L^2$ error between the solution 
           $\wex$ from \cref{eq:appendix-wex-bessel} and the 
           numerical solution as a function of the
           stabilization scaling factor $q_{\mathrm{tune}}$
           used in the HDG formulation.
           Here, $q_{\mathrm{tune}}$ varies in the 
           complex plane and the results are shown for 
           polynomial orders $3$, $4$, and $5$.}}
  \label{figure:appendix:tau-map-manufactured}
\end{figure}

The HDG discretization involves a stabilization parameter, 
as given in \cref{defNeutrace_spec}. Here we investigate the 
choice of the scaling factor $q_{\mathrm{tune}}$ that 
appears within this stabilization term, see \cref{stabtaud}.
In \cref{figure:appendix:tau-map-manufactured}, we plot 
the $L^2$ error between the manufactured solution $\wex$ 
\cref{eq:appendix-wex-bessel} and the numerical solution that
uses the right-hand side $\fex$ from \cref{eq:manufactured-rhs}
and the background parameters given \cref{eq:appendix:background-manufactured-a}.
Each numerical solution $w_\tau$ is associated with a choice
of scaling factor $q_{\mathrm{tune}}$. We let $q_{\mathrm{tune}}$
vary in the complex plane with real and imaginary parts between $\pm\num{e6}$,
excluding 0. We further consider polynomial 
orders \num{3}, \num{4}, and \num{5}.

We observe that the error decreases as the polynomial order
increases, as expected. Comparing among the different choices of $q_{\mathrm{tune}}$, the error increases
in the vicinity of the origin, particularly for values with a
positive imaginary part.
When either the real or imaginary part of the
scaling factor has a large magnitude ($> \num{e4}$), the error is
minimal and remains essentially unchanged regardless of the
choice of $q_{\mathrm{tune}}$.
In the following numerical experiments, we choose
$q_{\mathrm{tune}} = -\num{e6} \ii$.

\subsubsection{Convergence test}

\begin{figure}[ht!] \centering
  \includegraphics[scale=1]{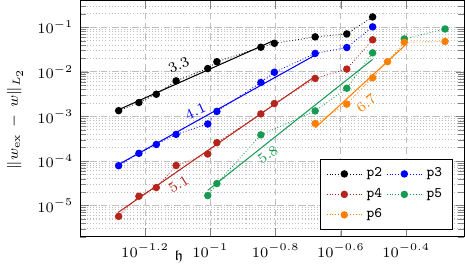}
  \caption{\modifadd{Evolution of the $L^2$ error between the manufactured
           solution $\wex$ from \cref{eq:appendix-wex-bessel}
           and the numerical solution $w$ as a function of the
           mesh size for polynomial orders from $2$ to $6$.
           The computed errors are indicated by the bullets, while the
           solid lines represent the fitted convergence rates,
           whose orders are indicated.}}
  \label{figure:convergence-solver}
\end{figure}

We now investigate the convergence rate of the method.
In \cref{figure:convergence-solver}, we plot the $L^2$ error
between the manufactured solution \cref{eq:appendix-wex-bessel}
and the corresponding numerical solution as a function of the
mesh size $\hmesh$, where $\hmesh$ denotes the mesh edge length.
We consider polynomial orders from $2$ to $6$ and
report the corresponding convergence rates by indicating the
slopes of the resulting curves.

We observe the expected convergence rate of $\mathfrak{h}^{p+1}$ for a
polynomial approximation of order $p$, confirming the
theoretical accuracy of the HDG discretization.
Moreover, for a fixed mesh size $\hmesh$, increasing the
polynomial order consistently reduces the $L^2$ error. 
These results confirm that the proposed formulation 
retains its convergence properties even in the co-existence of $\Bhyp$ and $\Bell$ regions.

\subsection{Propagation with localized sources near the sonic line}
\label{appendix:singularity}

In this section, we consider a configuration with a sharper
transition between $\Bell$ and $\Bhyp$, intended to be more representative
of the one occurring in the solar backgrounds (see \cref{figure:solar-models:N2Sl2}). 
The computational half-disk domain has radius $r_{\max}=\num{0.55}$, 
and the sonic line is located at $r=\num{0.5}$, so that the region 
$\Bhyp$ is relatively small and confined near the surface.
The synthetic background is defined as follows:
\begin{equation} \label{eq:appendix:background-manufactured}
  \text{$N^2$ from \cref{eq:appendix-N2-var} with $a=0.499$ and $b=0.501$}, 
   \qquad
   c_0 \,=\, 1 \,, \qquad 
   \omega \,=\, 40\,\pi\,, \qquad
   \gamma_{\mathrm{att}} \,=\, 1\,.
\end{equation}

\subsubsection{Solutions with different type of sources}

\begin{figure}[ht!] \centering
\subfloat[\modifadd{The center of the source is positioned in $\eta_{\mathrm{src}}=\num{0.48},\,z_{\mathrm{src}}=0$.}]
  {\includegraphics[scale=0.70]{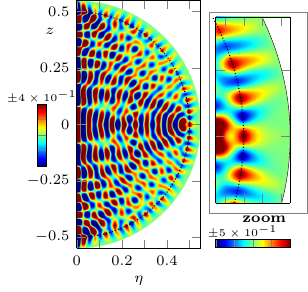}
   \includegraphics[scale=0.70]{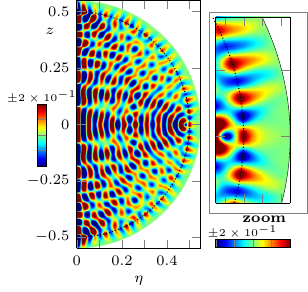}
   \includegraphics[scale=0.70]{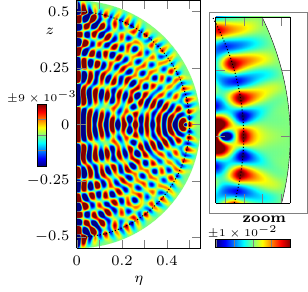}
   \includegraphics[scale=0.70]{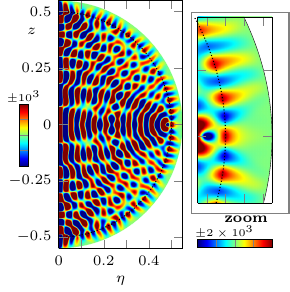}
   \label{figure:source-comparisonA}
  }
  
\subfloat[\modifadd{The center of the source is positioned in $\eta_{\mathrm{src}}=\num{0.499},\,z_{\mathrm{src}}=0$.}]
  {\includegraphics[scale=0.70]{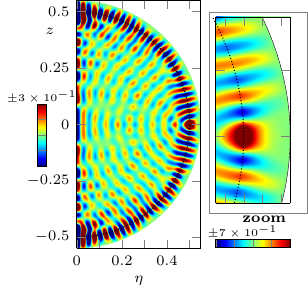}
   \includegraphics[scale=0.70]{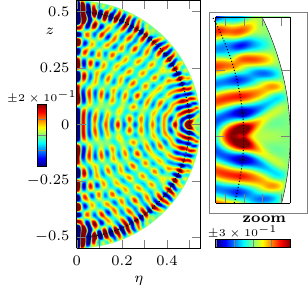}
   \includegraphics[scale=0.70]{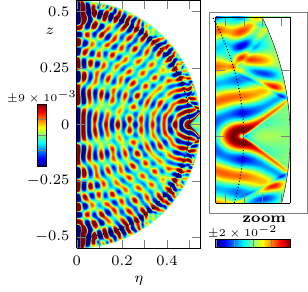}
   \includegraphics[scale=0.70]{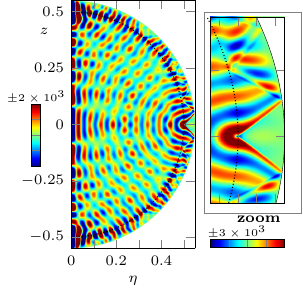}
   \label{figure:source-comparisonB}
  }

\subfloat[\modifadd{The center of the source is positioned in $\eta_{\mathrm{src}}=\num{0.505},\,z_{\mathrm{src}}=0$.}]
  {\includegraphics[scale=0.70]{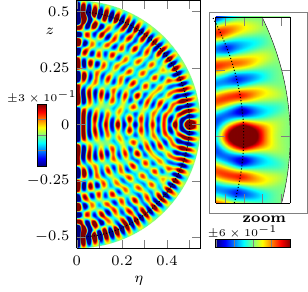}
   \includegraphics[scale=0.70]{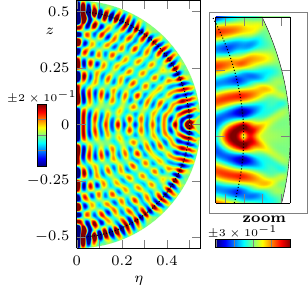}
   \includegraphics[scale=0.70]{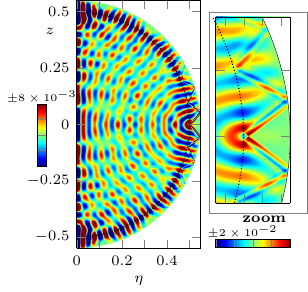}
   \includegraphics[scale=0.70]{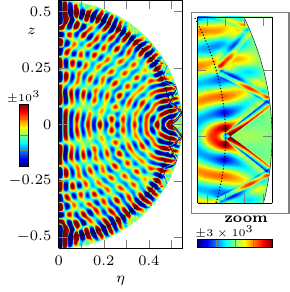}
   \label{figure:source-comparisonC}
  }

%
%
%

  \caption{\modifadd{Comparison of the real parts of the solutions depending on the source type.
           For each row, the three columns on the left employ Gaussian sources
           with variance $\sigma_{\mathrm{G}}=\num{e-2}$, \num{5e-3} and 
           \num{e-4}, respectively. The right column corresponds to a Dirac point source.
           The computational half-disk has radius \num{0.55} with 
           the sonic line between $\Bell$ and $\Bhyp$ at $r=\num{0.5}$
           (indicated by the dotted black line). Each panel 
           also shows a zoom of the solution near the source 
           and close to the surface, in the zone
           $(\num{0.47},\num{0.55})\times(-\num{0.10},\num{0.18})$.
           }}
  \label{figure:source-comparison}
\end{figure}

We consider a localized source for the right-hand side, which is either 
a Dirac source located at position $(\eta_{\mathrm{src}},z_{\mathrm{src}})$,
or a Gaussian-shaped function $h_{\mathrm{G}}$ centered 
at this position and with a variance $\sigma_{\mathrm{G}}$ such that,
\begin{equation}
  h_{\mathrm{G}}(\eta,z) \,=\, \exp\left( \dfrac{1}{2} 
  \left( \dfrac{\eta-\eta_{\mathrm{src}}}{\sigma_{\mathrm{G}}} \right)^2
  \,+\,
  \left( \dfrac{z-z_{\mathrm{src}}}{\sigma_{\mathrm{G}}} \right)^2
  \right) \,.
\end{equation}
In \cref{figure:source-comparison}, we show the numerical solutions 
depending on the position of the sources and of the variance of the Gaussian
source. The sources are positioned along $z_{\mathrm{src}}=0$, at different
locations near the sonic line (in $r=\num{0.50}$), namely 
with $\eta_{\mathrm{src}}=$\num{0.48}, \num{499} and \num{0.505}, 
corresponding respectively to the top, middle and bottom rows 
of \cref{figure:source-comparison}. The simulations use polynomial
order $4$ and a refined mesh with about \num{120e3} cells.

When the source is located in $\Bell$ and sufficiently away from the sonic line, 
with $\eta_{\mathrm{src}}=\num{0.48}$,  
the solution remains smooth across the sonic line, as shown in \cref{figure:source-comparisonA}. 
The wavefield is attenuated within the region $\Bhyp$,
and all sources (Dirac or Gaussian with different variances)
produce essentially the same oscillatory pattern.
When the source is moved closer to the sonic line, 
either below or above it, namely for
$\eta_{\mathrm{src}}=\num{0.499}$ in \cref{figure:source-comparisonB} 
and
$\eta_{\mathrm{src}}=\num{0.505}$ in \cref{figure:source-comparisonC},
the solution in the interior region $\Bell$ remains similar for all 
sources but the solutions in $\Bhyp$ have different behaviours. 
For Gaussian sources with a relatively large variance
(left column of \cref{figure:source-comparison}), the 
solution is smooth and strongly attenuated within $\Bhyp$.
As the Gaussian variance decreases, however, the solution develops
increasingly localized cone features. 
For the sharpest Gaussian and the Dirac sources, the 
propagation cone clearly emerges from the source location 
and propagates back and forth within the $\Bhyp$ region.
This phenomenon occurs when the source is sufficiently close to 
the sonic line and when the source is a sharp Gaussian 
or a Dirac. On the other hand, smoother source distributions 
suppress these features.

\subsubsection{Convergence with mesh refinement}

\begin{figure}[ht!] \centering
\subfloat[\modifadd{On a mesh with  \num{e3} elements: real part of the wavefield solution (left) and difference
          with a reference solution (right).}]
{\includegraphics[scale=0.70]{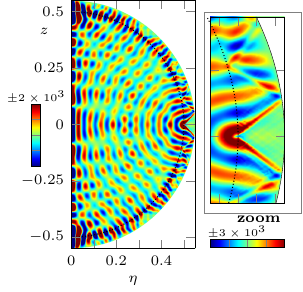}
 \includegraphics[scale=0.70]{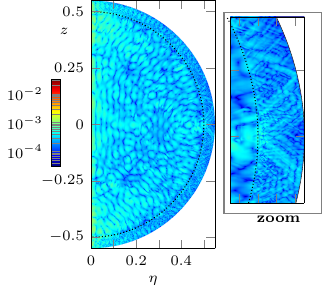}} \hspace*{1em}
\subfloat[\modifadd{On a mesh with  \num{5e3} elements: real part of the wavefield solution (left) and difference
          with a reference solution (right).}]
{\includegraphics[scale=0.70]{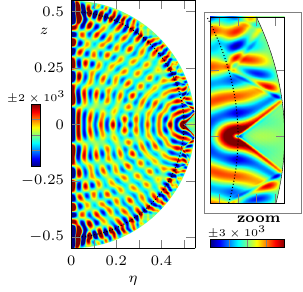}
 \includegraphics[scale=0.70]{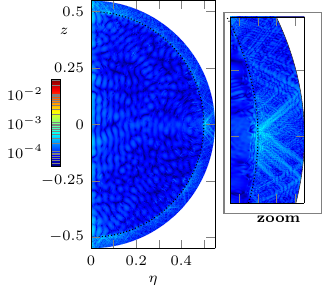}}
\caption{\modifadd{Simulations with different meshes for a source 
         at $\eta_{\mathrm{src}}=\num{0.499}$ and 
         polynomials of order $4$. The reference solution
         is computed with a mesh containing \num{120e3}
         elements.
         Zooms in the zone $(\num{0.47},\num{0.55})\times(-\num{0.10},\num{0.18})$
         are also provided.}
        }
\label{figure:appendix-pt-source-error-map}
\end{figure}

Without an analytic solution to evaluate the error, 
we rely on numerical solutions obtained with a refined mesh.
To obtain a reference solution $w_\mathrm{ref}$, we employ 
a mesh containing about \num{120e3} elements,
which is employed for the simulations in \cref{figure:source-comparison}. 
It corresponds to a characteristic mesh edge 
size $\hmesh=\num{3.5e-3}$.
In \cref{figure:appendix-pt-source-error-map}, we visualize the 
solutions obtained with coarser meshes, as well as the difference between 
the reference solution and the simulation. We consider the source that
is close to the sonic line, with $\eta_{\mathrm{src}}=\num{0.499}$,
and observe the persistence of the zigzag pattern with all meshes.

For the coarsest meshes, small differences are visible in the zoom 
near the source center, which disappear as the mesh is 
refined. The discrepancy is primarily localized around the source, 
in particular within the cone-feature that appears in $\Bhyp$. 
The expected convergence behaviour is observed: the error decreases 
under mesh refinement, and the numerical solution approaches the 
reference one.

\begin{figure}[ht!] \centering
\subfloat[\modifadd{With polynomial order $3$.}]
  {\includegraphics[scale=.7]{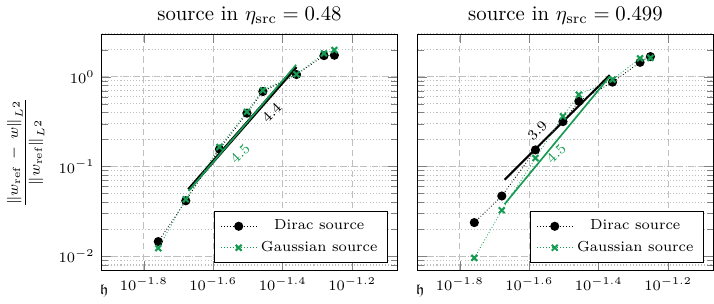}}
\subfloat[\modifadd{With polynomial order $4$.}]
  {\includegraphics[scale=.7]{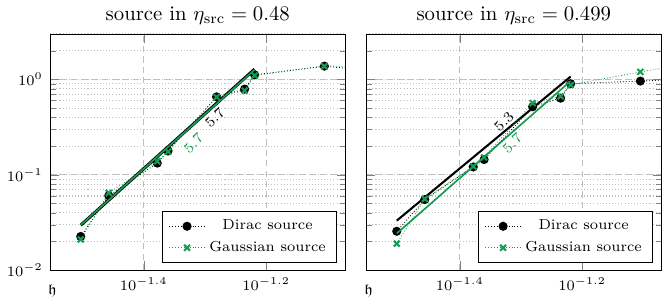}}
\caption{\modifadd{
         Evolution of the $L^2$ difference depending on the source
         position and polynomial orders for Dirac and Gaussian sources.
         Error is obtained with a reference solution computed 
         on a refined mesh with edge size $\hmesh=\num{3.5e-3}$.
         The Gaussian source has a variance of $\sigma_{\mathrm{G}}=\num{e-2}$. 
         The source position is excluded to compute the difference.}}
\label{figure:appendix-point-source-convergence}
\end{figure}

To further assess convergence, we present in \cref{figure:appendix-point-source-convergence} 
the relative $L^2$ error between the reference solution and simulations performed on 
successively refined meshes. We consider polynomial orders 3 and 4, and compare 
the Dirac and Gaussian source with $\sigma_{\mathrm{G}}=\num{e-2}$.
We show results for a source located near the sonic line, as well as for 
a source in $\Bell$.
Furthermore the error is computed after excluding the cells in the 
immediate vicinity of the center of the source.

For the case in which the source is sufficiently far from the sonic line, 
both source formulations have similar behaviour, recovering the expected 
convergence order of $p+1$, even slightly higher here. When the source is 
closer to the sonic line, both sources retain the 
behaviour in $\mathfrak{h}^{p+1}$, although we notice a slight degradation 
for the Dirac source compared to the Gaussian one.

\subsubsection{Scaling factor for the HDG stabilization}

\begin{figure}[ht!]\centering
  \includegraphics[scale=0.75]{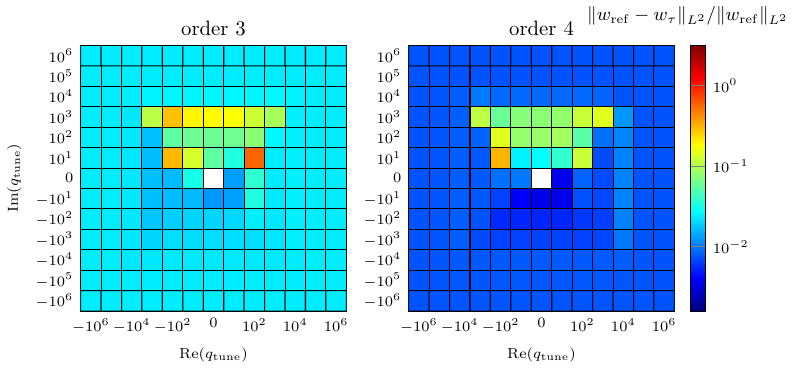}
  \caption{\modifadd{$L^2$ difference between the reference solution 
            obtained with a refined mesh and the numerical solution 
            as a function of the stabilization scaling factor 
            $q_{\mathrm{tune}}$ used in the HDG formulation.
           Here, $q_{\mathrm{tune}}$ varies in the 
           complex plane and the results are shown for 
           polynomial orders $3$ and $4$.
           The Dirac source is positioned at
           $\eta_{\mathrm{src}}=\num{0.499}$.
           }}
  \label{figure:appendix:tau-map-dirac}
\end{figure}

We next evaluate accuracy as a function of the stabilization 
coefficient $q_{\mathrm{tune}}$. Similarly to what is 
carried out in \cref{appendix:convergence}, 
we present maps of the relative error 
in \cref{figure:appendix:tau-map-dirac} with $q_{\mathrm{tune}}$ varying in the complex plane. 
Solutions are obtained with a mesh containing
about \num{5e3} elements, while the reference solution 
employs the finest mesh.

We observe a similar pattern to that obtained for 
the manufactured solution: When the stabilization magnitude
is sufficiently large, solutions have the 
smallest difference compared with the reference solution, 
and a specific choice of coefficient is 
relatively flexible. As the magnitude of the stabilization coefficient 
is reduced, the discrepancy increases, particularly 
for those with positive imaginary part.

\subsection{Attenuation level}
\label{appendix:attenuation}

\begin{figure}[ht!] \centering
\subfloat[With a Gaussian source of variance $\sigma_{\mathrm{G}}=\num{e-2}$.]
  {\includegraphics[scale=0.72]{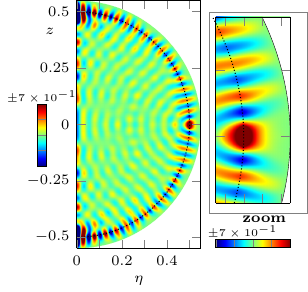}
   \includegraphics[scale=0.72]{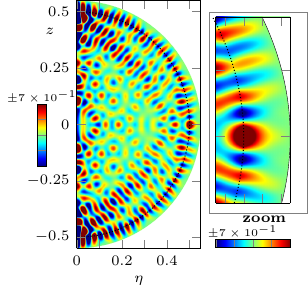}
   \includegraphics[scale=0.72]{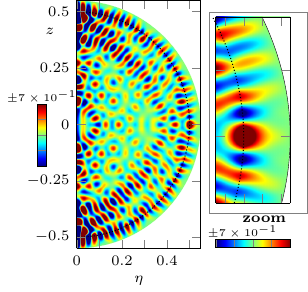}
   \includegraphics[scale=0.72]{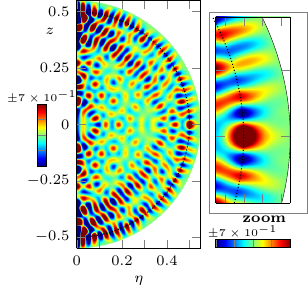}
  }

\subfloat[With a Dirac source.]
  {\includegraphics[scale=0.72]{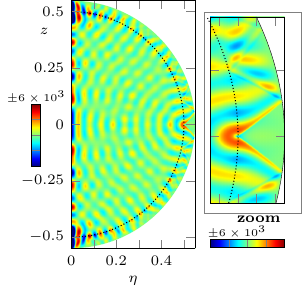}
   \includegraphics[scale=0.72]{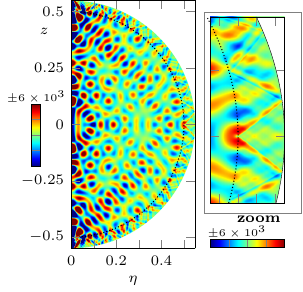}
   \includegraphics[scale=0.72]{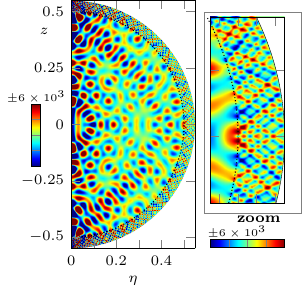}
   \includegraphics[scale=0.72]{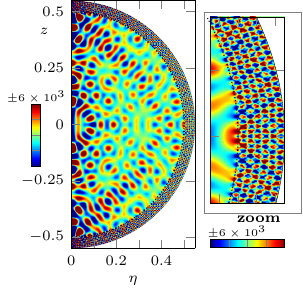}
  }
  \caption{\modifadd{Comparison of the simulations depending on the attenuation level and the source type.
           From the left to right columns, attenuation 
           $\gamma_{\mathrm{att}}=$\num{1}, \num{e-1}, \num{e-2} and \num{e-3} are employed.
           The source is positioned in $\eta_{\mathrm{src}}=\num{0.499}$ and  
           the sonic line between $\Bell$ and $\Bhyp$ at $r=\num{0.5}$
           (indicated by the dotted black line). Each panel 
           also shows a zoom of the solution near the source 
           and close to the surface, in the zone
           $(\num{0.47},\num{0.55})\times(-\num{0.10},\num{0.18})$.}
           }
\label{figure:appendix-attenuation}
\end{figure}

Attenuation is a key ingredient in the mathematical analysis of 
the problem particularly for well-posedness. 
Here, we investigate numerically the behaviour of the solutions as 
the attenuation is reduced. We consider the same configuration 
as in \cref{appendix:singularity}, with a source located below 
the sonic line, centered at $\eta_{\mathrm{src}}=\num{0.499}$, $z_{\mathrm{src}}=0$.
In \cref{figure:appendix-attenuation}, we present the resulting 
wavefields for four attenuation levels, $\gamma_{\mathrm{att}}=$1, 
\num{e-1}, \num{e-2}, and \num{e-3}. 
We compare between the Dirac and Gaussian source, 
the latter with variance $\sigma_{\mathrm{G}}=\num{e-2}$. 
All simulations are performed with polynomial order 4
and the aforementioned reference mesh.

For the Gaussian source, reducing the attenuation does not modify the 
structure of the solution, but only increases its amplitude. In contrast, 
for the Dirac source, the wavefield in $\Bhyp$ is significantly modified 
as the attenuation is reduced.
At higher attenuation, only a single cone from the source is visible. 
As the attenuation decreases, however, multiple lines emerge, forming 
a grid-like pattern within $\Bhyp$. 
To determine whether in the cases at very low attenuation, e.g. 
with $\eta=10^{-1}$ or $10^{-3}$, the characteristic paths are 
space-filling is a scope of future work.
A clearer understanding will also require devising new discretization methods
to treat the mixed-type closed BVP, granted its well-posedness at zero attenuation.


\bibliographystyle{siamplain}
\bibliography{sections/bibliography}
\end{document}